%% file: ms.tex
\documentclass[9pt,twocolumn,twoside]{pnas-new}
\usepackage[utf8]{inputenc}
\usepackage[english]{babel}
\usepackage{amsmath,amssymb,amsthm,amscd}

\usepackage{subfiles} 
\usepackage[]{hyperref}

\usepackage{subcaption}
\usepackage{placeins}  

\usepackage{outlines}
\usepackage{xr}

\def\reals{\mathbb{R}}

\def\x{\mathbf{x}}
\def\catm{\mathcal{M}}

\def\MS{\Omega}
\def\C{\mathcal{C}}
\newcommand{\lt}{\ensuremath <}

\usepackage{graphicx}
\graphicspath{{./figures_Main/}}
\usepackage{caption}
\usepackage{subcaption}
\usepackage{amsmath}
\usepackage{placeins}  
\usepackage{outlines}  
\templatetype{pnasresearcharticle} 

\newtheorem{theorem}{Theorem}
\newtheorem{lemma}{Lemma}
\newtheorem{corollary}{Corollary}

\newtheorem{proposition}{Proposition}

\newtheorem{termin}{Terminology}

\makeatletter
\renewcommand\appendix{\par
	\setcounter{section}{0}%
	\gdef\thesection{\@Alph\c@section}
	\renewcommand{\theHsection}{A\arabic{section}}
}
\newcommand\unappendix{\par
	\setcounter{section}{0}%
	\gdef\thesection{\@arabic\c@section}
	\renewcommand{\theHsection}{E\arabic{section}}
}
\makeatother

\setboolean{displaywatermark}{false}
\date{ \today}

\begin{document}





\subfile{intro5EDS}


\subfile{examples}



\clearpage
\newpage

\begin{flushleft}
\textbf{\LARGE Supplementary Information}
\end{flushleft}

\setcounter{equation}{0}
\setcounter{figure}{0}
\setcounter{table}{0}
\setcounter{page}{1}
\setcounter{section}{0}
\makeatletter
\renewcommand{\theequation}{S\arabic{equation}}
\renewcommand{\thefigure}{S\arabic{figure}}
\renewcommand{\bibnumfmt}[1]{[S#1]}
\renewcommand{\citenumfont}[1]{S#1}

In the following, we provide technical
details and various supplementary
illustrations to those provided in the
main paper that hereafter is referred as
${\bf I}$.

\babel@toc {english}{}
\contentsline {part}{I\hspace {1em}Mathematical concepts}{1}{part.1}%
\contentsline {section}{\numberline {1}Morse-Smale Systems}{1}{section.S1}%
\contentsline {subsection}{\numberline {A}Saddle-attractor configurations}{2}{subsection.1.1}%
\contentsline {section}{\numberline {2}Fold and cusp bifurcations}{2}{section.2}%
\contentsline {subsection}{\numberline {A}Center manifolds}{3}{subsection.2.1}%
\contentsline {section}{\numberline {3}Catastrophe Manifold}{3}{section.3}%
\contentsline {subsection}{\numberline {A}Fold curves and bifurcation sets}{4}{subsection.3.1}%
\contentsline {subsection}{\numberline {B}Heteroclinic connections and flips}{4}{subsection.3.2}%
\contentsline {section}{\numberline {4}Counting cusps on fold curves \& circles}{5}{section.4}%
\contentsline {subsection}{\numberline {A}The fold orientation}{5}{subsection.4.1}%
\contentsline {subsection}{\numberline {B}Saddle \& near-fold bundles}{5}{subsection.4.2}%
\contentsline {subsection}{\numberline {C}Fold circle bundle}{5}{subsection.4.3}%
\contentsline {subsection}{\numberline {D}Conditions for a fold circle to be M\"obius}{5}{subsection.4.4}%
\contentsline {subsection}{\numberline {E}Cusps and bifurcation diagrams}{6}{subsection.4.5}%
\contentsline {section}{\numberline {5}Characterising 3-attractor MS components}{7}{section.5}%
\contentsline {subsection}{\numberline {A}Crossings of bifurcation curves}{7}{subsection.5.1}%
\contentsline {subsubsection}{\numberline {A.1}Crossings near cusps}{8}{subsubsection.5.1.1}%
\contentsline {subsection}{\numberline {B}Heteroclinic bifurcations (flips) and flip curves}{9}{subsection.5.2}%
\contentsline {subsubsection}{\numberline {B.1}Target saddles}{9}{subsubsection.5.2.1}%
\contentsline {subsubsection}{\numberline {B.2}Source saddles}{9}{subsubsection.5.2.2}%
\contentsline {subsubsection}{\numberline {B.3}Crossings sequences}{9}{subsubsection.5.2.3}%
\contentsline {subsection}{\numberline {C}Simplest/minimal MS components}{10}{subsection.5.3}%
\contentsline {section}{\numberline {6}Topological charge and the elliptic umbilic}{10}{section.6}%
\contentsline {subsection}{\numberline {A}The compact elliptic umbilic}{10}{subsection.6.1}%
\contentsline {subsection}{\numberline {B}Boundary conditions}{11}{subsection.6.2}%
\contentsline {subsection}{\numberline {C}Using flip curves}{11}{subsection.6.3}%
\contentsline {subsubsection}{\numberline {C.1}Flips ending on $\ell _A$, $\ell _B$ and $\ell _C$}{11}{subsubsection.6.3.1}%
\contentsline {subsubsection}{\numberline {C.2}Algebra of three-way flips}{11}{subsubsection.6.3.2}%
\contentsline {subsubsection}{\numberline {C.3}Flips at $\partial \Delta $ imply a M\"obius fold circle}{12}{subsubsection.6.3.3}%
\contentsline {subsubsection}{\numberline {C.4}Flip curves in $\mathcal {M}$}{12}{subsubsection.6.3.4}%
\contentsline {section}{\numberline {7}Decision structures}{14}{section.7}%
\contentsline {section}{\numberline {A}Existence of decision regions $R$}{16}{section.A1}%
\contentsline {subsection}{\numberline {A}Four attractor configurations}{17}{subsection.A1.1}%
\contentsline {section}{\numberline {B}No folds at cuspoidal joining attractors}{17}{section.A2}%
\contentsline {section}{\numberline {C}Proof that the connected components of $\mathcal {M}$ are disks}{17}{section.A3}%
\contentsline {subsection}{\numberline {A}The interior of a fold circle is a disk}{17}{subsection.A3.1}%
\contentsline {subsection}{\numberline {B}Proof of Prop.\ \ref {prop:disk}}{18}{subsection.A3.2}%
\contentsline {section}{\numberline {D}Bifurcation curves: Proof of Prop.\ \ref {prop:BsubC}}{18}{section.A4}%
\contentsline {section}{\numberline {E}Proof of Prop.\ \ref {prop:remove}}{18}{section.A5}%
\contentsline {section}{\numberline {F}Justification of Condition (*) for Theorem \ref {corollary:1}}{19}{section.A6}%
\contentsline {part}{II\hspace {1em}Examples}{20}{part.2}%
\contentsline {section}{\numberline {1}Three state system has elliptic-umbillic parameter space}{20}{section.E1}%
\contentsline {section}{\numberline {2}A flow defines a potential}{20}{section.E2}%
\contentsline {section}{\numberline {3}Patterning by lateral inhibition}{20}{section.E3}%
\contentsline {section}{\numberline {4}The Turing Model in potential form}{22}{section.E4}%
\contentsline {section}{\numberline {5}Spreading the pattern by a wave}{24}{section.E5}%
\contentsline {section}{\numberline {6}Morphogen models}{25}{section.E6}%
\contentsline {section}{\numberline {7}Adaptive systems}{26}{section.E7}%

\vspace{1cm}

\part{Mathematical concepts}
\subfile{SI_Maths_current.tex}

\clearpage  

\subfile{DecisionStructures}

\clearpage

\appendix
\subfile{Appendices_current.tex}
\unappendix

\clearpage  
\newpage
\part{Examples}

\subfile{examples_SI}


\end{document}

%% file: intro5EDS.tex
\title{Geometry of Gene Regulatory Dynamics}

\author[a]{David A. Rand}
\author[b,c]{Archishman Raju} 
\author[a,d]{Meritxell S\'aez}
\author[e]{Francis Corson}
\author[c,1]{Eric D. Siggia}

\affil[a] {Zeeman Institute for Systems Biology and Infectious Epidemiology Research, University of Warwick, Coventry, CV4 7AL, UK}
\affil[b] {Simons Centre for the Study of Living Machines, National Centre for Biological Sciences, Tata Institute of Fundamental Research, Bangalore, India}
\affil[c]{Center for Studies in Physics and Biology, Rockefeller University, New York, NY 10065}
\affil[d] {The Francis Crick Institute, 1 Midland Road, London, NW1 1AT, UK}
\affil[e]{Laboratoire de Physique de l'Ecole Normale Sup\'erieure, CNRS, ENS, Universit\'e PSL, Sorbonne Universit\'e, Universit\'e de Paris, 75005 Paris, France}

\leadauthor{Rand} 

\significancestatement{ Genetic screens have enumerated the genes that control the process of self-organization that converts a featureless fertilized egg into an embryo. Even the elementary steps may involve ten genes so models that attempt to represent each gene contain a plethora of unmeasured parameters. Mathematics has largely categorized the types of solutions that can arise from the equations governing gene networks. These representations are well suited to modern time-lapse imaging where a limited number of genetic markers are followed in time. Models with minimal parameters that focus on the nonlinear regime from inception to pattern maturation, simplify data fitting, and provide an intuitive and transparent representation for the dynamics of development. 
}

\authorcontributions{All authors contributed to the calculations and composition of the paper.}
\authordeclaration{Please declare any conflict of interest here.}
\correspondingauthor{\textsuperscript{1}To whom correspondence should be
addressed. E-mail: siggiae\@rockefeller.edu}

\keywords{Morse-Smale $|$ Waddington landscape $|$ bifurcation $|$ Turing model $|$ gene network} 

\begin{abstract}

Embryonic development leads to the reproducible and ordered appearance of complexity from egg to adult. 
The successive differentiation of different cell types, that elaborates this complexity, result from the activity of gene networks and was likened by Waddington to a flow through a  landscape in which valleys represent alternative fates. 
Geometric methods allow the formal representation of such landscapes and codify the types of behaviors that result from systems of differential equations. 
Results from Smale and coworkers imply that systems encompassing gene network models can be represented as potential gradients with a Riemann metric, justifying the Waddington metaphor. 
Here, we extend this representation to include parameter dependence and enumerate all 3-way cellular decisions realisable by tuning at most two parameters, which can be generalized to include spatial coordinates in a tissue. 
All diagrams of cell states vs model parameters are thereby enumerated.
We unify a number of standard models for spatial pattern formation by expressing them in potential form. Turing systems appear non-potential yet in suitable variables the dynamics are low dimensional, potential, and a time independent embedding recovers the biological variables. Lateral inhibition is described by a saddle point with many unstable directions.  A model for the patterning of the Drosophila eye appears as relaxation in a bistable potential. Geometric reasoning provides intuitive dynamic models for development that are well adapted to fit time-lapse data. 

\end{abstract}

\dates{This manuscript was compiled on \today}

\maketitle
\thispagestyle{firststyle}
\ifthenelse{\boolean{shortarticle}}{\ifthenelse{\boolean{singlecolumn}} {\abscontentformatted}{\abscontent}}{}

\section{Introduction}

Much of classical physics, chemistry,
and by extension biology is represented
by differential equations.  Particularly
in biology, the precise form of these
equations and their parameters is poorly
known.  In addition, the long time
behavior of these systems is typically
opaque and one resorts to a case by case
numerical solution.  Mathematics
sidesteps the question of solving a
particular equation by using geometric
methods to enumerate the discrete types
of solutions.  Then qualitative features
of the biology may select a type and
mathematics in some instances then
supplies a minimal parameterization.

A most striking aspect of development is
the extreme fidelity of the output in
response to insults, which nicely aligns
with the mathematical notion of
genericity, i.e., simply by insisting
all nearby systems are equivalent,
eliminates exceptional cases that require parameter tuning
and are unlikely to be relevant to biology.
The assumption of genericity strongly constrains
the dynamical behavior.
Although the dynamics appropriate for
development are simple, expressing a
model in geometric language classifies
its essential features.  In particular
Smale and his school have shown that
models that plausibly encompass
development all admit a potential that
decreases as the system evolves.  This
potential derives from a graph
representing which critical or decision
points can flow to which others.  Thus
the potential is implicit in, and derived from, the dynamics.  Then
the differential equation model can be
written effectively as a metric times the gradient
of this potential all defined on a Riemann manifold.  Thus the Waddington
metaphor for development, as flow down a landscape with bifurcations signifying decisions between alternative cell
fates, is literally true. 

Geometric reasoning will typically
reduce a genetic model to a few
variables per cell, and thus the
explicit connection of variables to
genes is lost.  But the typical network
for cell communication involves tens of
genes, and most gene centric models include only a subset of these, thus are to
some degree phenomenological.  Geometry
takes this reduction to the extreme, and
will deliver the minimum number of
parameters that can not be eliminated by
variable redefinitions, and the minimal
phase space for representing the
dynamics surrounding a cellular
decision.

The results of Smale and colleagues do not apply through bifurcation points which is essential for applications.
Thus we extend their results to encompass bifurcations, both local and global. 
We limit ourselves to at most
three possible states per
cell to capture a progenitor cell giving rise to two alternative fates.
Within this class we enumerate
all ways the cellular states can be
arranged by two parameters.  
Our topological arguments yield global results in parameter and state spaces that link
together locally described bifurcations.


Geometric methods are well suited to fitting
time lapse data obtained when cells are poised among competing fates.
A minimal parameterized topographic model with the correct geometry 
is essential to extract model parameters from the cells as they
are transitioning. Merely enumerating terminal fates looses this information.
When dealing with
many cells, focusing on the saddle
points that represent cellular decisions
quantifies similarities among all models
that use inhibition to define a pattern.
Turing showed how chemical reactions plus diffusion
can generate patterns in an otherwise uniform system.
In geometric terms the Turing instability is represented as
a saddle point and
the trajectory from it to the terminal pattern can be
represented by gradient dynamics in the
tangent space of the saddle composed
with a time independent sigmoid, thus
revealing similarities to models of lateral inhibition
by long-range contacts or
diffusing factors.  Geometric methods
are optimal for bridging the time from
the initiation of a pattern to its
saturation, and thus extracting the
essential dynamics of cell specification.

\section{The mathematics of gene network models}

A gene network model defines a differential equation that describes the changes on the system with time. 
The time
integration of a differential equation
or, equivalently a vector field defines a
flow $x\rightarrow \phi^t(x)$ which tells us the 
state $\phi^t(x)$ at time $t$ if the
initial state at time $t=0$ was $x$.  
Off of the bifurcation set,  the
restpoints $p$ of these systems have
well-defined stable and unstable
manifolds i.e.\ (stable manifold)
$W^s(p)=\{ x | \phi^t(x)\to p \text{ as
} t\to\infty\}$ and (unstable manifold)
$W^u(p)=\{ x | \phi^t(x)\to p \text{ as
} t\to -\infty\}$, Fig.\ref{fig:one}.
Moreover, these restpoints $p$ are of three types:
attractors ($W^u(p)$ empty, they attract
all nearby points), saddles ($W^s(p)$ \&
$W^u(p)$ not empty) and repellors
($W^s(p)$ empty).  We say a saddle has
index $\lambda$ if its unstable manifold
has dimension $\lambda$.

If a system is not on the bifurcation
set then small perturbations of it do not change
its qualitative form (i.e.\ 
there is a homeomorphism of phase space sending
the trajectories of one system onto the other).
We say that such a system is \emph{structurally stable}.


Relatively simple sufficient conditions
have been found \cite{Smale62} for
systems with a finite number of
restpoints and periodic orbits to be
structurally stable and these are called
\emph{Morse-Smale (MS) systems}  \cite{GuckHolmes}.
These assumptions exclude chaos but
include models of development.  In what
follows, we exclude periodic orbits and
insist that trajectories can not escape
to infinity.  This is entirely plausible
for development and makes the point at
infinity a repeller which is needed for
certain topological arguments.  In our
applications the phase space $M$ is the
$n$-dimensional Euclidean space
$\reals^n$


\subsection*{The downhill structure of generic
landscapes} 
\begin{figure}[H]
\centering
\includegraphics[width=8cm]{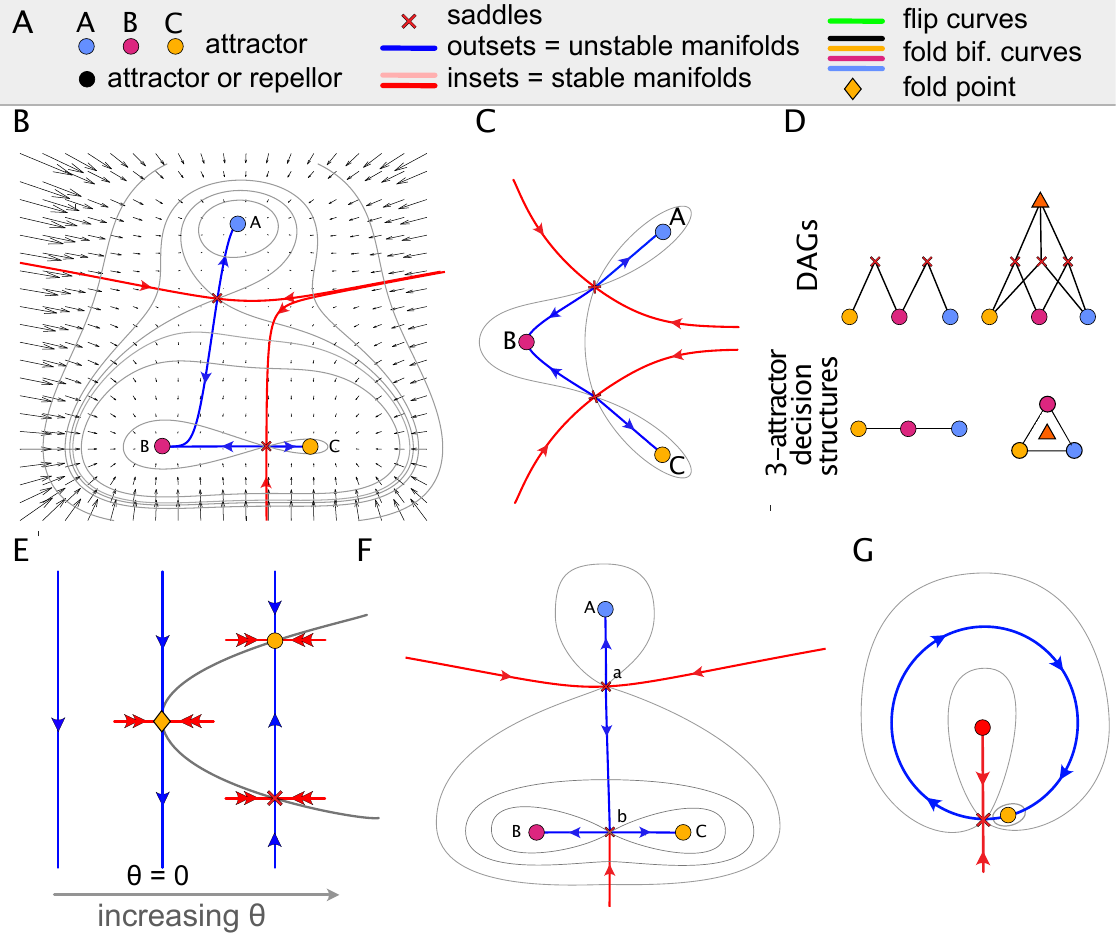}
\caption{Representations of Morse-Smale systems.
{\bf A.} The symbol set used consistently to describe elements of 
the phase space or parameter space (flip and fold bifurcation curves).
%
{\bf B.} A Morse-Smale
system with three attractors and two
saddles showing the stable (red) and unstable (blue)
manifolds of the saddles. Contours
of a potential are shown in grey.  In
this system the unstable manifolds meet
at the attractor $B$ in a cusp like shape.
{\bf C.} In this Morse-Smale system
which also has 3 attractors, the unstable
manifolds of the saddles make up a
smooth curve.  
{\bf D.}
(Top row) Two examples of directed acyclic graphs (DAG) corresponding
to the system
in {\bf B} (left) and the
compactified elliptic umbilic of Fig.\ref{fig:globalbifns}D
(right).
(Bottom row) The decision structures
associated with the DAGs above them.  The
filled circles represent the attractors
and a connection between attractor $A$ and
$B$ means that a cell whose state sits at
$A$ (resp.\ $B$) can transition to $B$ (resp.\
$A$) via a saddle-node bifurcation that
destroys $A$ (resp.\ $B$).  Thus the
connections characterize the escape
routes and possible decisions.  The
connections also correspond to the index
1 saddles in the system that connect $A$
and $B$.  Some escape routes wrap around an
index 2 saddle as shown and indicated by a triangle.  
To minimize the numbers of decision diagrams we do not distinguish 
cases where multiple saddles connect the same two fixed points (see SI).
%
{\bf E,F.} The two simplest bifurcations:
the local saddle-node and the global heteroclinic flip.
These are the main events underlying decision-making
in our dynamical systems.
{\bf E.} A saddle-node or fold
bifurcation.  As $\theta$
increases through $0$
a saddle and an attractor are
born which then separate
with a distance of order
$\sqrt{\theta}$. 
{\bf F.} The
configuration shown in {\bf A} can flip
to one where the saddle $a$ is
connected to $C$ instead of $B$ via a
{\it heteroclinic flip}.  To do this it
passes through the intermediate state
shown where there is a heteroclinic
connection in which the unstable
manifold of the saddle $a$ connects to
the saddle at $b$.  
%
{\bf G}.  An example
of a compact landscape involving a
repellor.  We have taken a case
where the attractor is close to the
saddle to illustrate that the
attractor can move around the circular
unstable manifold and collide undergoing
a saddle-node bifurcation and turning
the unstable manifold into a limit
cycle.  This is called a SNIC
bifurcation.  Although it is important
to be aware of such bifurcations we do
not consider them anymore since the
existence of a limit cycle moves us out
of the gradient-like restpoint-only
systems.  
}\label{fig:one}
\end{figure}

The Waddington analogy of development to
flow in a topography can be formalized
mathematically.  
All MS systems possess a Liapunov function
(a.k.a.\ potential function) defined on the phase space 
which decreases along trajectories and for which
restpoints (and periodic orbits) are critical points.  
This formalizes the notion of height in a topography.
We call such
dynamical systems \emph{gradient-like}.
It is commonly thought that this is
enough to specify dynamics but the
Liapunov function is not enough to
determine where a cell will go when it
escapes an attractor, ie a small perturbation to Fig.\ref{fig:one}F will direct the unstable manifold from the upper saddle to either state B or C.
Therefore extra information about the dynamics is
necessary.

There are two essentially equivalent
ways to specify the missing information.  
One way is to supply the stable and unstable manifolds
of the saddle points. The unstable manifolds of the 
index one saddles for instance describe the 
transition routes between attractors.

The other way to augment the Liapunov function,
is to note that
MS systems are nearly \emph{gradient systems}.
In a gradient system a
potential $f$ together with a Riemannian
metric $g_{ij}$ completely defines the
dynamical system:
\begin{equation}
\label{eqn:gradient}
	\dot{x}_i =-\sum_j g^{ij}\frac{\partial f}{\partial x_j} 
\end{equation}
where $(g^{ij})=(g_{ij})^{-1}$. 
The restpoints of such a vector field
are the critical points of $f$
($g_{ij}$ is positive definite), and
the metric rotates and stretches the
potential gradient so
it coincides with the vector field.
Gradient-like MS systems are also nearly
gradient and the difference is just at
the restpoints.  By making adjustments
of the restpoints in arbitrarily small
neighbourhoods one can convert the
system into a topologically equivalent
gradient system
(\cite{NewhousePeixoto76}, SI App.\ A).
The need to adjust the restpoints is
because at such points the gradient
system has a special structure (e.g.\
the eigenvalues are real), while the MS
system allows complex eigenvalues with
nonzero real parts.
In applications, the model with its metric abstractly represents how signals distort the landscape and
direct cells to the available fates or attractors.  

\subsection*{A directed acyclic graph organizes cellular decisions}

Now consider a MS system and
 the graph $G$ whose vertices
correspond to the restpoints and where
two vertices $\beta$ and $\beta^\prime$
are connected with a (downward)
directional edge (denoted $\beta\succ
\beta^\prime$) if there is a trajectory
going from $\beta$ to
$\beta^\prime$, Fig.\ref{fig:one}D.
Then, the following non-trivial results
hold \cite{Smale60}: (i) it is never true that
$\beta\succ \beta$; (ii) if $\beta\succ
\beta^\prime$
 and $\beta^\prime\succ
 \beta^{\prime\prime}$ then $\beta\succ
 \beta^{\prime\prime}$
(thus $\succ$ is a partial ordering)
(iii) if $\beta\succ \beta^\prime$ then
$\dim W^s(\beta) \leq \dim
W^s(\beta^\prime)$.  The conditions (i-ii) that
exclude $\beta\succ \beta^\prime$ and
$\beta^\prime\succ \beta$ are crucial and
eliminate
heteroclinic cycles (these are not
generic).  By (iii) we can naturally
attach levels to the nodes of $G$ using
the \emph{index} $d=\dim W^u(\beta)$.
The attractors are at the bottom level
($d=0$), the repellors at the top and
the index $d$ saddles at level $d$.  The
graph $G$ is a directed acyclic graph
which we shall refer to as the {\it
DAG}.

Using this we can allocate heights $h_i$
to the restpoints which for consistency
must satisfy that $h_i > h_j$ whenever
$\beta_i\succ \beta_j$ and then find a
Liapunov function $h$ that takes these values at
restpoints. Moreover, this function
can be taken to be a Morse function
\cite{Meyer68}.  This means
that near each restpoint $\beta_i$ there is a
coordinate system $x_1,\ldots , x_n$
such that in these coordinates  $ h = h_i+
\sum \pm x_i^2$ with the number of
minuses equalling the index of $\beta_i$.

This height function is unique in the
following sense \cite{Meyer68}.  If
$\hat{h}$ is another Liapunov function
such that the heights
$h_i^\prime=\hat{h}(\beta_i)$ are
ordered in the same way as those of $h$,
then $h$ and $\hat{h}$ are qualitatively
equivalent in the sense that there are
homeomorphisms, $\phi$ of the phase
space and $\psi$ of the reals $\reals$
such that $\hat{h}(x) = \psi
(h(\phi(x)))$.


As we will see below a number
of advantages flow from these results.
Since motion is always downhill 
it gives a hierarchical structure 
to the dynamics and easier understanding
of the eventual fates. Secondly, there is
a powerful classification theory for the
bifurcations of such systems that is aided by
the existence of the potential, and, finally, 
it allows a better understanding of how
complexity of such systems can be built up.

\subsection*{Parameterised landscapes}
Our dynamical systems depend upon
parameters that will be changed by the
signals received by the cell.  Changes
in the parameters may cause bifurcations
where the qualitative nature of the
dynamics undergoes a discontinuous
change and these lead to transitions
between cell states.

For a given number of attractors or
saddles there are a large number of
topologies of landscape potentials.  
However, if we
define a decision as the eventual
attractor $A$ reached after the state is
freed from its initial attractor $B$
either by a bifurcation destroying $B$
or by stochastically escaping the basin
of $B$ then we can associate to a
DAG a unique 
connected simple graph that we call 
\emph{decision structure} 
(see Fig.\ref{fig:one}D) which encodes the possible decisions. 
The number
of these is much smaller than the number of landscapes.  All
three and four-attractor decision structures
are shown in Fig.\ \ref{fig:one}D and
the SI Sec I.7 respectively.

For applications to biology, we consider situations where a precursor cell
decides between two alternative fates.
Thus we classify decisions among three or fewer
states that depend on two or fewer
parameters.  The fold bifurcations, Fig.\ref{fig:one}E, divide
the parameter space up into components
in which the restpoints vary smoothly
with parameters. We call
these \emph{MS-components}.

In what follows, we leverage the continuity of MS dynamics within a MS component
to arrive at global representation of the parameter space. 
The component boundaries consist of smooth fold bifurcation segments 
that meet in special points that we can
enumerate under assumptions of
genericity. 
The sequence of these
points around the boundary of the
MS-component classifies the component.
This classification for gradient-like
MS systems is global and,
not only tells us what to expect in
detailed mechanistic models, but also
provides a class of archetypal models
that can be parameterised and compared to
experimental data 
\cite{Corson2012,camacho2021quantifying,Saez21} 
as described below.

\subsection*{Universality and normal forms}

\begin{figure}[H]
\centering
\includegraphics[width=.9\linewidth]{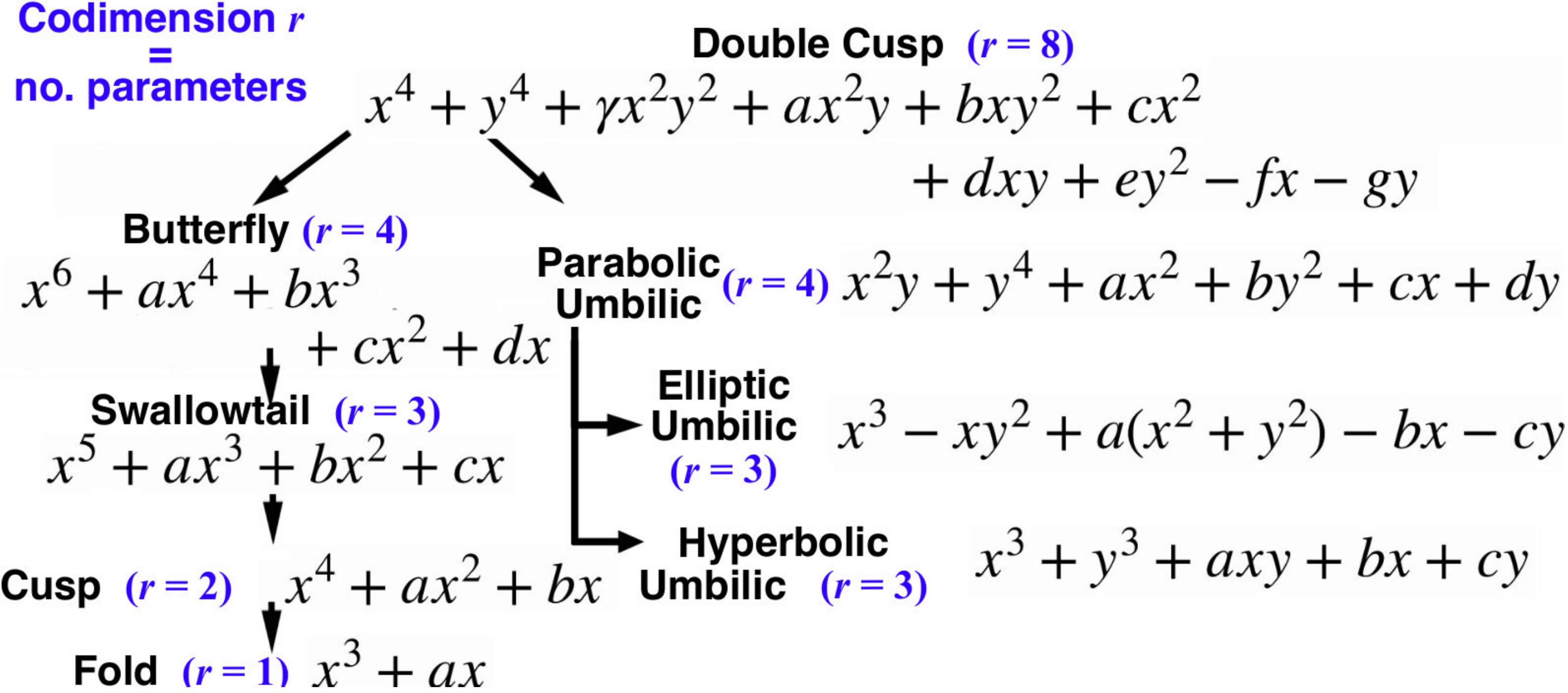}
\caption{The hierarchy of elementary
catastrophes obtained by coupling two bistable systems.
These describe all generic local
bifurcations of potentials depending on
$r \leq 4$ parameters.  If $h_\theta$ is
such a generic family then there are
coordinates $(x_1,\ldots ,x_n)$ on phase
space such that $h_\theta = f_\theta (y)
+\sum_{\rho+1 }^n \epsilon_i x_i^2$
where $f_\theta (y)$ represents one of
the elementary catastrophes in the
figure, $\rho = 1\ or\ 2$ and the
remaining phase space variables can be
reduced to a diagonal quadratic form
where $\epsilon_i = \pm 1$.  
Several of these polynomials are prototypes for the parameterized landscapes below
For a precise statement see
\cite{zeeman1976c}.
}
\label{fig:catlist}
\end{figure}

It is very useful to contrast our
proposed construction of parameterized
landscapes with the ideas of a universal
unfolding and normal forms that are
derived from Thom's theorem
\cite{Thom69,zeeman1976c} that gave rise
to Catastrophe Theory
(Fig.\ \ref{fig:catlist}). This subject
assumes strictly gradient dynamics.  It
then classifies, irrespective of the
dimension of state space, the generic
local bifurcations in parameterised
dynamical systems where the number $r$
of parameters $\theta$  (or codimension) is $\leq 5$ ,Fig.\ \ref{fig:catlist}.  By
smooth variable changes any function
near the bifurcation can be reduced to a
normal form polynomial in one or two
dimensions ,Fig.\ \ref{fig:catlist}.  The dynamics in the other
dimensions is just contraction onto the
one or two effective dimensions, thus
proving strong dimensional reduction.
The parameters in the normal form are
the minimum necessary to represent the
bifurcation to within variable changes.
On the other hand, in parallel with our
discussion above about needing to know
more than the Liapunov function, it does
not classify the global bifurcations in
such systems arising from heteroclinic
connections
\cite{guckenheimer1973,Khesin90,GuckHolmes}
which need to be handled separately.
Moreover, the results are local in that
they only apply close to a bifurcation
point.

We aim for results
that are global in phase
and parameter space and apply to
generic parameterized families with a finite number of restpoints and no periodic orbits.
Nevertheless the
catastrophes of higher codimension are
biologically relevant because even if
the parameters that we can manipulate to control fates 
(morphogens) are fewer in number than the codimension,
two-dimensional cuts through complete parameter space will
occur in our enumeration.  Figure
\ref{fig:globalbifns}A occurs as part of
the butterfly (codim.\ $r=4$), and Fig.
\ref{fig:globalbifns}D is essentially
related to the elliptic
umbillic (codim.\ $r=3$).

%

In the following discussion we focus on generic
systems and bifurcations and do not
treat non-generic situations that occur because
the system is restricted by, for
example, invariant boundaries (e.g.\ the
Lotka-Volterra systems) or symmetries.

\subsection*{Building up complexity:
the simplest bifurcations} 

Any gradient-like MS system can be built
up by using just 
three
simple bifurcation types
\cite{Newhouse76}. Only two of these
are relevant to our discussion:
the
\emph{fold} (Fig.\ \ref{fig:one}E), and
the simplest global bifurcation, the
\emph{heteroclinic flip} (Fig.\
\ref{fig:one}F).  As we will see, these
bifurcations are particularly important
for development and using them provides
a powerful method to build complexity
that is available for evolution to work
on (SI Sec I.2).

The \emph{fold} (a.k.a.\ the
\emph{saddle-node}) bifurcation results in the appearance 
of an attracting or repelling restpoint and an
index 1 saddle as a parameter $\theta$ is changed
(supercritical case) or the disappearance of the same (subcritical case).
More generally a fold results in the creation/destruction of a pair of
saddles of index $i$ and $i+1$.
The fold is local in that all the action takes place around a
point in parameter and phase space.


When a heteroclinic flip occurs, the
outset of a saddle flips between two
different attractors by passing through
a state where the unstable manifold of
one saddle is the inset of the other
i.e.\ we have a heteroclinic
connection  (Fig.\ref{fig:one}F).  

Figure \ref{fig:2} shows the generic
local ways we can build a
system with three attractors starting
from a bistable one within a two
dimensional parameter space.
Specifically we show all the ways that
folds (codimension-1) defined by a
smooth curve in the parameter space can
intersect in codimension-2 cusps or
crossings Fig.\ref{fig:2}C-F.  Then
Fig.\ \ref{fig:2}G-H shows the two
ways in which a curve of heteroclinic
flips can terminate in a fold bifurcation curve,
defining a second class of codimension-2
points in parameter space.  From these
locally described bifurcation sets, we
can construct all 2d phase diagrams of
MS systems.  (Catastrophe theory
enumerates higher codimension points,
but all the codimension-2 points are
contained in our list.)


\subsection*{Dimensional reduction characterizes transitions among landscapes}

\begin{figure}[H]
\centering
\includegraphics[width=\linewidth]{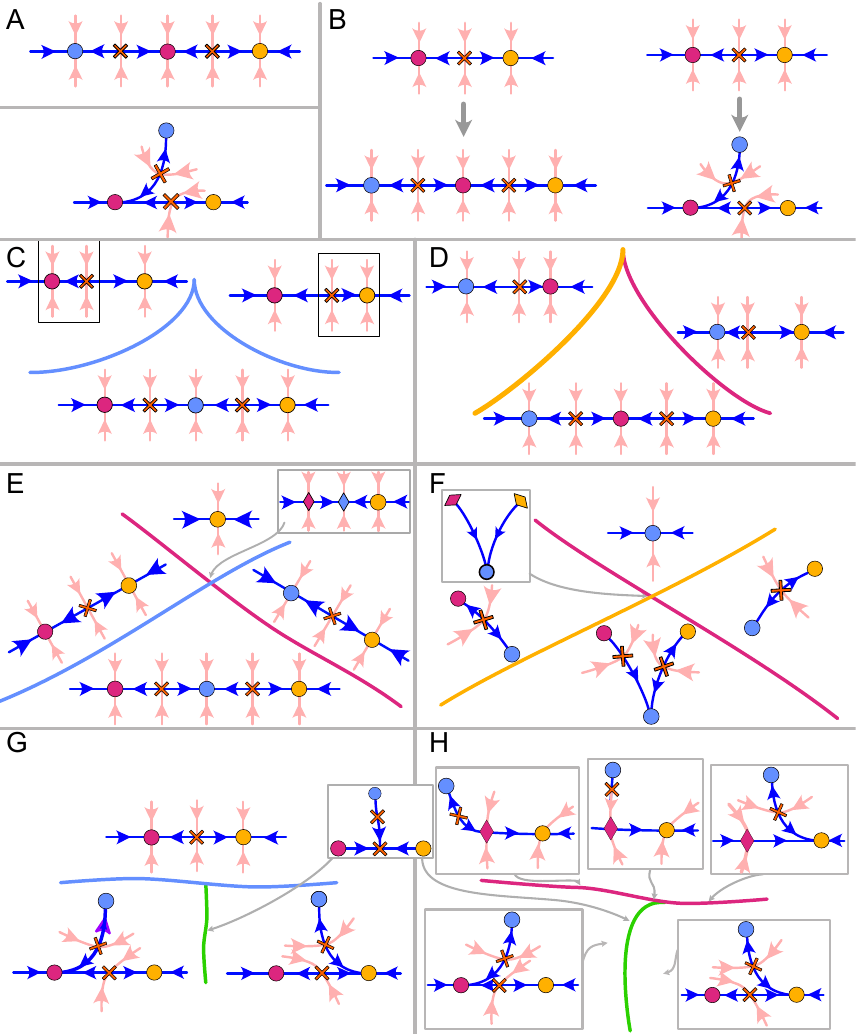}
\caption{{\bf Building complexity from
bistable landscapes using one and
two-dimensional perturbations.} 
Symbols follow the key in Fig.\ref{fig:one}A.
(A)
We distinguish between these two
3-attractor systems even though they are
topologically conjugate because in applications the allowed transitions are different.  
The red attractor can not undergo a saddle-node bifurcation 
when the outsets of the saddles merge in a cusp. 
The attractors represent biological states that are not equivalent, as they would be under topological conjugacy. 
(Top) All the
restpoints lie on a smooth curve defined
by the outsets of the saddles.  (Bottom)
The outsets of the saddles meet at a
central sink in a cusp with a
common tangency. 
(B) Fold
bifurcations add an extra saddle-sink
pair to the bistable system.  The new
attractor is marked in blue. 
(C-H)
The allowed bifurcation sets for three
and fewer fixed points when two
parameters are varied.  The colored curves
denote fold bifurcations where the like colored attractor disappears.
(C)
Dual cusp.  The
new attractor in blue and saddle are
inserted in the middle.  For this to
occur the fold bifurcation curve must contain
at least one cusp point.  
(D)
Standard cusp.  A bistable system has a
third attractor added. The cusps in (C,D)
are defined by the discriminant of the 
same 3rd order polynomial representing 
either a source and two saddles or a saddle and two attractors (cf Fig.\ref{fig:catlist}).
(E,F)
Crossing points.  Two fold bifurcation curves
can intersect transversally creating a
region with a single attractor and a
region where there are three attractors.
(E). One of the two
bifurcation curves involves the central
attractor and then the attractors must
lie on a smooth curve. 
(F). In this case
the curves correspond to peripheral
attractors and then both smooth and
cusped unstable manifolds are possible.  
(G,H).
There can be a (codimension 2) point on
the fold bifurcation curve where the outset
of the top saddle undergoes a
heteroclinic flip as shown.  Generically
these heteroclinic flips occur on a
curve (green).  
(G). When the source saddle bifurcates away
the flip curve in green intersects the fold bifurcation curve transversely.
(H). When the target saddle is
destroyed in a fold bifurcation the flip curve
generically terminates in a cusp (see SI)  
}
\label{fig:3attractor}
\label{fig:2}
\end{figure}

Repellors and higher index saddles can complicate
the classification in Fig.\ref{fig:2} e.g., each index-1 saddle
can be replaced by two index-1 and one index-2 saddles.
This is
handled using the following result (SI Appendix A).
As we have seen above, the developmental
transitions and decisions are determined
by the disposition of the unstable
manifolds of the index 1 saddle points
relative to the attractors.  Thus a
first useful result (see SI) is that for
a MS system in our context one can
always find an attracting ball $B$ in
phase space with
smooth spherical boundary $\partial B$ that
contains all the $N$ attractors, $N-1$ index 1
saddles connected to them and also captures
almost all the trajectories. 
There may be choices of $B$ depending on which index 1 saddles
are included which in turn define which decision structures are
allowed within $B$.

Under these conditions,
we can find $B$ above so that
the stable manifolds of the index 1 saddles
transversally intersect the spherical boundary of $B$ in
disjoint spheres of dimension $\text{dim}B - 1$
and divide $B$ into regions each of which is
the basin of one of the attractors (see SI).
The stable manifolds of the saddles define the boundaries of the
basins of attraction for each attractor. Two attractors in
adjacent basins are joined by the unstable manifold of the
saddle. Thus the entire MS component is controlled by its boundary

As a corollary, for such systems with 3
attractors, there are two index 1
saddles that connect them and they have
a simple disposition: the unstable
manifolds of each saddle each connect to
two of the attractors and they join at
one of the attractors (called the
\emph{central attractor}).  It is fixed
at crossings and cusps Fig.\
\ref{fig:2}C-F, but interchanges at flip
points.  The other attractors are called
\emph{peripheral}.  The union of the
unstable manifolds is called the
\emph{connector}.  At the central
attractor, the unstable manifolds either
join smoothly or in a non-smooth
fashion.  If non-smooth and this
attractor is close to undergoing a fold
bifurcation then the join has a
cuspoidal shape as in Fig.\ \ref{fig:2}
(SI Sect.\ I1).  In either case the DAG
linking the attractors is shown in
Fig.\ref{fig:one}D (left).

\begin{figure}[h]
\centering
\includegraphics[width=\linewidth]{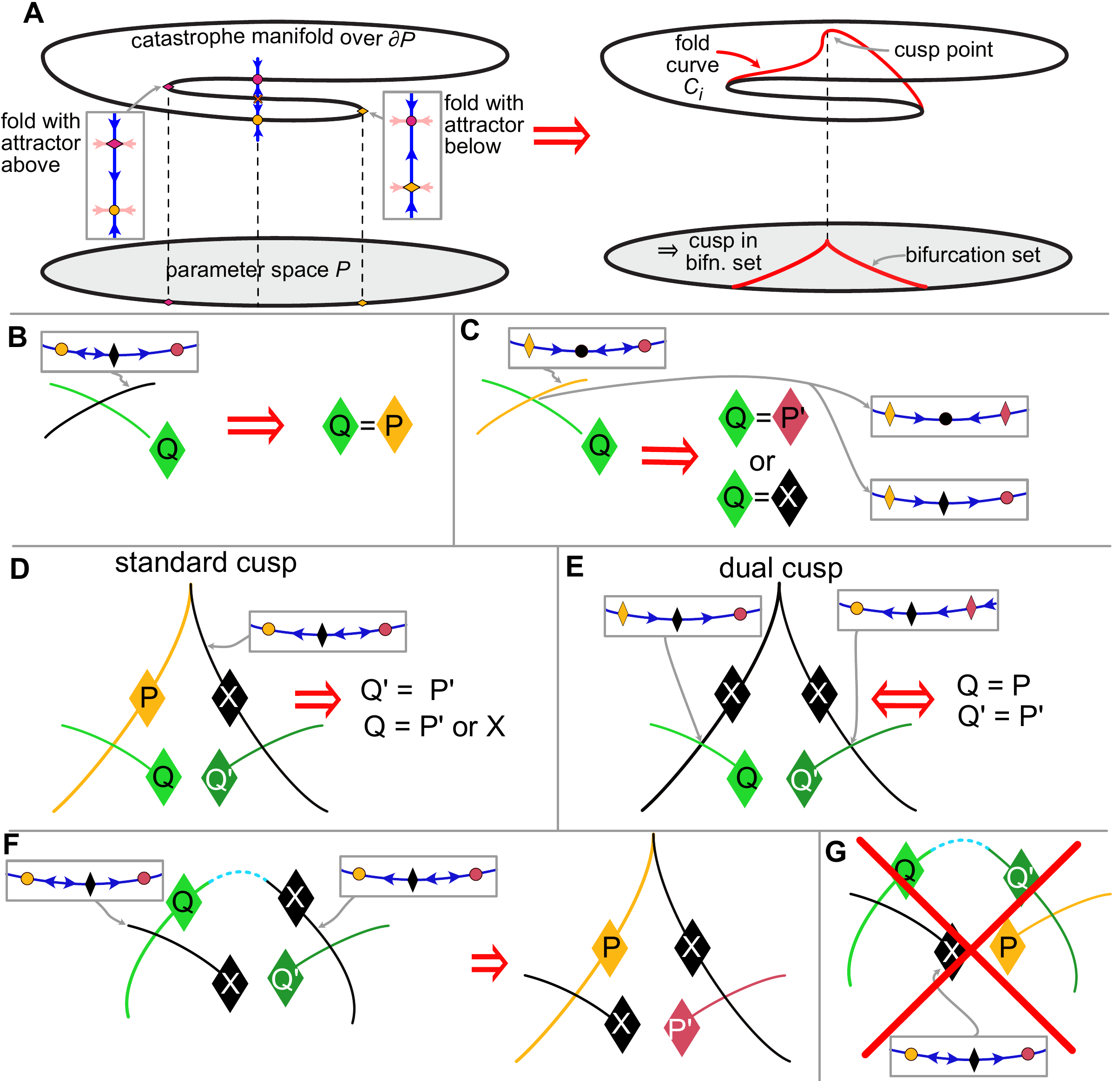}
\caption{Legend as in Figs.\ref{fig:one}A and \ref{fig:2}. 
{\bf A.} Over part of the boundary $\partial P$
of a 2-dimensional domain of parameter
space $P$ we assume we have an S-shaped
bistable section and that the system is
monostable over the rest of $\partial P$.
Then it follows that 
under very mild
conditions (SI Sect.\ I4E)
inside $P$ there
must be at least one cusp (see 
\cite{stewart1980catastrophe}
for $n=1$ case).  This is
because there must be a fold curve
joining the two folds shown and on this
fold curve the two folds shown have
opposite orientations.
(B-G) Local rules governing the
disposition of cusps and crossings in the
boundary of a 3-attractor MS component.
$X$ is always the central attractor, $P$, $P'$ peripheral.
{\bf B.} The orientation of the fold 
bifurcation involving the central 
attractor determines the possible transverse 
crossings of the fold bifurcation curve. 
On the black fold bifurcation curve the central attractor bifurcates with the saddle to its right (connected to red), 
thus the only allowed crossing is the bifurcation of the yellow attractor with its saddle.
{\bf C.} The peripheral attractor 
involved in the fold constrains 
the possible crossings of the fold bifurcation curve.
{\bf D.  Standard cusp.} Fold bifurcation curves 
crossing the cusp's branches satisfy 
the conditions in (B) and (C).
{\bf. E. Dual cusp.} Fold bifurcation curves 
crossing the cusp's branches satisfy 
the conditions in (B) and (C) 
and the crossings imply the existence 
of the cusp.
{\bf F.} The opposite orientations of the folds of 
the central attractor on the two fold bifurcation curves as shown imply the existence of a cusp.
{\bf G.} If $Q-Q'$ is a single fold bifurcation curve, certain consecutive crossings are not possible.
By (B) $Q=P$ but $P$ can not intersect itself, so if $Q'=X$ then the next intersection can only be with $P'$ via (D).
}
\label{fig:cuspexists}
\end{figure}

The universal dimension independent topological structure of the attractors, index-1 saddles, and their unstable manifolds
that comes from the existence of $B$ is a statement of dimensional reduction derived purely from the dynamics.
The restrictions on the orientation of the central attractor that derive from $B$ lead to the
results in Fig.\ \ref{fig:cuspexists} and severely constrain 
the bifurcations that can occur in the
boundary of a MS component.

\subsection*{Global bifurcation structures}

 \begin{figure}[h]
\centering
\includegraphics[width=\linewidth]{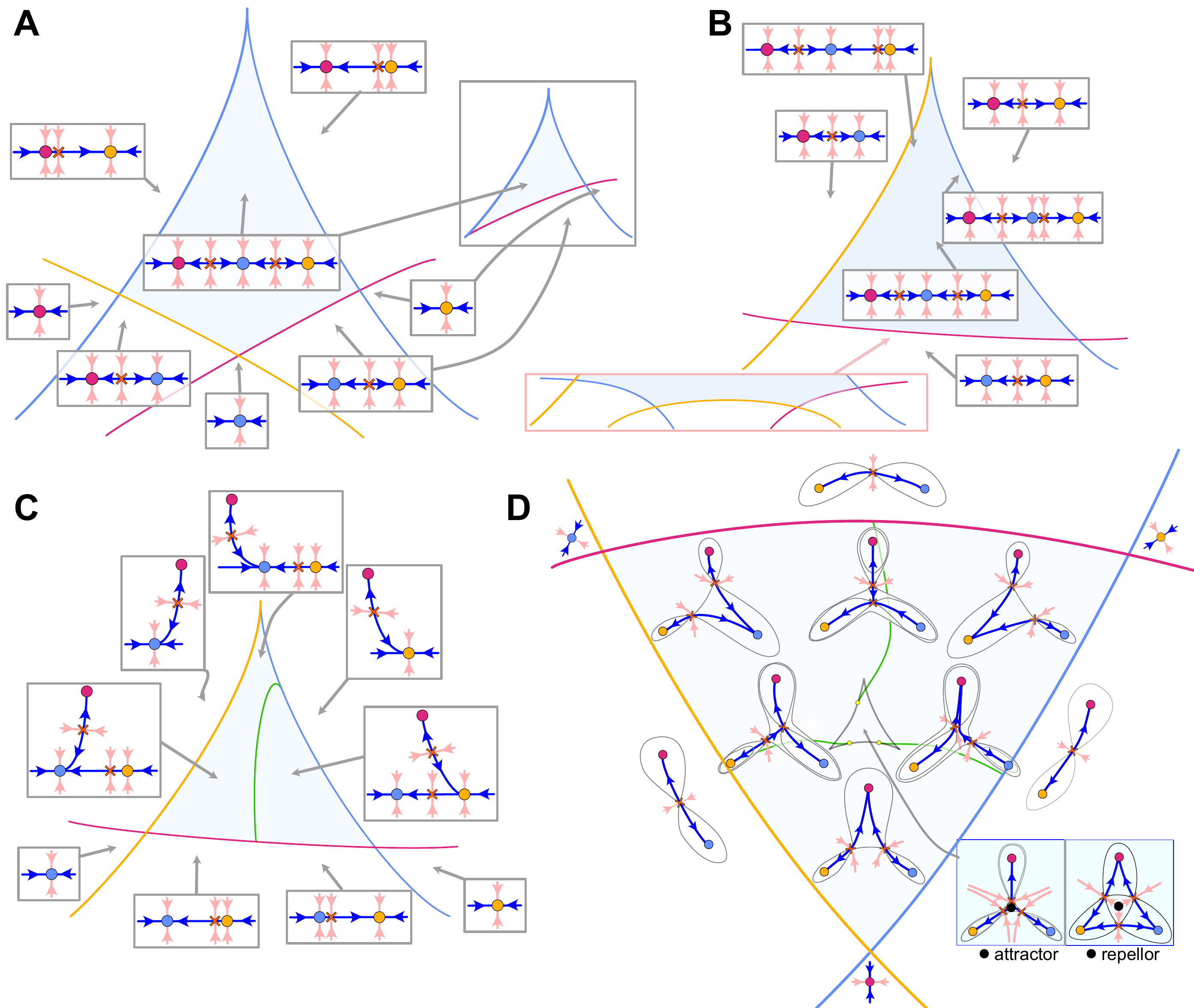}
\caption{
{\bf Minimal
3-attractor MS-components (light
blue).} Legend as in Fig.\ \ref{fig:one}A or \ref{fig:2}.
The color of a fold bifurcation curve
indicates the attractor that bifurcates
on it.  
(A) The minimal
MS-components containing a dual cusp.
The inset figure shows the minimal
system involving a dual and standard
cusp.  The MS components with $\leq 2$
attractors are a subset of this for the
dual cusp case.  
(B) Minimal
MS-component for a standard cusp. Inset in pink box shows
a simple alternative configuration replacing the
red curve at the bottom of the cusp.
(C) Ditto for a standard cusp and a flip curve.  
Note that for parameters near a
cusp point the connector among the three attractors will be smooth.
The intersection of the flip curve (green) is cusped when it
hits the fold for the central attractor and 
transverse when the peripheral, red, attractor disappears.
There is a second curve emanating from the cusp in the flip curve where the connector 
changes from smooth to cusped. 
(D) The compactified
elliptic umbilic.  Inside the MS component boundary
and outside the triangular hypocycloid
$B_\C $ there are three attractors and two
saddles.  This dynamical system can be
obtained by taking the gradient
dynamical system of the potential
associated with the potential $f_\theta
(x,y)=x^4+y^4
+x^3-2xy^2
	+ a (x^2+y^2)+\theta_1
	x+\theta_2 y$.
with respect to a generic Riemannian
metric.  The term $x^4+y^4$ is added to
the usual polynomial for Thom's elliptic
umbilic in order to make the dynamical
system compact.  It has the effect of
adding the three attractors and the outer fold bifurcation curves
to the 3 saddles. The sign of the parameter $a$
controls whether the central rest point is an
attractor or repellor.  By the MS
inequalities \cite{Smale60} this is a minimal
effect. 
}
\label{fig:globalbifns}
\end{figure}

We  want to understand the
structure of the bifurcations and
transitions as the parameters pass into
and out of  MS-components with
$\leq 3$ attractors.  We start by
discussing the compact ones $\MS$
i.e.\ ones which do not intersect the
boundary of the parameter domain.  To
get the corresponding characterisation
for noncompact versions one just opens
up the boundary of the MS component, $\partial \MS$ at a
crossing point.

For our discussion it is necessary to
consider the so-called \emph{catastrophe
manifold} (SI Sec I.3),
\begin{equation*}
	\catm=\{ (x,\theta ) | x \text{ is a
	restpoint of } \phi^t_\theta \}.
\end{equation*}
Then, a key point is that, if the
parameter space is $d$-dimensional and
the state-space $n$-dimensional, $\catm$
is generically a submanifold of
$\reals^n \times \reals^d$ having the
same dimension $d$ as the parameter
space \cite{arnold2013,zeeman1976c}. 
Thus, here it is a surface in
$n+2$ dimensions. For our systems the
topology of $\catm$ is trivial (it is
diffeomorphic to a disk, see SI Sec I.3) but the
way it sits in $\reals^n \times \reals^d$
and the non-trivial structure of the projection 
$\chi (x,\theta)\mapsto \theta$
from $\catm$ to parameter space
is hugely informative.

A point  $\x=(x,\theta )$ in $\catm$
corresponds to a restpoint $x$ of
the parameterised landscape at $\theta$
and we will be particularly interested
in the subset where $x$ is either a
fold or cusp point.  Generically this
consists of smooth curves $\C $ (called
\emph{fold curves}) which can either be
open and connecting to the boundary of
$\catm$ or they can be circles
(called \emph{fold circles}).  The part of the
bifurcation set corresponding to folds and cusps
consists of the sets 
$B_\C = \chi (\C ) $ in parameter space, 
called fold bifurcation curves,
where $\chi$ is the projection defined above.
Although $\C $ is smooth, $B_\C $ will
typically have cusp points as in Figs.\ref{fig:2}
and \ref{fig:globalbifns}. A smooth piece of a 
fold bifurcation curve is called a fold bifurcation segment. 
If $\C $ is
a fold circle then $B_\C $ is a closed
curve without self-intersections (SI Sec II.3). 

Now a key result that will lead to our
characterisation of MS-components is the
following: If $\x=(x,\theta)\in\catm$ is
on a fold curve $\C $ then through $x$
there is a center manifold \cite{GuckHolmes} and as $\x$
moves along $\C $ the tangent $\ell (\x)$
to this center manifold varies in a
smooth fashion.  At $x$ the time
evolution determines a definite positive
direction and orientation on $\ell (\x)$ (Fig.\
\ref{fig:cuspexists}B).  
Generically, as $\x$
moves along a fold curve $\C$ this
orientation flips only at cusps and 
so counting orientation flips gives the number
of cusps in $\C$
(SI Sec. I.4).

From this observation we readily
get all the results in Fig.\
\ref{fig:cuspexists} and these provide
tight constraints on what can happen in
the boundary of a 3-attractor MS-component ($\Omega$ in parameter space)
as we now explain.
By definition the attractors
vary smoothly in $\Omega$ with
the parameters and maintain their
identity.

\def\cp{\theta}
As one proceeds around $\partial \Omega$ there is
a sequence of codimension 2 points
$\cp_1\,\ldots \cp_N$ with $\cp_N=\cp_1$
as described in Fig.\ref{fig:2}, namely cusps,
dual cusps, crossings and the two
types of endpoints of
flip curves.  If we know these then we
know all the bifurcation structures
associated with $\Omega$ and $\partial \Omega$.  The
points $\cp_i$ and $\cp_{i+1}$ are
joined by a smooth fold bifurcation
segment $\ell_i$.  The bifurcations on
this segment involve one of the
attractors $X_i$ and we say that the 
attractor identity of 
$\ell_i$ is $X_i$.

Figure \ref{fig:globalbifns} contains the simplest MS-components
that involve all three attractors and are not decomposable into two attractor systems.
Panel A is the minimal system with a dual cusp (SI Sec. I.5).
In this case, if the parameters are inside $\Omega$ and
the state starts in the central
attractor (blue), then it stays there unless
the parameters cross the
curves through the cusp. Then it
transitions to either peripheral attractor depending upon the branch crossed.
For this reason we regard this as an \emph{all
or nothing landscape} because a
population of cells with their initial
state at the central attractor
would all transition to the
same new state upon bifurcation. An
intimately related minimal MS component
is shown in the Fig.\
\ref{fig:globalbifns}A(inset). This involves
a dual and a standard cusp joined together.

The simplest MS-components containing a
standard cusp are those in Fig.\
\ref{fig:globalbifns}B,C.  The
difference between them is the existence
of the flip curve (green) shown in Fig.\
\ref{fig:globalbifns}C.  The transitions
allowed using the landscape in Fig.\
\ref{fig:globalbifns}B are rather
limited.  If parameters start in $\Omega$ one can only
transition between a peripheral attractor and
the central one or visa-versa.  

The
transitions are richer with the flip
curve as it enables bifurcations
that can change the central attractor and
the escape routes available to cells.  
On this curve
there is a heteroclinic connection from
the saddle connected to the red peripheral attractor
(the source) to the other saddle (the target).
The flip curve joins tangentially to the side of the cusp 
where the central attractor (blue)
and target saddle disappear in a fold.
The flip curve must either terminate on the same fold bifurcation segment or
end on the fold bifurcation curve where the source saddle and red peripheral 
attractor disappear as in Fig.\ref{fig:globalbifns}C. 
If one considers populations of cells in each of the three attractors,
then on the boundaries of $\Omega$ transitions out of red are possible and between blue and yellow
but not into red. 
This landscape illustrates a \emph{downhill flexible choice}. 
Near the lower, red, fold bifurcation curve, a signal that repositioned the
flip curve would control the fractions that populated the other two attractors (SI Sec. 5). 



More complex structures involving
combination of these cusps and more
crossing points are possible but these
also are highly constrained as
described in the SI.  
These more complex versions are unlikely to occur in experiment since 
morphogens are apt to act monotonically on
cell fates, and the new complexity is just
multiple appearances of the same fold bifurcation.


Finally, we need to consider the case
where there are no cusps in the MS component boundary
$\partial \Omega$; then only crossing points are
involved.  The case of two crossing
points is trivial and is decomposable in
the sense that it is a 2-attractor
MS-component with a non-bifurcating
attractor added.  However the case of
three crossing points is already
extremely interesting.

This gives the configuration shown in
Fig.\ \ref{fig:globalbifns}D with three smooth fold bifurcation curves 
corresponding to each of the attractors.
The simplest example with such a MS component boundary
is what we call the 
\emph{compactified elliptic umbilic} (CEU)
(Fig.\ \ref{fig:globalbifns}D) and one can show that
any 3-attractor MS component with such
a boundary consisting of  
smooth fold bifurcation curves is
just a more complex version of the CEU.
It automatically has non-trivial monodromy
because if one traces around a simple closed curve $\gamma$ just inside 
of this boundary the two saddles are interchanged.
Together with an analysis of the
curves of heteroclinic connections
that must occur, this implies that 
$\gamma $ contains a bifurcation curve $\Omega_\C$
where $\C$ is a fold circle
containing an odd number of cusps in three groups
corresponding to the three cusps in the CEU (SI I.6).

We emphasize that these results follow from the properties of a gradient-like MS system, we do {\emph not}
assume a gradient system. 

Note that biology takes place around $\catm$. 
Thus in the vicinity of the cusp in Figs.\ref{fig:cuspexists}A, \ref{fig:globalbifns}B it is  possible to find a path in parameter space 
along which a population of cells in one attractor transitions smoothly to a second attractor without crossing a fold bifurcation curve. 
Thus although one is crossing a fold point in parameter space, if there are no cells at the attractor that disappears, it is invisible in an experiment.


%% file: examples.tex
\section{Models for fate specification and embryonic patterning}

The parameter spaces of a number of published 3-state models exemplify the categories in Fig.\ref{fig:globalbifns}.
A model for the mouse blastocyst places a weakly stable inner cell mass state between the primitive endoderm and epiblast \cite{de2016cell}. They find a dual cusp in the parameter space comprising the growth factor FGF that can be tuned externally, and one of many parameters defining the mutual inhibition between the two terminal states (their Fig. S4). The development of the vulva in {\it C.elegans} is a clear example of three cell fates and an elliptic umbillic, \cite{Corson2012}, and also fit as Fig.\ref{fig:globalbifns}C in \cite{camacho2021quantifying}. 
The specification of embryonic stem cells into neural and mesodermal fates was described by Fig.\ref{fig:globalbifns}A coupled to a sector of Fig.\ref{fig:globalbifns}D \cite{Saez21}. 

We next explicitly solve a gene network model for mutual inhibition among three states to show specifically that it conforms to Fig.\ref{fig:globalbifns}D.
We then recast in potential form, a number of models for patterning an array of cells, to illustrate how our results extend to interacting cells with details in SI Sec II.


\subsection{Three state system has elliptic-umbillic parameter space}
\begin{figure}[h]
\centering
	\includegraphics[width=.8\linewidth]{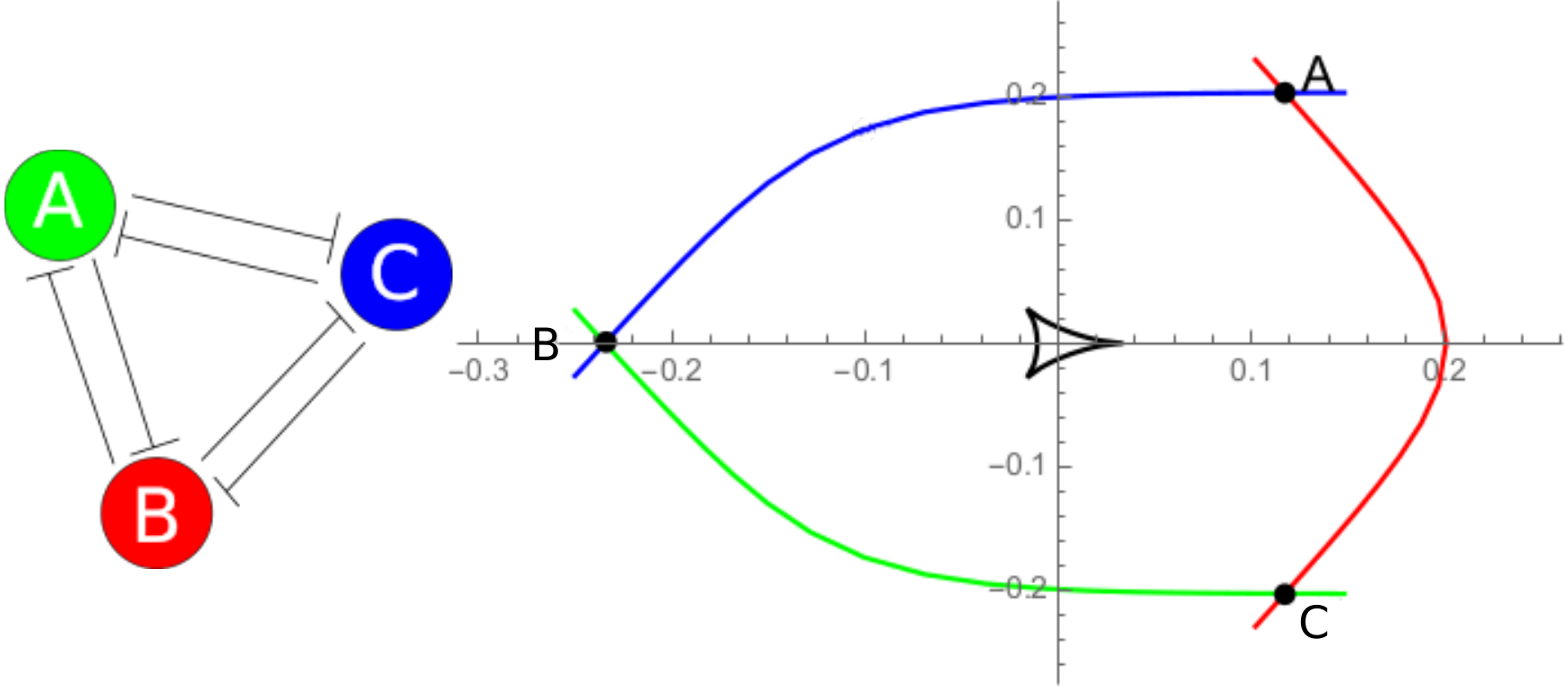}
\caption{A symmetric three cell or gene network with mutual repulsion and the elliptic umbillic conformation Fig.\ref{fig:globalbifns}D. The outer colored lines are the fold curves when one of the states disappears (three attractors become two). At the points A,B,C there is one stable restpoint and four with eigenvalues $\sim (-1,0,-1)$ (ie the intersection of two fold lines). The inner black contour consists of three fold lines and
inside it is an index 2 saddle, three index 1 saddles and three attractors.
}
\label{parameterspace2D}
\end{figure}

The discussion surrounding Fig.\ref{fig:globalbifns}, envisioned multiple states for one cell. But the examples would apply equally well to a model of three interacting cells each with a single binary state with mutual repression among the active forms. The mathematics is agnostic to the biological interpretation, Fig.\ref{parameterspace2D}. We chose to work around a point in parameter space with permutation symmetry in the three cells to simplify the parameter search, but the result persists so long as the three boundary curves have the topology shown, Fig.\ref{fig:globalbifns}D. 

\subsection{A flow defines a potential}
\begin{figure}[h]
\centering
	\includegraphics[scale=0.40]{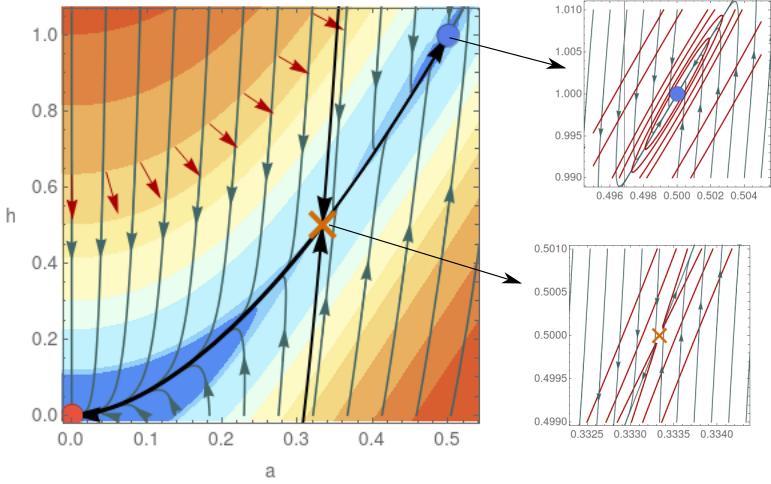}
\caption{The system has two stable fixed points and a saddle in between with flows shown in green. Contours of the potential, which is defined as the integral of the vector field along the trajectory to the stable fixed point, are shown going from low (blue) to high (red) potential. The red arrows on one of the contour lines show the flow from the potential gradient. Evidently, the inverse metric is needed to rotate the flow which derives from the potential into the actual flow. The construction does not work near the fixed points but the potential can be defined there locally and glued to the global potential. Plots of the contour lines in red and flow magnified near the upper fixed point and the saddle are also shown.
}
\label{flowtopotential}
\end{figure}

Consider a typical two variable activator-inhibitor system, $\dot a = f_a(a,h) - \nu a, \dot h = f_h(a) - h$, which are ubiquitous in signaling pathways. They pose an ostensible challenge to represent in potential form since we have $\partial_a f_h > 0$, the activator turns on $h$; while $\partial_h f_a < 0$, the inhibitor represses $a$. Thus the cross derivatives have opposite sign, so it's of interest to reduce these to potential form. The system has two stable fixed points and a saddle. There is a general construction which defines the potential as Euclidean distance 
near the fixed points, insets and outsets of saddles and then glues the various functions together~\cite{nitecki1975filtrations}. A more intuitive construction in 2D is given by defining the potential to be the vector field integrated along the trajectory to the stable fixed points and then making the two functions smooth across the inset to the saddle. 
This is shown in Figure~\ref{flowtopotential}. The inverse metric is defined as the sum of two projection operators, one which projects the potential stream lines along the direction of the flow and one which projects it in a direction tangent to the contour (SI Sec II.2). 
Together they define a proper metric that aligns the potential gradient to the flow, except near the restpoints where a separate construction is necessary and the two pieces then glued together.

\subsection{Patterning by lateral inhibition}
Inhibition is ubiquitous in biology and is manifest on the epidermis of many animals as a sparse array of hairs, feathers, or sensory bristles . How are these patterns established? A recent model for sensory bristle patterns in a fly \cite{corson2017self} nicely illustrates the organization of restpoints into the DAG of Fig.~\ref{fig:one}D, and the process of gluing local potential representations around critical points to model the complete system. 

Each cell $i$ is described by a single variable $0 \leq a_i \leq 1$ where the value $1$ corresponds to the neural fate that will develop into a bristle. In the terminal state the bristles are spaced by 4-5 cell diameters in all directions. Cells are initialized with $a_i \sim 0$ and are confined to a strip by inhibition from the surroundings so that the initial pattern formation is one dimensional along the strip and then extends laterally. We generalize Fig.\ref{parameterspace2D}, three cells that mutually inhibit, to a periodic ring of $N$ cells with long range inhibition represented as:   

\begin{equation}
\begin{aligned}
	\dot a_i &= \sigma(a_i-h_i) - a_i, \\
	h_i &= \sum_{j \ne i} K_{i,j} d(a_j)
\end{aligned}
\label{eqn:staticInhibition}
\end{equation}

\noindent
where $\sigma(a)$ and $d(a)$ are sigmoidal functions of their arguments varying between 0 and 1. The kernel $K_{i,j}$ is a Gaussian function of the distance between cells $i,j$, so the sum $h_i$ is the net inhibition seen by cell $i$.

\begin{figure}[h]
\centering
	\includegraphics[width=.99\linewidth]{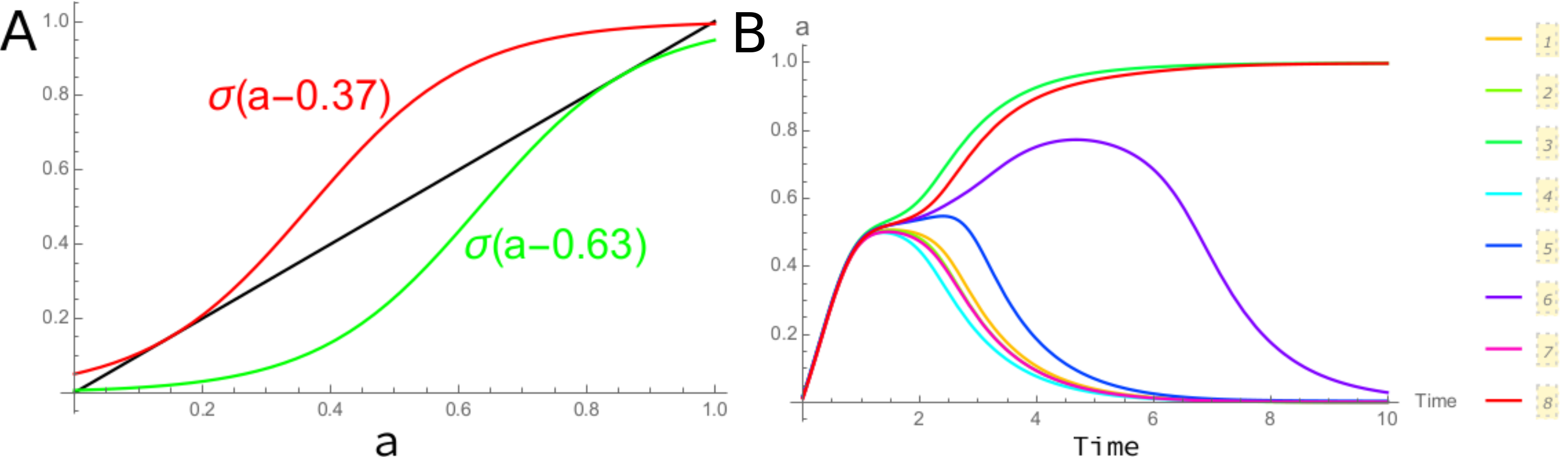}
\caption{Behavior of Eq.~\ref{eqn:staticInhibition}. (A) Values of inhibition $h$ for which the $\dot a$ equation has saddle-node bifurcations. When $h$ from neighboring cells is $\lt 0.37$ only the $a=1$ state is stable, bistability persists for $0.37 \lt h \lt 0.63$, and only the $a=0$ state exists for larger $h$. The diagonal is shown in black. (B) Eight cells on the circle with a kernel chosen to allow only two cells with $a=1$ at the end. The time to reach steady state can vary by 2x depending on whether a third cell hangs close to the saddle point as seen here vs Fig.\ref{staticInhibitionPot}
}
\label{staticInhibition}
\end{figure}

The qualitative behavior of the model can be understood from Fig.\ref{staticInhibition}a. When $h<0.37$ only $a=1$ is stable while for $h>0.63$ only $a=0$ is stable, with bistability in between. Initially $h=0$, all the $a_i$ grow from $0$, and cells inhibit their neighbors with the result that trajectories peal off from a common envelope. Which cells turn on is a sensitive function of initial conditions and noise, but the kernel will keep neural precursor cells well separated. 

The first `decision' is defined by the principal saddle point on the diagonal $a_i = a$, where $a$ varies from $\sim 0.9$ for $N=2$ cells to $\sim 0.51$ for $N$ large. At this point the number of unstable directions scales $\sim N$. For $N=8$ in Fig.~\ref{staticInhibition}B, there are 5 unstable directions with nearly identical eigenvalues, 3 stable ones and the Jacobian matrix is symmetric. Proceeding down the DAG from the most unstable saddle, we find a state $\sim (1,1,1,0,1,1,1,0)$ that has 4 unstable directions, as does a second less symmetric state with only cells $4,7$ off and the others on to varying degrees. States with fewer unstable directions and various patterns of ON/OFF cells can all be found. Thus for very physical reasons this model generates a rich DAG. There are also stable terminal states that are not reflection invariant on the circle such as $\sim (1,0,0,1,0,0,0,0)$, but these were not found as the endpoint of the dynamics when initializing with small randomized $a_i$. 

To reproduce the flow around the principle saddle point on the diagonal from a potential, we propose what is essentially an antiferromagnet but with a diagonal inverse metric (n.b. $d(a)$ in Eq.~\ref{equ:staticInhibitionPot} is identical to Eq.~\ref{eqn:staticInhibition})
  
\begin{equation}
\begin{aligned}
	\dot a_i &= -g(a_i) \nabla_{a_i} F , \\
	F &= \sum_i V(a_i) + \frac{1}{2} \sum_{i \ne j} d(a_i) K_{i,j} d(a_j) .
\label{equ:staticInhibitionPot}
\end{aligned}
\end{equation}

\begin{figure}[h]
\centering
	\includegraphics[width=.99\linewidth]{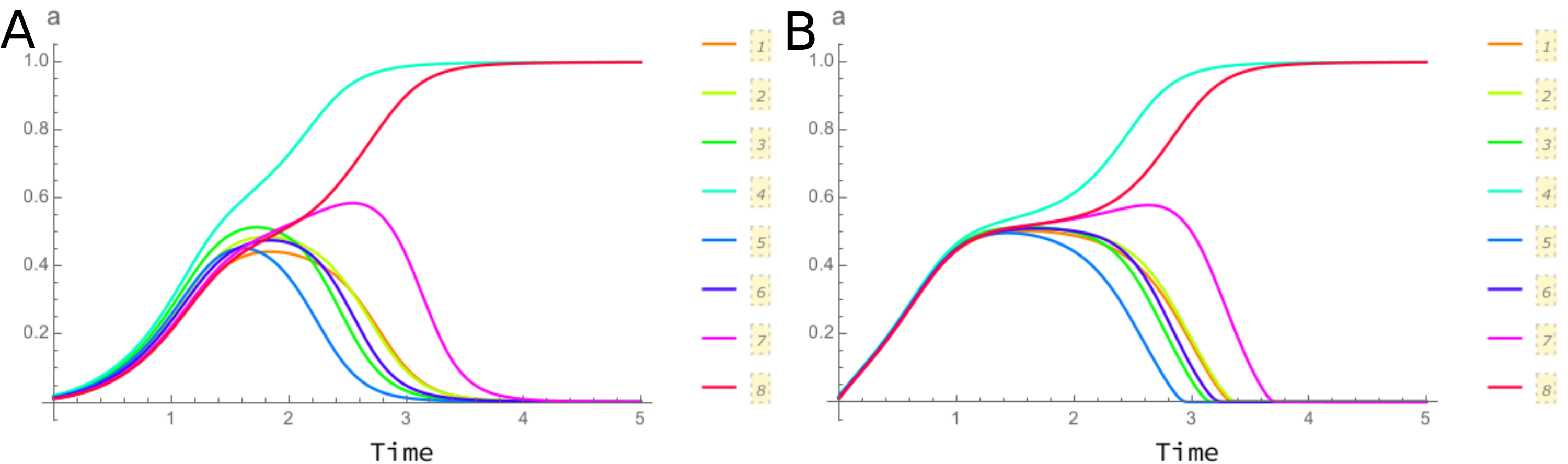}
\caption{Two potential models for Eq.~\ref{eqn:staticInhibition} run with the same initial conditions. (A) The potential in Eq.~\ref{equ:staticInhibitionPot} with the inverse metric extending to the origin. (B) The inverse metric near the origin now matches Eq.~\ref{eqn:staticInhibition}.
}
\label{staticInhibitionPot}
\end{figure}

In the SI we show that we can match both the location of the principle saddle point on the diagonal, $a_i = \bar a(N)$, and the Jacobian for any number of cells $N$ by fixing two functions of a single variable, $g$ and $V$. The solution for $g$ near the origin behaves as $g(a) \sim a$ in contrast to the Eq.~\ref{eqn:staticInhibition} where the velocity is $\mathcal{O} (1)$ (compare Fig.~\ref{staticInhibition}b with Fig.~\ref{staticInhibitionPot}. But since the saddle point is restricted to $a > 0.51$ we can easily correct the inverse metric for smaller $a$.

The dynamics derived from these potential models are shown in Fig.~\ref{staticInhibitionPot} where it is obvious comparing with Fig.~\ref{staticInhibition}b that we have corrected the dynamics around the origin with the two part inverse metric, at the expense of now making the decrease of single $a_i$ in response to the antiferromagnetic repulsion too abrupt. But that can plausibly be fixed by adjusting the metric in a separate region of the phase space. 

Putting the original model in potential form shows the extent to which the entire pattern formation process is controlled by the saddle point on the diagonal. Analytic calculations for how small differences amplify are facilitated by knowing the stable and unstable manifolds of the saddle. 

We started from a very idealized model, but if it were elaborated to include a more realistic description of the Notch-Delta signaling that is responsible for the inhibition or a more complete account of the neural fate, we suspect that the principal saddle point will retain its unstable directions but add many more stable ones. The best estimate for how much those additional details disappear from the dynamics is in fact the stable eigenvalues at the saddle point. 

One may object that any model with contact inhibition is unrealistic since signals have to travel diffusively. That point is somewhat debated in {\it Drosophila} after cellularization \cite{alexandre2014patterning,kornberg2019scripting}, but nevertheless we show in the next section, when diffusion is responsible for communication between cells,  the underlying potential models have the structure discussed here. Finally representing a biological `decision' as a saddle point in a high dimensional space applies during mesoscopic times, from when the pattern first emerges, to when it is close to saturation.  

\FloatBarrier

\subsection{The Turing Model in potential form}
While the terminology `Turing system' and reaction-diffusion are used somewhat interchangeably, we focus in this section on the narrow use of the term to mean an activator inhibitor pair $a,h$ where the inhibitor diffuses much more rapidly than the activator. The system is formally infinite dimensional since the fields are defined in continuous space, so not strictly a MS system, but practically so, as we will see. The crucial property is existence of a saddle point for uniform non-zero values of $a,h$ that evolves into a nonuniform state where concentrated regions of $a$ produce inhibitor that spreads rapidly and confines the activator. In a truly homogeneous-isotropic system there is a huge degeneracy of patterns all with a common characteristic scale, with very slow rearrangements among them. 
This space-time scale is of no biological relevance, and again we wish to focus our modeling effort on mesoscopic times, after some small biases trigger the basic instability and until the localized patches of activator are stably formed, i.e., the unstable manifold or Waddington valley leading from the initial stability to the terminal patch of activator. Thus it suffices to consider Turing systems on a circle with parameters that give a single localized state, modulo symmetries. The question is how to describe the dynamics on the unstable manifold of the saddle point as it tends to the stable state.  

Consider a system of the form,

\begin{equation}
\begin{aligned}
	\dot a(x,t) &= f_a(a,h) - \nu a + D_a \partial_x^2 a , \\
 	\dot h(x,t) &=  f_h(a) - h + D_h \partial_x^2 h .
\label{equ:turing}
\end{aligned}
\end{equation}

\noindent
The spatially uniform version of this model was the subject of the example in Sec 3B that derived a potential from the flow in spite of the fact that the cross derivative have opposite sign. 

Any matrix with real eigenvalues can be represented as the product of two symmetric matrices one positive definite~\cite{bosch1986factorization}. However for Eq.~\ref{equ:turing} at the saddle point, that would require the inverse metric to contain the Laplace operator which we consider overly cumbersome (SI Sec II.4). Fortunately the resolution of the cross derivative problem in the continuum is both more interesting and useful in fitting data. 

To be most explicit, consider a discrete version of Eq.~\ref{equ:turing} with $N$ cells and a Laplacian obtained by differentiating an expression of the form $\sum_i D_y (y_i - y_{i+1})^2/2$, where $y = a,h$. With $D_{a,h} = 0$ the phase space for fixed $x$ was shown in Fig.\ref{flowtopotential}, where we made the origin stable by assuming $f_{a,h} \sim a^2$ in Eq.~\ref{equ:turing}.
This definition separates the linear Turing instability caused by the diffusive coupling when $D_h \gg D_a$ from the patterning system defined by the lateral inhibition model where the activator in all cells increased  from $0$ at a constant rate.) 

Now linearize around the saddle point of the discrete form of Eq.~\ref{equ:turing}. There will be $2N$ modes that generally occur in pairs when reflection symmetry connects two inequivalent states that vary in space (i.e., with cell index $i$). Due to the large value of $D_h / D_a$ there will be $N-1$ very stable modes centered on the $h_i$, one modestly stable one for uniform $h$. In the $a$ sector of the spectrum, the uniform mode is stable by the definition of a Turing system (the first unstable mode is nonuniform and defines the scale of the terminal pattern) and among the remaining, a few will be positive in pairs. The most stable eigenvalue for both $a$ and $h$ may be a singleton or in a pair depending on whether $N$ is even or odd.

\begin{figure}[h]
\centering
	\includegraphics[scale=0.40]{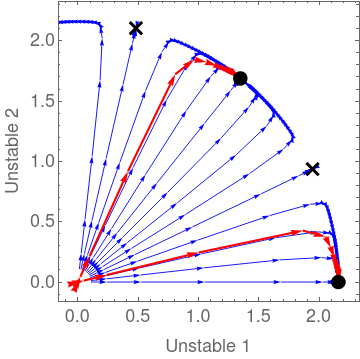}
\caption{A sector of the tangent plane to the unstable manifold at the saddle for a discrete version of Eq.~\ref{equ:turing} with $N=7$ cells. The blue arrows show the trajectories of the potential fit with different initial conditions chosen at equally spaced angles. The red arrows show the projection of the actual equations on the tangent plane. Arrows are plotted at equal times and so give the approximate velocity.  
}
\label{turingModelN}
\end{figure}

Figure \ref{turingModelN} shows the case of $N=7$ with one unstable pair of modes. The dynamics can be projected on the (linear) tangent space of the unstable modes at the saddle point. The unstable manifold is 7-fold symmetric, compact, and flat enough that its projection onto the tangent space is 1:1. At the boundary of the unstable manifold there are 7 fixed points, related by rotational symmetry, and between them 7 saddles each with one unstable direction Fig.~\ref{turingModelN}.

The dynamics on the tangent plane can be accurately parameterized in polar coordinates relative to the central saddle by the system:

\begin{equation}
\begin{aligned}
	 \dot r &= \lambda r - \beta r^3 , \\
 	\dot \theta &=  g(r) \sin(7(\theta - \theta_0)) ,
\label{equ:turingModelN}
\end{aligned}
\end{equation}

\noindent
where $g(r)$ is a Gaussian tightly centered at $\sqrt{\lambda/\beta}$, the system is $2 \pi/7$ periodic, and $\theta_0$ is a phase shift. The Eqs.\ref{equ:turingModelN} are potential with $g(r)$ acting as an inverse metric. However, the precise form of $g(r)$ is not crucial since as $N$ increases, the number of fixed points increases accordingly and the angle dependence disappears leaving an invariant circle on which the radial motion terminates. 

\begin{figure}[h]
\centering
	\includegraphics[width=.99\linewidth]{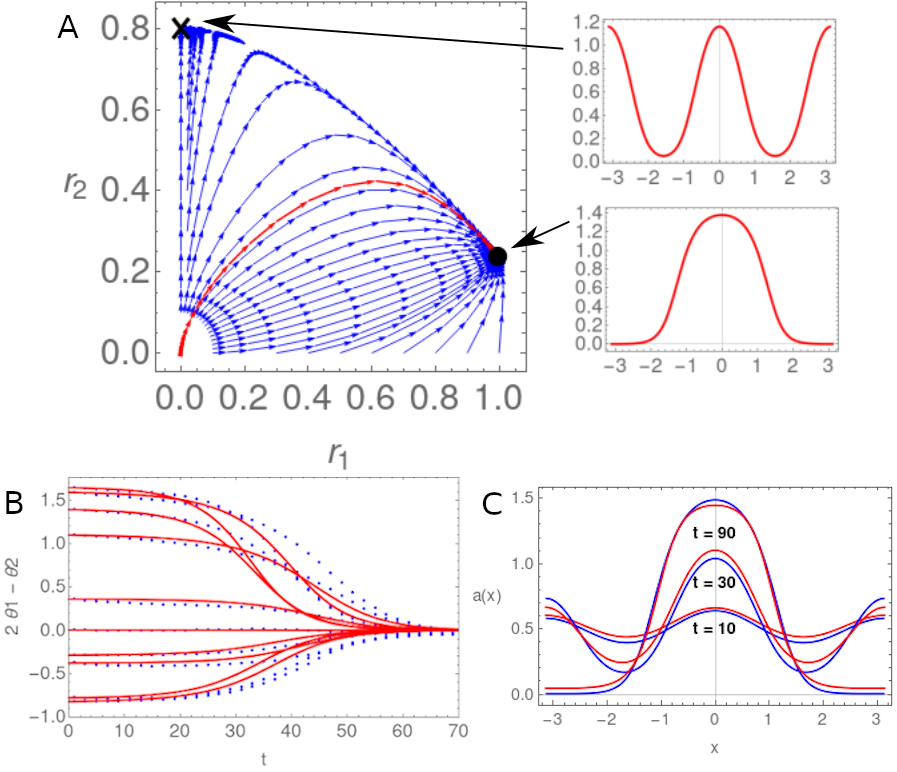}
\caption{The continuous Turing PDE is solved and projected on to the variables $z_1$ and $z_2$ in Eq.~\ref{equ:turing2modes} by simply doing a linear projection onto the Fourier modes ($z_i = r_i e^{i \theta_i}$). 
(A) The streamlines of Eq.~\ref{equ:turing2modes} in the space of the two radii with the angles set to 0 showing a 2 dimensional projection of the 4 dimensional unstable manifold. The call outs show the terminal profiles of $a(x)$ at the saddle with two peaks and the fixed point with one.  The initial condition is $a(x) = 0.01 \cos(2 x) + 0.001 \cos(x)$ 
(B).  Equation.~\ref{equ:turing2modes} implies that the angles stably lock together with $2 \theta_1 - \theta_2 = 0$ . This is shown with blue dots for the actual Turing model and red for the potential fit for various initial conditions.
(C) A saturation function is required to match the solution in the tangent space in (a,h) to $a(x)$ in physical space $x$. The solution of the continuous system (blue) is compared with a sigmoidal saturation of the tangent space flow (red) for the red cross-over trajectory in (a).
}
\label{turingtwomodes}
\end{figure}

Of greater interest is that we have reduced the dynamics from the $2N$ dimensions to the two dimensions of Fig.~\ref{turingModelN}, with a rate parameter $\lambda$ and a limiting radius. All the complexity of $f_{a,h}$ in Eq.~\ref{equ:turing} is reduced to these two numbers, and manifestly could not be made any simpler. This describes the entire unstable manifold from the linear instability to the finite amplitude stable state. Still missing is the embedding of the radius variable $r$ in the tangent space into the fields $a,h$ that we will deal with after properly treating the continuous case. However this embedding is by definition time independent, all the dynamics lives in $r$, and we thus have a very compact representation of a time dependent field with accuracy at the few percent level, more than adequate for biology.

Guided by the discrete case we represent the continuum equations around the saddle point as a linear term that we can Fourier transform on the circle and a nonlinear term local in position that we expand to third order. Our restriction to mesoscopic times makes it entirely reasonable to work on a compact space. Call the most unstable Fourier mode $k_1$ and if we close the system with cubic nonlinearity within this space we must obtain an equation of the form $\dot z_1 = \lambda_1 z_1 + f_3 z_1 | z_1|^2 $ where $z_1$ is the complex coefficient of that Fourier mode.  If a second unstable mode $| k_2 | = 2 | k_1| $ exists then we find couplings between the modes as in:

\begin{equation}
\begin{aligned}
	\dot{z}_1 &= \lambda_1 z_1 - \beta z_1 (|z_1|^2 + 2 |z_2|^2) - 2 \gamma  z_1^* z_2 , \\
	\dot{z}_2 &= \lambda_2 z_2 - \beta z_2 (|z_2|^2 + 2 |z_1|^2) - \gamma z_1^2 .
\label{equ:turing2modes}
\end{aligned}
\end{equation}

The $k_2$ mode describes two incipient blobs on the circle that under our assumptions are unstable to a single blob (i.e., the $k_1$ mode grows faster). The unstable manifold is now four dimensional and goes to a fixed point through a series of saddle points each with one less unstable mode. The first and second of these saddles correspond to a state with two blobs and is inaccessible for typical initial conditions. Thus we have described the dynamics of the cross over between two blobs and one on the circle with four modes and dynamics derived from a potential. One consequence of these equations is that the angles of the two modes are locked as shown in Fig~\ref{turingtwomodes}. We note that the model is potential with trivial metric only to cubic order and higher order Taylor series would induce a nontrivial metric.

To map the modes $z_{1,2}$ back to the continuum $a,h$ we have to interpolate between the Fourier modes for the linear system and the static solution for $a$ which is a single blob. 
The simplest way to model this constraint is to simply add the Fourier modes linearly and saturate them with a sigmoidal function $\sigma(x)$ at medium and late times reflecting the high and low values of $a$. Thus one needs to fit the parameters in Eq.~\ref{equ:turing2modes} as well as the sigmoidal function shape as a function of $z_1$ and $z_2$ to map our potential to continuum $a$. The comparison is plotted in Fig\ref{turingtwomodes}c with details in SI. 



\begin{figure}[h]
\centering
	
		\includegraphics[width=.99\linewidth]{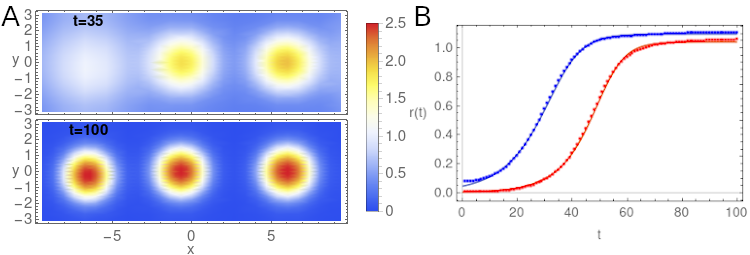}
\caption{(A) A model in 2 spatial dimensions leads to the formation of three blobs each of which are slightly different because of randomness in the parameters and initial conditions. The blobs are shown at two different time points demonstrating that the left and right end form at different times. Since the blobs have spherical symmetry, the profile is first averaged over angles and then projected on to the Bessel Function $J_0(R)$ being the first term in a Fourier-Bessel series, where $R$ is a scaled radial coordinate in space.  (B) The simulated 2D dynamics is projected onto the tangent plane (dots) for the left (red) and right (blue) blobs in (A). The data is fit very well by the radial part of Eq.~\ref{equ:turingModelN} (curve) with parameters specific to each blob. The space-time profiles in (A) are then fit by a sigmoid depending on $r(t)$ from Eq.~\ref{equ:turingModelN} and $J_0(R)$ (SI Sec II.4)
}
\label{turingspotsfit}
\end{figure}

To illustrate the utility of this reduced representation for fitting experimental data consider a 2-d continuum periodic rectangular system which leads to the formation of three blobs. 
Each of these blobs are slightly different because of a long range gradient as well as randomness in the microscopic equations. However, our reduced representation captures the essential nature of the dynamics and the different blobs can be fit to with different parameter in the radial mode in Eq.\ref{equ:turingModelN} as shown in Fig.~\ref{turingspotsfit}.

When presented with time-lapse data for a Turing system, we would take the radial profile of each blob as a function of time and fit to a single time independent sigmoidal function composed with a solution to $\dot r$ in Eq.~\ref{equ:turingModelN}, with parameters and initial conditions specific to each blob. We expect this procedure will collapse considerable variability among the dynamics of the blobs to universal functions with minimal parameters as in Fig.\ref{turingspotsfit}.

We conclude this section by generalizing the lateral inhibition model, Eq.~\ref{eqn:staticInhibition} to one with a diffusing inhibitor which is instructive to compare with the Turing system. Consider the system for $N$ cells periodically continued (suppressing constants),

\begin{equation}
\begin{aligned}
	\dot a_i &= f_a(a_i) - h_i d^\prime (a_i) , \\
	\dot h_i &= d(a_i) - h_i + D_h (h_{i+1}+h_{i-1} - 2h_i).
\label{equ:dynamicInhibition}
\end{aligned}
\end{equation}

\noindent
where $d(a)$ is again a sigmoidal function mapping $a$ into the interval $[0,1]$. There is no loss of generality in making the coupling between $h$ and $a$ in $\dot a_i$ the derivative of the source of $h$ in $\dot h_i$ since we can redefine $a$ to make this true. For illustration we assume $f_a(a) \sim a$ around the origin to make that point unstable. 
If we solve $ \dot h_i = 0$ for $h_i$, substitute into the $\dot a_i$ equation, 
we obtain the contact (rapid diffusion) form of the equations with the Greens function appropriate to the circle i.e., Eq.~\ref{equ:staticInhibitionPot}. 
The cross derivatives $ \partial_h \dot a $ and $\partial_a \dot h $ are equal and opposite in sign as we expect for inhibition. 

For an example with all constants order one, and the $D_h$ adjusted to produce the desired spacing between the active cells; there is again a saddle point along the diagonal with $\sim N$ stable directions corresponding to the $h_i$ and a few additional ones with weight in the $a_i$ directions. A few of the stable eigenvalues can be complex reflecting the nonzero antisymmetric terms in the Jacobian matrix. The unstable eigenvalues are numerically very close to those of the system with $h_i$ eliminated and the dynamics within the $a_i$ subspace is identical for practical purposes (SI).

So in comparison with the Turing system, there are now many unstable modes all deriving from the same saddle point along the diagonal in $a_i$. The origin is unstable rather than stable and the dynamic competition that generates the pattern derives from the saddle point on the diagonal. We believe the geometric view point provides the clearest characterization of these two systems with diffusing inhibitors. 

\FloatBarrier

\subsection{Spreading the pattern by a wave}
When a two dimensional area is to be patterned, the process often proceeds from a boundary which provides the template. 
One such system is the morphogenic furrow in the fly eye. It moves as a wave with constant velocity from posterior to anterior across the eye imaginal disk, leaving behind the hexagonal crystal of ommatidia that characterize the insect eye \cite{baker2007patterning}.

We follow the ideas of \cite{lubensky2011dynamical} who modeled this system with effectively three modes: a cell localized activator $a$, an intermediate range inhibitor that is responsible for the spacing between successive rows of ommatidia, and a long range activator, $b$, that destabilizes the $a=0$ state ahead of the furrow to allow the cells to compete as described in Eq.~\ref{equ:dynamicInhibition}. We will replace the dynamic inhibitor in our reformulation with static lateral inhibition which we just showed to be functionally equivalent.

\begin{figure}[h]
\centering
	\includegraphics[scale=0.60]{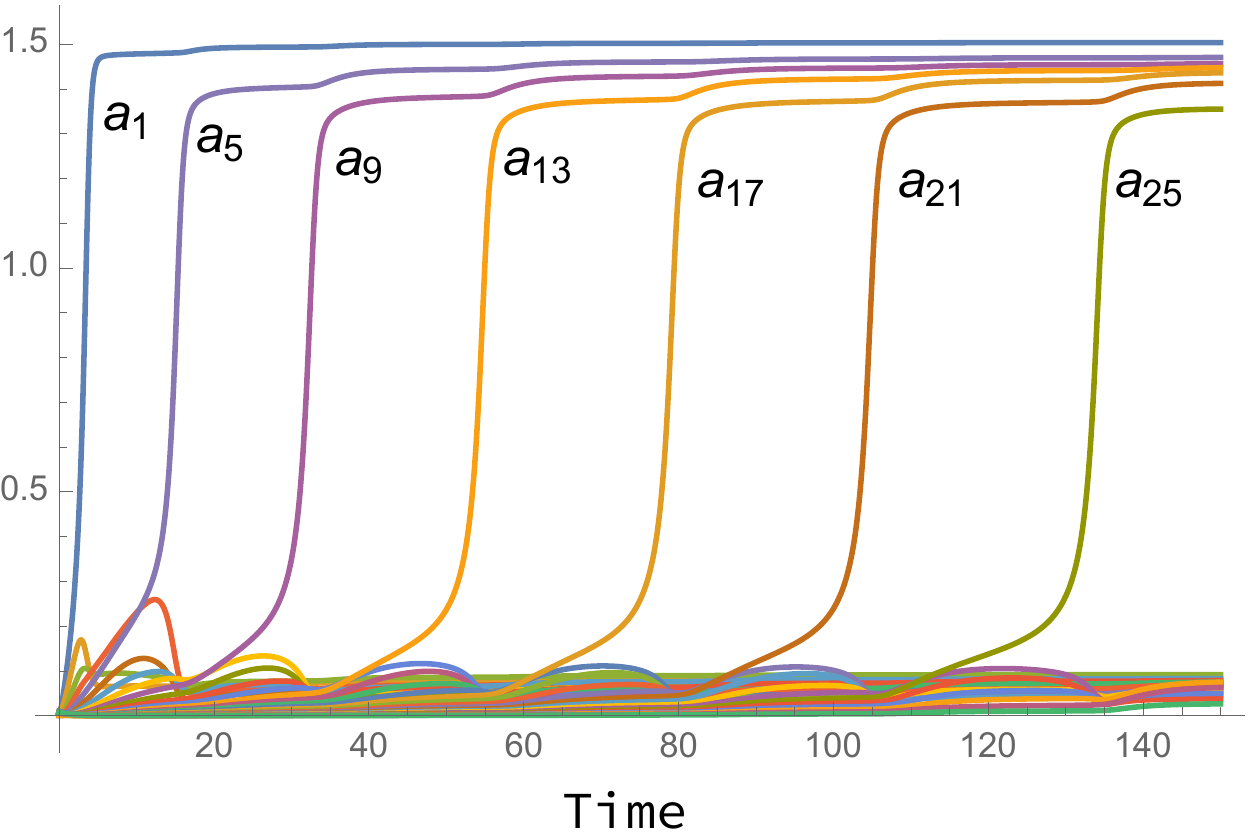}
\caption{Dynamics for the 1D morphogenic furrow from Eq.~\ref{equ:flyeye} for N=30 where every 4th cell is active. The wave appears to slow towards the end since the long range activator $b$ is diffusing out the right hand boundary. For the same reason the plateau in $a_i$ is a bit lower at the end. 
}
\label{flyeye}
\end{figure}

We can directly formulate our model as a potential since there is no sign conflict when integrating two activators. We thus propose in one dimension,

\begin{equation}
\begin{aligned}
	& u(a,b) = v(a) + (b - a^2)^2/2 - a b, \\
	& U = \sum_{i=1}^N u(a_i,b_i) + D_b (b_i - b_{i-1})^2/2 +  \frac{1}{2} \sum_{i \ne j} d(a_i) K_{i,j} d(a_j)
\end{aligned}
\label{equ:flyeye}
\end{equation}

\noindent
where constants are omitted,  $v(a)$ is a bistable potential, and the double sum copies Eq.~\ref{eqn:staticInhibition}. The boundary conditions are open, and we fix $b_0 = 1$ so the left end of the lattice is activated and the wave propagates towards $i=N$. The only condition on the one cell potential $u(a,b)$ is that the origin $a,b = 0$ is stable and the saddle point occurs around $a,b \sim 0.1$ in units where the  final state shows $a \sim 1$ in the active cells, zero elsewhere, and $b \sim 1$, i.e., the value of the saddle point is small but not tiny and stabilizes the cells at $a=0$ until the leading edge of the wave of $b$ arrives by diffusion and kicks them over the saddle. The static inhibition controls the spacing between the active cells, and the diffusion constant governs the speed within limits, Fig.~\ref{flyeye}.

The potential formulation makes it obvious that the moving solution in Fig.~\ref{flyeye} is just an instance of relaxational dynamics for a field with two minima, the simplest instance being, $\int_x(s a - a^2/2 + a^4/4 + D (\partial_x a)^2/2) , s \ll 1$ where an interface between locally stable solutions at $a \sim \pm 1$ moves to replace the higher potential solution with the lower potential one, and does so with constant velocity. For the parameters of Fig.~\ref{flyeye} all solutions of period 2 to 5 have negative potential $U$ and thus lie below the initial conditions with $U>0$. However only the period 4 solutions is realized as the wave spreads.

\subsection{A static morphogen gradient and adaptive systems}

We note in closing two instances where the components are few enough in number and well enough characterized that gene centric models are informative and quantitative, yet the geometric viewpoint poses questions that are often neglected (see SI). 

A morphogen is a diffusible factor whose level defines cell fate. 
Morphogen systems like any other dynamical model may be multistable implying hysteresis in the pattern. An informative geometrical representation is given in the SI.

An adaptive system has one variable that responds to an external signal, yet whose value is signal independent under static conditions. Adaptation is common in biology from sensory systems to signaling pathways where its a more rapid way to connect position to fate than reading a static signal \cite{sorre2014encoding}. This property is succinctly expressed in the form of a potential and metric (SI Sec II.7).

\section{Discussion}
We have given a tight argument that in
the absence of periodic behavior, robust
(i.e.\ structurally stable) gene
regulatory dynamics satisfy the
Morse-Smale assumptions and thus can be
described by a downhill gradient-like
system where the restpoints can be
organized into a directed acyclic graph (DAG)
defining the allowed transitions between
them.  Moreover, we observed that such
dynamics can reasonably be represented
as a true gradient system given by a
potential and a Riemannian metric.
The mathematics achieves such sweeping results by imposing genericity: system properties are unchanged in a neighborhood of the system in question. In analogy, embryonic development is robust against environmental noise. 

In geometric terms, the inhibition among cells that produces a sparse array of neural progenitors, 
is simply a high index saddle point that initiates the DAG. 
Turing pattern formation begins from the linear instability of a uniform state, making it a saddle point.
Its nonlinear development is just the unstable manifold of the saddle that we represented as a MS system.
There was a cascade of saddle points with decreasing index to the terminal pattern. 

A representation of the flow as a metric and potential continues
to work around saddle node bifurcations
and heteroclinic flips, the two
operations needed to grow any MS system
from the trivial one \cite{Newhouse76}.
These operations are thus the only
topology changing 'mutations' that are
needed to computationally evolve gene
regulatory networks\cite{franccois2014evolving}.

The results of Catastrophe Theory (CT)
are clearly of relevance to our analysis
but they are local and, since they only
study the potential, they do not capture
the stable and unstable manifolds of the
dynamics or the heteroclinic flips.  Our study shows that the
latter are crucial to understanding
developmental dynamics and the study of
them opens up new ways to extend the
theory from local to global.

In particular, we borrowed from the
mathematics that Smale developed for his
proof of the high-dimensional Poincare
conjecture, to rigorously deduce a new form of dimensional
reduction in which for
generic systems like ours the
topological relationship between index 1
saddles and attractors is effectively
independent of dimension and can be
represented by 2-dimensional systems.
The {\it canalization} of Waddington is simply the focusing of the dynamics onto the unstable manifolds of the saddles. 
The results in Fig.\ref{fig:globalbifns}
depend upon this understanding which is
very different to the dimension
reduction that CT supplies which only
applies locally in state and parameter
space.

A network model represented by
activating and inhibitory arrows among
genes has little relation to the
geometry of the flow, that is the
ultimate representation of the dynamics.
We expect that such models with
two parameters and $\leq 3$ attractors
will produce the bifurcation diagrams in
Fig.\ \ref{fig:globalbifns}.

Recent examples of modeling development have taken a more geometric approach from the start.
The first several steps in the differentiation of mouse
embryonic stem cells were decomposed into two
coupled three way decisions.  The first
involving an all or nothing decision
that deposits all cells into either
adjoining state depending on a morphogen
used the landscape in Fig.\
\ref{fig:globalbifns}A.  The second used a flip
bifurcation to flexibly allocate cells
to one of two attractors \cite{Saez21},
corresponding to part of
\ref{fig:globalbifns}D.
In \cite{corson2017gene} the so called
triple points in the phase diagram of
the {\it C.  Elegans} vulva plausibly
correspond to Fig.\ref{fig:globalbifns}D, 
and a related study \cite{camacho2021quantifying} used Fig.\ref{fig:globalbifns}C.
If embryonic stem cells can transition directly to 
ectoderm, mesoderm, and endoderm, then Fig.\ref{fig:globalbifns}D is a possible model with the pluripotent state being the fourth attractor in the center. 

Cartoons of development commonly represent the
succession of fates as a binary tree,
suggesting the existence of three
attractor systems at the nodes.  Left
implicit is a developmental clock that
biases fate progression in one
direction. This could be a
competence window that limits the
response to signals to a temporal
interval combined with epigenetic
changes. 


There is no reason that our analysis
cannot be generalized to periodic systems
where the structurally
stable dynamics are also MS. All the key
ingredients regarding MS systems are
available (with obvious changes such as a constant potential on the periodic orbit) \cite{Meyer68}, but the generic bifurcations
are more numerous and complex and that
makes this a substantially more
challenging task. When a developmental system
is subject to an autonomous periodic signal,
such as might be the case with the cell cycle,
the relevant dynamics is represented by a map rather than by
a flow as we have assumed so far, and opens up a richer corpus of behavior.



In applications the states in question are rarely terminal, thus parameters are not static. 
When systems are not steady, do all the other degrees of freedom have time to relax to the two dimensions necessary to represent all three way decisions? This question can only be resolved by fitting data, and in examples we have seen that high index saddles can also have multiple stable directions, which is then the recipe for quantifying the neglected variables. 

The geometric description is compact since all reference to specific genes is lost, yet experiments commonly mutate genes and observe changes in patterning dynamics and outcomes. Clearly each mutant requires at least one parameter to fit, but is there any way to predict a double mutant from each of the single ones? An interesting ansatz was used in \cite{Corson2012}, where a single vector sigmoid function was wrapped around the non-compact polynomial equations for the dynamics. A linear force added to the sigmoid rendered the dynamics bounded, but the dependence on the two morphogens was entirely linear inside of the sigmoid. The double mutant was thus the sum of the single ones. Nevertheless the model captured the essential genetic interactions. While the qualitative details of the problem dictates the geometry, how it is parameterized is very germane for applications.

We have concerned ourselves exclusively with representing gene networks as potentials, but not with morphogenesis for which forces and potentials are natural.
Morphogenesis and fate assignment are tightly coupled in development as exemplified the ordered expression of the HOX genes during the unfolding of the anterior-posterior axis \cite{iimura2007hox}, but it is as yet unclear how to unify potentials for fates and forces.

\section*{Acknowledgements}
The authors thank James Briscoe, John
Guckenheimer, and David Lubensky for
discussions.  EDS was supported by NSF
grant No.  2013131, DAR thanks Ian
Stewart for very helpful early
discussions.  DAR and MS were funded for
this work by EPSRC grant EP/P019811/1.
MS was also supported by the Francis
Crick Institute, which receives its core
funding from Cancer Research UK, the UK
Medical Research Council and Wellcome
Trust (all under FC001051).  FC was
supported by the ANR grant
ANR16-CE13-0003-02.  DAR, FC and EDS
participated in the 2019 Kavili
Institute Program on quantitative
biology supported by NSF Grant No.
PHY-1748958, NIH Grant No.  R25GM067110,
and the Gordon and Betty Moore
Foundation Grant No.  2919.02 .  AR
acknowledges support from the Simons
Foundation.



%% file: SI_Maths_current.tex
\setcounter{page}{1}

Throughout this note we restrict
attention to the generic case.  A
generic property is one that holds for
almost all systems.  This 
means that the systems for which the
property holds form an open and dense
subset of the relevant function space.
We will be a little cavalier in not
mentioning the function spaces we are
using but generally it should be clear.

\section{Morse-Smale Systems}

The time integration of a differential
equation $\dot{x}=X(x)$ or, equivalently
the vector field $X$ defines a flow
$x\rightarrow \phi_X^t(x)$ on the phase
space $\phase$.  If $x\in P$ then
$\phi^t (x)$ is the solution with
initial condition $x$.  

Throughout we shall be considering vectorfields
$X$ and their associated flows $\phi_X^t(x)$ 
on a domain $P\subset \reals^n$ in which there
is a globally attracting ball $B$ i.e.\ 
a region $B$ which is a $n$-dimensional ball
(i.e.\ diffeomorphic to
$\{ x\in\reals^n : ||x||<1\}$)
with a boundary $\partial B$ that is a 
smooth\footnote{Throughout smooth means C$^2$.}
sphere (i.e.\ diffeomorphic to
$\{ x\in\reals^n : ||x||=1\}$) 
such that every trajectory of $X$ starting in $P$ outside
$B$ enters $B$ eventually and $X$ is transverse to $\partial B$.


A point $x$ of $\phase$ is called chain
recurrent for the flow provided that
for any $T, \epsilon > 0$ there are
points $x_i\in \phase$ and real numbers
$t_i > T$, $i = 0,1,\ldots ,N$, such
that $x = x_0 = x_N$, and
$d(\phi^{t_i}(x_i), x_{i+1}) <
\epsilon$.

A Morse-Smale (MS) flow is one satisfying the
following properties: 
\begin{enumerate}
	\item[(a)] the set of
	chain recurrent points consists of a
	finite number of hyperbolic rest points
	and hyperbolic closed orbits (see
	\cite{guckenheimer2013nonlinear} for definitions);
	\item[(b)] the
	unstable manifold of any closed orbit or
	rest point has transversal intersection
	with the stable manifold of any closed
	orbit or rest point.
\end{enumerate}

We study generic parameterised 
families in which the structurally
stable systems are 
MS flows $X$ that only have
rest points.  We call these
\emph{gradient-like} Morse-Smale flows because
by \cite{Smale61gradient}
they always have a smooth Liapunov
function i.e.\ a function $L:
\phase\rightarrow \reals$ such that
its derivative is zero at
rest points and $X(L)<0$
otherwise, where $X(L)=dL\cdot X$.

\subsection{Saddle-attractor configurations}\label{sec:sad-att}
\begin{figure}[H]
\centering
\includegraphics[width=0.80\linewidth]{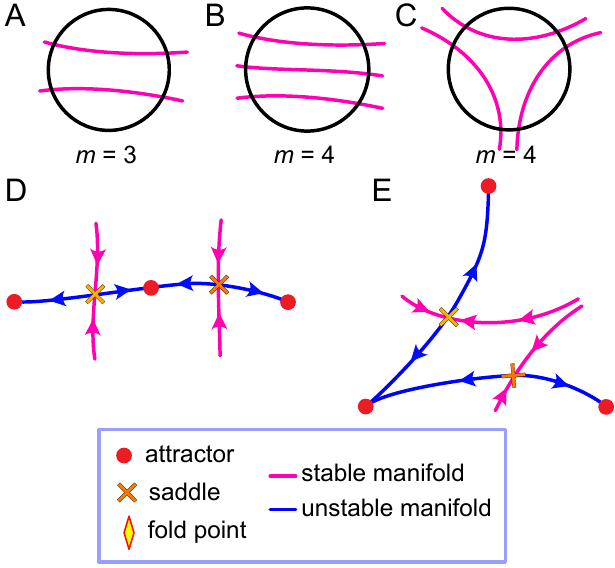}
\caption{(A-C) These illustrate some of
the different possibilities (the most
important for us) for the configuration
of the stable manifolds of the saddles
in the attractor disk region $R$ defined
above which is represented by the
interior of the circle.  All cases for 3
and 4 attractors are shown.  (D-E) These
illustrate the smooth (D) and cuspoidal
(E) versions of the connector (blue
curve) which is made up of the unstable
manifolds of the two saddles in each
case. Other non-smooth configurations are possible
such as with a spiral due to non-real eigenvalues
at the central attractor.
}
\label{fig:connectors}
\end{figure}

Consider for a moment the planar case where
$n=2$ and the above conditions apply
so that the boundary of $P$ is a
smooth simple closed curve and the flow
is inwardly transverse to it.
Then one can always find a
smooth simple closed curve $\gamma$ such
that the flow is inward transverse to
$\gamma$ and so that the interior
$\Omega$ of $\gamma$ contains all the
$m$ attractors of the system, $m-1$
saddles and no other rest points.  
Moreover, $\gamma$
can be chosen so that the stable
manifolds of the saddles inside $\gamma$
intersect the boundary in two points and
do so transversally.  They thus divide
$\Omega$ into $m$ components each of
which is the basin of one of the
attractors and, moreover, the unstable manifold of
each saddle intersects the basin of two
attractors.  
The possibilities with $m=3,4$
are shown in Fig.\ 
\ref{fig:connectors}(A-C) and we see that the 
$m=3$ case is particularly simple.

There is a corresponding decomposition in $n$ dimensions.
In Appendix \ref{sect:decision} we
explain the following result: For a
gradient-like MS system on a region 
$P$ in $\reals^n$ with attracting 
ball $B$ as above one can always
find an attracting $n$-dimensional disk
$R$ in $P$ with smooth spherical boundary
$\partial R$ that contains all the
$m$ attractors plus $m-1$ index 1 
saddles but no other rest points.  It
captures almost all the trajectories 
in the following sense: if $W^s$
is the union of all stable manifolds of
rest points outside $R$ and $x\in R\setminus W^s$
then for $t$ large $\phi^t(x)\in R$.
Moreover, $R$ has the property that the
stable manifolds of the index 1 saddles
inside $R$
transversally intersect the spherical
boundary of $R$ in disjoint spheres of
dimension $n - 2$ and divide
$R$ into regions each of which is the
basin of attraction of one of the attractors.

We are interested in regions like $R$
above because in applications to
developmental biology the index 1
saddles play a particularly important
role: their unstable manifolds determine
the escape routes by which cells change
their state by escaping one attractor
and moving to another.

As a corollary to the above result, for
such systems with 3 attractors, associated with
the 3 attractors are two saddles $s_1$ and $s_2$
with a disposition as in Fig.\ \ref{fig:connectors}D,E
i.e.\ we can label the attractors $A$, $B$
and $C$ so that $s_1\succ A,B$
and $s_2\succ B,C$ where
$s\succ X$ means that the unstable manifold of
$s$ intersects the stable manifold of $X$.

Thus the unstable manifolds of these two
saddles join at one of the attractors.
The attractor at the joint is called the
\emph{central attractor} and the other
attractors are said to be
\emph{peripheral}.  The union of the
parts of the
unstable manifolds which connect the
attractors is called the
\emph{connector}.  The joint can either
be smooth (Fig.\ \ref{fig:connectors}D)
so that the connector is a smooth curve
or it can be non smooth (e.g.\ as in Fig.\
\ref{fig:connectors}E).

Consider such a situation where we have
two index 1 saddles $s_1$ and $s_2$ whose
unstable manifolds intersect the stable
manifold of an attracting rest point at
$x$.  If the latter is close to
undergoing a fold or cusp bifurcation
then there is a line $\ell$ through $x$
and a $n-1$-dimensional submanifold
$W^{ss}$ (the strong stable manifold)
intersecting $\ell$ tranversally at $x$
such that all trajectories starting near
$x$ and not starting in $W^{ss}$
approach $x$ in a direction tangent to
$\ell$.  Consequently, it is generic
that the unstable manifolds of the two
saddles in the connector approach $x$
tangent to $\ell$.  If they approach in
opposite directions relative to $\ell$
then the connector is smooth (Fig.\
\ref{fig:connectors}D) but if they
approach from the same direction
relative to $\ell$ then the connector is
not smooth and has what we call a
\emph{cuspoidal} structure (Fig.\
\ref{fig:connectors}E).

If the attractor $x$ is not close to a
fold then there can be other non-smooth
connectors.  One obvious example is
where the linearisation of the system
about $x$ has eigenvalues with non-zero
imaginary part so that the unstable
manifolds spiral into $x$.  The reader
should note that in order to keep our
figures to a minimal size we never
illustrate such cases but we do take
them into account in our arguments.

In Lemma \ref{lem:nofold} of Appendix
\ref{App:no_fold} we show that if the
joint is cuspoidal at an attractor then
that attractor cannot undergo a fold
bifurcation involving either of the
saddles $s_1$ and $s_2$.  For the other non-smooth
cases this is obvious. Thus such a joint
involving $s_1$, $s_2$ and $x$ 
can only undergo a fold
bifurcation involving $x$ and $s_1$ or $s_2$
if the joint is smooth.

\section{Fold and cusp bifurcations}
We consider the bifurcations between
gradient-like MS systems in
families of differential
equations given by a family of vector fields
$X_\param$ parameterised by $\param\in
\reals^c$.  Throughout this discussion
we restrict to the case $c=2$.

When $c=2$ for gradient-like systems 
the only local bifurcations
that can occur generically are the
\emph{fold} (or \emph{saddle-node}) and
the \emph{cusp} bifurcation.  There are
other codimension 2 bifurcations for dynamical systems
\cite{arnold2013dynamical,guckenheimer2013nonlinear}
such as the Takens-Bogdanov bifurcation
but these introduce periodic orbits and
therefore move us outside gradient-like
systems that we focus on here.

At such a fold or cusp bifurcation
point $\param$ there is an invariant
1-dimensional smooth center manifold $W^c
(x_0)$ through the point $x_0$ in phase
space where the bifurcation is taking
place and the system on this submanifold
may be transformed into one of the
following families via a $C^{N(r)}$
change of coordinates
(\cite{arnold2013dynamical} Chap.\ 1,
Sect.\ 3 and Chap.\ 2, Sect.  5.7):
\begin{equation}\label{eqn:fold}
	\text{(fold) } \dot{x}=\pm x^2 + \param_1
\end{equation}
\text{ or }
\begin{equation}\label{eqn:cusp}
	\text{(cusp) } \dot{x}= \pm x^3 +  \param_1 x +\param_2.
\end{equation}
In
fact, if the original system is
$C^\infty$ then for each $r>1$ one can
find a neighbourhood $U$ of $x$ such that there
is a $C^r$ center manifold of $x$ in
$U$. 


If the system is $n$-dimensional
then the dynamics near $x_0$ are topologically conjugate to
\begin{align}\label{eqn:fold2}
	\text{(fold) } & \dot{x}=x^2 + \eps_1(\param);\\
	 &\dot{v}=-v;\,
	\dot{w}=w; \nonumber
\end{align}
or
\begin{align}\label{eqn:cusp2}
	\text{(cusp) } &
\dot{x}= \pm x^3 +  \eps_1(\param) x +\eps_2(\param);\\
&\dot{v}=-v;\,
\dot{w}=w\, \nonumber 
\end{align}
where $v\in\reals^{n_s}$,
$w\in\reals^{n_u}$ and
$n_s$ (resp.\ $n_u$) is the
number of eigenvalues of the linearisation
of the system at $x_0$ with negative 
(resp.\ positive) real part.

Consider the set of eigenvalues
$\lambda_i$, $i=1,\ldots ,n$, of the
linearisation $\dot{x}=Ax$ of the system
$\dot{x}=X(x)$ at an equilibrium $x_*$
and let $r_i$ be the real part of
$\lambda_i$.  A \emph{resonance} is a relation
of the form $r_s=\sum_i m_i r_i$ with the
$m_i$ positive integers with $\sum_i
m_i\geq 2$.  We say this equilibrium is
non-resonant if no such resonance
exists. 

Consider a fold $x_0$ as above in $n$
dimensions. Then if it is non-resonant
   for $r>2$ one can find $C^r$
coordinates $(x,y)\in\reals\times
\reals^{n-1}$ on some neighbourhood $U$
of the fold point in which the differential
equation takes the form
\begin{align}\label{eqn:ndim}
	\dot{x}&=\pm x^2 +
	a(\param)x^3+\param_1\\
	\dot{y}&=A(x,\param )y\nonumber
\end{align} 
(\cite{arnold2013dynamical}
Part I, Chap.\ 2, Sect.\ 5.7). Without loss of generality 
we can assume that the bifurcation 
takes place at $\param = 0$. 
In the systems we study
the $y$ direction is stable and the
matrix $A(0,0)$ has all its eigenvalues
with negative real parts.

%
%

\subsection{Center manifolds}
Center manifolds will play a key role in our
considerations. Our use of the term \emph{center
	manifold} will be a little more general
than usual as normally it is discussed
when the systems is at a bifurcation and
we will want to use it when we are
only near a bifurcation.  For example,
we want to be able to associate a center
manifold to an attractor that is close
to undergoing a fold bifurcation.  Also
we need to consider the smoothness of
the variation in the center manifold as
parameters are changed.

For relevant
information about center manifolds see
\cite{hirsch2006invariant} Sect.\ 5A.  In
particular note that by Theorem 5A.3 of
\cite{hirsch2006invariant}, if $W^c$ is a
center manifold through a rest point $x$
and $W$ is a backward invariant set
containing $x$ then, near $x$, $W$ is
contained in $W^c$.  Thus, for example, if the
unstable manifold of a saddle is
asymptotic to a fold point, then close
to the fold point it is in the center
manifold.
Center manifolds are not necessarily
unique but their tangent space is.  We
will use this fact below.  

We now consider what we call
\emph{pseudo-hyperbolic} rest points
$x$.  At such rest points $x$ there is $a>b>0$
such that the Jacobian of the
vector field at $x$ has eigenvalues $\lambda$
that either have their real part $\leq
-a$ or $\geq -b$.  Then
pseudo-hyperbolic index 1 saddles and
attractors have 1-dimensional center
manifolds $W^c(x)$ that vary smoothly
with parameters (Sect.\ 5
\cite{hirsch2006invariant}, especially
Theorems 5.1, 5.5 and 5A.1).  If $\varphi^{t}$
is the flow, this
manifold is characterised by the fact
that $z\in W^c(x)\iff ||
\varphi^{-t}(z)-x||/e^{ct} \rightarrow
0$ as $t\rightarrow \infty$ for any $c$
with $a>c>b$.  There is a complementary
submanifold $W^{ss}(x)$ transversal to 
$W^c(x)$ at $x$ characterised by
$z\in W^{ss}(x)\iff ||
\varphi^t(z)-x||/e^{-ct} \rightarrow 0$ as
$t\rightarrow \infty$ for such a $c$.
This we call the \emph{strong stable
	manifold}.  Note that our use of the
term \emph{center manifold} is a little
more general than usual as in that case
one commonly takes $b=0$.

Index 1 saddles are always
pseudo-hyperbolic and attractors are if
they are close to having a fold
bifurcation.  For an index 1 saddle,
part of the unstable manifold containing
the saddle can be taken for a center
manifold.\label{para:pseudo-hyperbolic}

According to Theorems 5.1 of
\cite{hirsch2006invariant}, $W^c(x)$ has
$C^r$ dependence upon parameters
provided $e^{jb-a}<1$ for $1\leq j\leq
r$.  Thus the center manifold for
saddles always is smooth and that for
attractors is smooth provided they are close
enough to having a fold bifurcation.
The later point is true because
the closer an attractor is to being
a fold, the closer one can take $b$
to zero.

%
%

\section{Catastrophe Manifold} 

\begin{figure}[H]
	\centering
	\includegraphics[width=.75\linewidth]{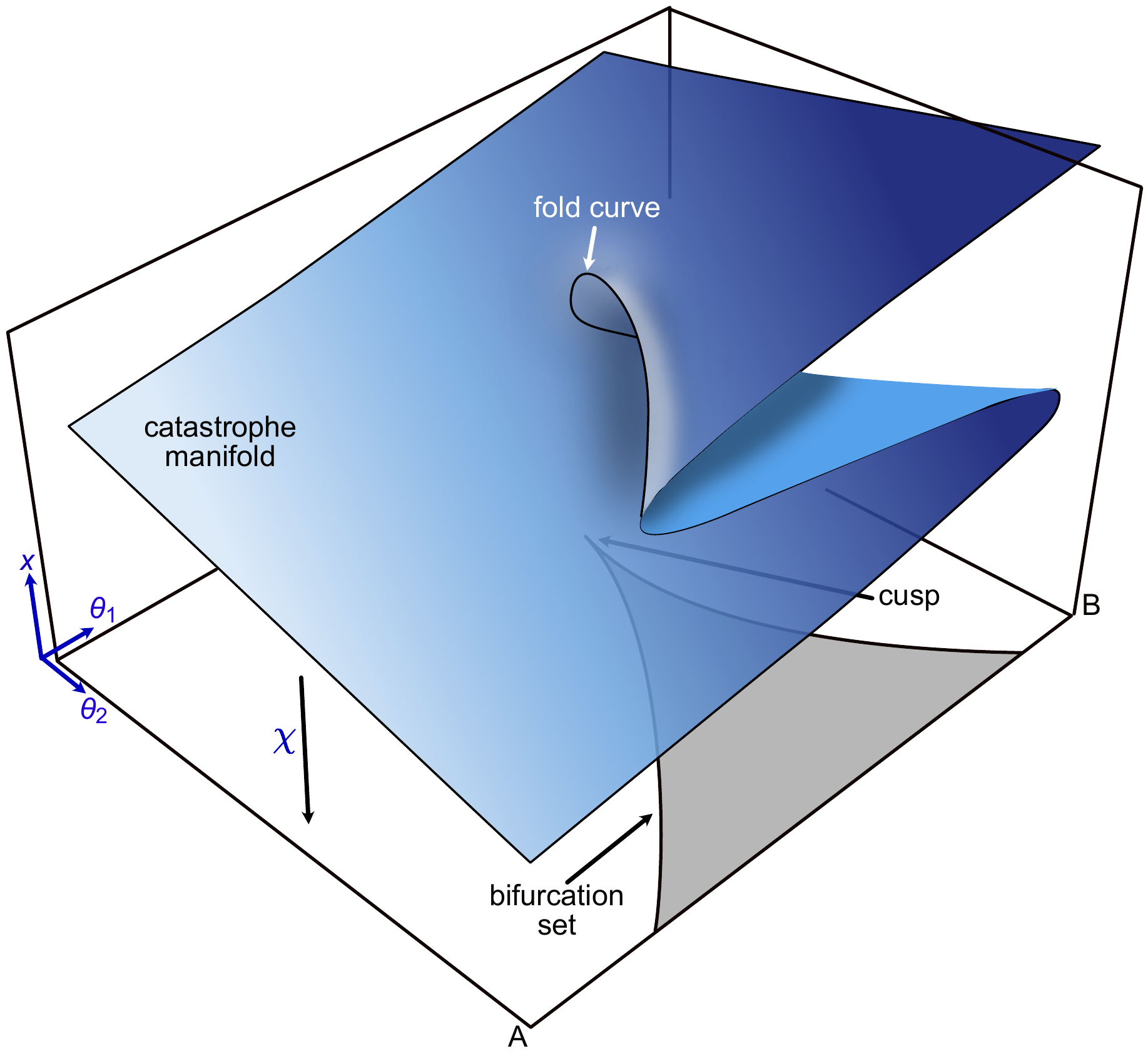}
	\caption{The catastrophe manifold $\catm$ for the
		cusp catastrophe.  Notice that although
		the bifurcation set has a cusp and hence
		is not smooth the corresponding fold
		curve in $\catm$ is smooth.}
	\label{fig:bluecusp}
\end{figure}

We now
consider a family of flows parameterised
by $\param\in\reals^c$ given by the
family of vector fields
$X_\param=X_\param (x)$.  The
catastrophe manifold is defined as
\[
\catm =\{ (x,\param): x \text{ is a rest point of } X_\param\}.
\]
Generically, (see \cite{arnold2013dynamical}, I Sect.\
1.4),
this is a $c$-dimensional submanifold of
$\reals^{n}\times\reals^{c}$.  We denote
by $\chi : \catm \rightarrow \reals^c$
the projection $\x=(x,\param )\mapsto
\param$ and let $\mathcal{S}$ denote the
set of singularities of $\chi$ i.e.\ the
set of points $\x\in \catm$ where $\chi$
is not a local diffeomorphism.  Then
$\mathcal{B}=\chi(\mathcal{S})$ is the set of
local bifurcation points. In the generic 
case this consists of fold and cusp points
and the only other bifurcation points
are where heteroclinic connections occur.
The latter are considered to be global
bifurcations because their effect is 
not restricted to a small neighbourhood.

Since in our discussion $c=2$, $\catm$
will be a surface in $\reals^{n+2}$ and
$\chi$ will be a projection into the
plane.  Fig.\ \ref{fig:bluecusp} is
useful to see the form of cusps and
folds in this case.

\def\u{\param_1}
\def\v{\param_2}
If $\param\in \reals^2$ is a fold
bifurcation point and $x\in\reals^n$ is
the corresponding rest point then $\x
=(x,\param)$ is in $\catm$ and, by
\cite{arnold2013dynamical}, I Sect.\ 1.4
(cf.\ Theorem 8.1 of
\cite{trotman_zeeman}), generically
there is a smooth curve $C$ in $\catm$
through $\x$ that consists of fold
points.  At $\x$ there are local
coordinates $(x_1,x_2)$ on $\catm$ and
$(\u,\v)$ on $\reals^2$ such that $\chi$
is locally given by $(\u,\v)=(x_1^2,x_2)$.

A universal unfolding\footnote{See
	\cite{trotman_zeeman} for a definition
	of universal unfolding.} of the cusp is
$\dot{x}=\pm x^3+\u x+\v$ and, in the minus sign
case the catastrophe manifold $\catm$ is
given by $x^3-\u x-\v=0$.  Therefore, the
map $(\u,x)\mapsto (x,\u ,\v=-\u x+x^3)$ from
$\reals^2$ to $\catm$ parameterises
$\catm$ in terms of $x$ and $\u$.  Thus,
in this parameterisation $\chi$ is given
by $(\u,x)\mapsto (\u,\v=-\u x+x^3)$ and this
is singular when $\u=3x^2$ which defines
a smooth curve $C$ in $\catm$.  The
bifurcation set $B_C$ is its image under
$\chi$, which is the set of points given
by $\u=3x^2,\v=- 2x^3$ i.e.\ $4\u^3=27\v^2$.
The $+$ case is entirely analogous.

In Appendix \ref{app:2} we prove the
following result telling us that in a
broad range of cases of interest to
us the catastrophe manifold has trivial
topology. On the other hand, as we will
demonstrate, the mapping $\chi$ has a
highly non-trivial structure.

\begin{proposition}\label{prop:disk}
	Suppose there are no bifurcation points
	near the boundary of the parameter
	domain $P$ and that $P$ is diffeomorphic to
	a 2 dimensional disk.  Then the connected components
	of $\catm$ are diffeomorphic to the disk
	$D^2$.
\end{proposition}

\noindent{\bf Standing hypothesis 1.}
We therefore assume throughout that
the connected components of $\catm$
are diffeomorphic to 2 dimensional disks. 
This is the case for a far more general set of circumstances
than those given in Prop.\ \ref{prop:disk}.

It follows from this hypothesis that any
simple closed curve $\Gamma$ in one of these
components $\catm_0$ separates $\catm_0$ into
two connected components, one of which
contains the boundary of $\catm_0$.
This component we call  the \emph{exterior} of 
$\Gamma$ and the other component the \emph{interior}.

In the discussion below we often lift to
the catastrophe manifold $\catm$ smooth
curves of the form $\param=\gamma (t)$,
$0\leq t\leq \tau_0$ in the parameter
space that do not meet any bifurcation
points.  A lift is a curve $\x=\Gamma
(t)$ such that $\chi (\Gamma (t))=\gamma
(t)$ and to ensure it is uniquely
determined we will need to fix a point
$\Gamma(t_0)$ such that $\chi
(\Gamma(t_0))=\gamma(t_0)$.  Often the
curve will be closed so that $\gamma
(0)\equiv \gamma (\tau_0)$ and then
$\Gamma $ can be taken to be closed
although one may have to extend the
domain of $t$ because the period of
$\Gamma$ is a multiple of $\tau_0$.  The
lifts exist because $\gamma$ meets no
bifurcation points.  

In some other cases we are able to lift
curves from parameter space to $\catm$
even though they intersect bifurcation
curves.  An example is the lift of
curves around a cusp in Sect.\
\ref{condsMob}, p.\ \pageref{condsMob}
and Fig.\ \ref{fig:around_cusp}.  These
lifts can be carried out because the
process is local and we can use the
universal unfolding of the bifurcation
to check that the lift is possible.

\subsection{Fold curves and bifurcation sets}

It follows from the above (cf.\
\cite{trotman_zeeman}, Theorem 8.1,
\cite{arnold2013dynamical}, I Sect.\
1.4) and the fact that our parameters
are 2-dimensional that generically
the set of singularities of 
$\chi$ (i.e. $\sings$) consists of a finite number of
1-dimensional submanifolds of $\catm$.
Moreover, if $C$ is one of these
submanifolds then every point on $C$ is
either a fold point or a cusp point.  We
call these submanifolds \emph{fold
	curves} and those that are circles
\emph{fold circles}. Clearly, since the closure of
$\catm$ is compact, a fold curve
that does not meet the boundary of $\catm$
must be a fold circle.

\noindent{\bf Standing hypothesis 2.}
We always assume that our parameterised system
is generic in the sense discussed above i.e.\ that
the fold curves are smooth curves in $\catm$.

\begin{proposition}\label{prop:BsubC}
	Generically, if $C$ is a fold circle then the
	associated bifurcation set $B_C =
	\chi(C)$ is a closed curve without any
	self crossings and $\chi |C :C
	\rightarrow B_C$ has degree one.  It is
	smooth except possibly at a finite
	number of cusp points and $B_C$
	separates $\reals^2$ into exactly two
	components, one of which (denoted
	$B_C^\text{int}$) is a
	disk.
\end{proposition}

\proof
See Appendix \ref{app:1} for
details.

\begin{termin}
	The \emph{bifurcation set} is the subset
	in parameter space of all bifurcation
	points.  A \emph{bifurcation curve} is
	any curve contained in the bifurcation
	set.  A \emph{fold curve} is a smooth
	curve of fold points in the catastrophe
	manifold $\catm$ and this is called a
	\emph{fold circle} iff it is a closed
	curve.  A \emph{fold bifurcation curve}
	is any set of the form $B_C=\chi(C)$
	where $C$ is a fold curve.  A
	\emph{bifurcation segment} or
	\emph{segment of a bifurcation curve} is
	a connected smooth curve contained in a
	set of the form $B_C=\chi(C)$
	where $C$ is a fold curve.
\end{termin}

A very useful construct associated
with fold curves and circles that
we will use a number of times involves
an analysis of their neighbourhoods.
Suppose $C$ is a fold circle. Consider a
small tubular
neighbourhood  $N$ of $C$. 
By definition of a tubular neighbourhood
there is a retraction $q: N\->C$ making 
$(q, N,C)$ a vector bundle whose zero 
section is the inclusion $C\-> N$. 
Then $N\setminus C$ has two components
one of which is contained in the interior of
$C$. We denote this component by $N^\text{int}_C$.
If $N$ is sufficiently small then
$N^\text{int}_C$ meets no other fold circles
and $\chi | N^\text{int}_C$ is injective.
This injectivity follows from the fold structure
given by equation (\ref{eqn:ndim}).
Let $N^\text{ext}_C$ denote the other
component of $N\setminus C$.

The following result
follows from  this discussion.

\begin{lemma}\label{lem:degree}
	If $C$ is a fold circle then $\chi$
	restricted to the interior $\breve{C}$
	of $C$ has degree 1 and at all points in
	$\breve{C}$ near $C$, $\chi$ is
	injective.
\end{lemma}

\subsection{Heteroclinic connections and flips}

We will be interested in systems where
there are two or more index 1 saddles.  In such
a system a codimension 1 global
bifurcation that can occur involves a
heteroclinic connection.  This occurs
when the unstable manifold of one of the
saddles $s_1$ intersects the stable
manifold of another saddle $s_2$.  In this case
we call $s_2$ the \emph{target saddle}
and $s_1$ the \emph{source saddle}.  If
this occurs at a parameter $\param$ then
we call this parameter value a \emph{flip
	point} and we refer to such bifurcations
as \emph{flip bifurcations} or just flips.

Generically, in 2-parameter families
these bifurcations occur on smooth curves
which we call \emph{flip curves} and
these curves end on fold bifurcation
curves in ways described below in Sect.\
\ref{sect:flipends}.

\section{Counting cusps on fold curves \& circles}

\subsection{The fold orientation}\label{sect:orient}

Consider the fold equation
(\ref{eqn:fold}).  Close to $x=0$ the
$\dot{x}$ is always $\geq 0$ or $\leq
0$.  Thus close to a fold point there is
a definite direction of flow on a center
manifold and this induces an orientation
on the center manifold and its tangent
space $\ell (\x )$. Moreover, $\ell (\x )$
varies smoothly with $\x\in C$.

Using this smoothness we have that this 
orientation is locally consistent
on any segment of a fold curve $C$ which
contains no cusp points but it switches
at cusp points as can be verified by
looking at the normal form
(\ref{eqn:cusp}).  Thus we have the following
proposition.

\begin{proposition}\label{lem:cusp2}
	Consider a fold curve $C$ whose endpoints are
	not cusps. Then, for a generic parameterised system, 
	the number of cusps in $C$
	equals the number of orientation switches in $C$.
\end{proposition}

\subsection{Saddle \& near-fold bundles}
Now consider a smooth closed curve
$\Gamma$ in $\catm$ such that, for
each point $\x=(x,\param)\in\Gamma$, $x$
is either an index 1 saddle point
or a pseudo-hyperbolic attracting rest point
(Sect.\ \ref{para:pseudo-hyperbolic}).
In this case we say that such a curve $\Gamma$ is
\emph{pseudo-hyperbolic}.

Then we can
associate to such a point $\x$ the
1-dimensional tangent space $E^c(x)$ to
the center manifold $W^c(x)$ of the
rest point at $x$.  These vary smoothly with
$\x$ and so we have a bundle
$\pi:E^c_\Gamma\rightarrow \Gamma$ over
$\Gamma$ with fibre the line $E^c(x)$.
Clearly this bundle $E^c_\Gamma$ must be
either a cylinder or a M\"obius band and
the topology of this bundle encodes
important information.

\subsection{Fold circle bundle}
For the special case where $\Gamma$ is
a fold circle $C$ we call the bundle
$E^c_C$ the \emph{fold circle bundle}
and denote it $\mathcal{C}_C$. 

Now if $C$ is a fold circle and we start
at a point $\x\in C$ which is not a cusp
point and traverse the circle, when we
arrive back at $\x$ the orientation will
have changed a number of times.  If this
number is odd when we arrive back at
$\x$, the orientation will be the
opposite to when we started out.  Thus
for the fibres of
$\mathcal{C}_C$ to glue smoothly at $\x$ and
for the orientation to be constant 
on fibres near to $\ell (\x )$
(since $\x$ is not a cusp point), $\mathcal{C}_C$ must
be a M\"obius band.  Conversely, if
$\mathcal{C}_C$ is a M\"obius band then
the number of switches must be odd.
Thus, for a generic parameterised system,
we have the following.

\begin{proposition}\label{prop:4}
	If $C$ is a fold circle,
	$\mathcal{C}_C$ is a M\"obius band if and only 
	if $C$ contains an odd number of cusps.
\end{proposition}

We call such fold circles \emph{M\"obius}
fold circles.

\subsection{Conditions for a fold circle
	to be M\"obius}\label{condsMob}

We now investigate the immediate neighbourhood of a
M\"obius fold circle in $\catm$. In particular, 
we demonstrate that a
simple closed curve $\Gamma$ in $\catm$ has
the bundle $E^c_\Gamma$
a M\"obius band
if and only if it contains an odd number of 
M\"obius fold circles,
and we show that,
if $C$ is a M\"obius fold circle, then on the
exterior of $C$ close to $C$,  $\chi$ has degree two.
As a corollary we see that although a
fold circle can contain simple fold circles,
it cannot contain a M\"obius one.

\begin{proposition}\label{prop:remove}
	Suppose that $\Gamma$ is a smooth
	simple closed pseudo-hyperbolic curve in $\catm$ 
	with $E^c_{\Gamma}$ a M\"obius band.
	Then in the generic situation,
	$\Gamma$ contains an odd number
	of M\"obius fold circles
	in its
	interior.  Moreover, if
	$C$ is one of these M\"obius fold circles
	one can find another such curve
	$\Gamma'$ such that $C$ is
	the only fold circle contained
	in the interior of $\Gamma'$
	except possibly for simple fold circles
	contained in the interior of $C$.
\end{proposition}

\proof
See Appendix \ref{sect:E}.
\qed

Now consider a M\"obius fold
circle $C$.
Then $C$ contains an odd number
of cusps.  Suppose that the folds in $C$
involve a saddle and an attractor with
the interior of $C$ consisting of points
$\x=(x,\param)$ where $x$ is an
attractor.  Let $B_C^\text{int}$ be the
open disk bounded by $B_C$.

\begin{figure}[H]
	\centering
	\includegraphics[width=\linewidth]{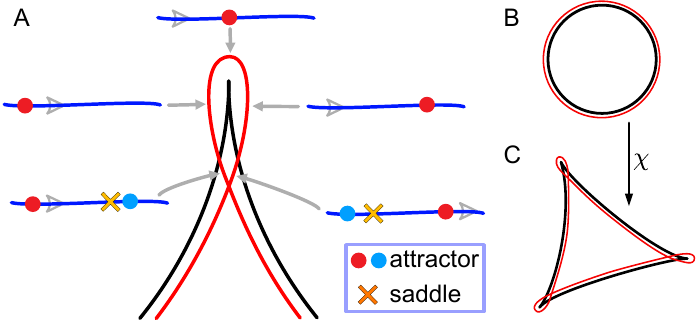}
	\caption{This shows how the disposition
		of the rest points involved in a
		standard cusp vary as one follows the
		projection into parameter space under
		$\chi$ of a closed curve (red) of
		attractors in $\catm$ which approximates
		the fold curve giving the cusp.  Since
		the red curve is close to the
		bifurcation set there is a center
		manifold through the red attractor.  The
		grey arrow shows the orientation on the
		center manifold. The cusp shown is a
		standard one and going round the loops
		swaps the attractors; for a dual cusp
		the roles of saddles and attractors are
		interchanged.}
	\label{fig:around_cusp}
\end{figure}

Now consider a closed curve $\gamma$
that has one self intersection inside
$B_C$ near each cusp (temporarily called nodes),
loops around the cusps as shown in Fig.\
\ref{fig:around_cusp}C, and otherwise 
lies entirely in the interior
$B_C^\text{int}$ of $B_C$ near to $B_C$. 
Provided it is sufficiently close
to $B_C$  
this curve can be lifted
to a smooth curve $\Gamma$ in
$\catm$ so that (i) $\chi(\Gamma)=\gamma$,
and (ii) if $(x,\param)\in\Gamma$, $x$ is an
attractor of $X_\param$ which we denote by $x(\param)$.

Since the curve $\gamma$ lies close to
$B_C$, the attractor $x(\param)$,
$\param\in\gamma$, lies on a smoothly
changing center manifold.  Since the
center manifold varies smoothly, we can
give it a consistent orientation
$\mathcal{O}$ for all $\param\in\gamma$.  
At each node and
at all points of $\gamma$ in
$B_C^\text{int}$, the unstable manifold
of the saddle intersects the stable
manifold of the attractor created in the
fold bifurcation of $B_C$.  

We define
the \emph{relative orientation} of the saddle at
$\param\in\gamma\cap B_C^\text{int}$ to
be the relative orientation of the
saddle to the attractor using the
orientation $\mathcal{O}$.  
Note that although the
orientation $\mathcal{O}$ is constant
the relative orientation of the
saddle can change.
In fact, this
relative orientation is constant along any
segment of $\gamma$ wholly inside
$B_C^\text{int}$ but flips as $\param$
moves around the loops going from the
node around the cusp (Fig. \ref{fig:around_cusp}A).  As a
consequence, if there are an odd number
of cusps then starting from a node and
doing one revolution around $\gamma$
back to that node results in a flip in
the relative orientation.  But at the node
$\param$, since it is in
$B_C^\text{int}$ and is close to the
cusp, the system has two attractors and a
saddle.  Thus traversing $\gamma $
results in the attractors swapping.

A similar picture holds for dual cusps
with the roles of the attractors and
saddles interchanged and with the
unstable manifold of the saddle
replacing the role of the center
manifold. 

Now let $U_C$ be an open set
in $\catm$ that contains $C$ in its interior
and meets no other fold curves in
the exterior of $C$
and consider
$\chi_1=\chi|U_C$.  The choice of nodes
above was arbitrary apart from being in
$B_C^\text{int}$ close to a cusp. Consequently,  we
deduce that there is a neighbourhood $U$
of the node such that $\chi_1^{-1}(U)$
consists of two disks on each of which
$\chi_1$ is orientation preserving (for
reasons given below).
Consequently, $\chi_1$ has degree 2.
This gives the following lemma and its
very useful corollary.

\begin{theorem}\label{thm:1}
	Consider the following conditions
	on a fold circle $C$.
	(i) $C$ is a M\"obius fold circle;
	(ii) $C$ contains an odd number of cusps;
	(iii) there is a simple closed
	pseudo-hyperbolic curve $\Gamma$ in
	$\catm$ that contains in its interior $C$ 
	and no other fold circles that are
	in the exterior of $C$ such that (a) under $\chi$,
	$\Gamma$ double covers a curve $\gamma$
	in the parameter space and (b)
	$E^c_\Gamma$ is M\"obius; and (iv) $C$
	is contained in an open disk
	$\mathcal{G}$ in $\catm$ on which $\chi$
	has degree 2.
	Generically, (i) $\iff$ (ii) $\iff$ (iii)
	$\Rightarrow$ (iv).
\end{theorem}

\proof
We already proved (i) $\iff$ (ii) in Prop.\ \ref{prop:4}.
That (i) implies (iii) follows from the above
discussion. The curve $\Gamma$ discussed there
double covers the curve $\gamma$ (cf.\ Fig.\ \ref{fig:around_cusp}).
Moreover, since $\gamma$ can be taken arbitrarily close to
$B_C$, $\Gamma$ can be taken arbitrarily close to
$C$. Thus the fibres of the bundles $\mathcal{C}_C$
and $E^c_\Gamma$ will be close and consequently 
$\mathcal{C}_C$
and $E^c_\Gamma$ will both have the same M\"obius band topology.

To see that (iii) implies (i) we need to
use Prop.\ \ref{prop:remove}.
This asserts that we can modify the
curve $\Gamma$ in (iii) so that it
contains in its interior the fold
curve $C$ and no other fold points in the exterior of $C$.
Then the modified $\Gamma$ is homotopic
to curves $\Gamma'$ arbitrarily close to
$C$ but exterior to $C$.  Since they are
connected by a homotopy, $E^c_\Gamma$ and
$E^c_{\Gamma'}$ are homeomorphic, and
since $\Gamma'$ is close to $C$,
$E^c_{\Gamma'}$ and $\mathcal{C}_C$ are
homeomorphic.

To see that (iii) implies (iv) note that
$\chi$ is a local diffeomorpism at
points of $\Gamma$ and, moreover, since
there are no fold points on $\Gamma$ the
sign of the Jacobian of $\chi$ at two
points on $\Gamma$ is the same.  It
follows immediately that if $\Gamma'$ is
any simple closed curve containing
$\Gamma$ in its interior then $\chi$ has
degree 2 on the interior of $\Gamma'$
which is a disk.
\qed

\begin{corollary}
	If $C$ is a fold circle
	then there
	are no M\"obius fold circles
	contained in its interior.
\end{corollary}

\proof
This follows from part (iv) of the
Theorem and the fact that the
restriction of $\chi$ to the interior of
a fold circle has degree 1 by Lemma \ref{lem:degree}.
\qed

More generally, if $\mathcal{G}$ is
a subset of $\catm$ diffeomorphic to a disk and
$\chi |\mathcal{G}$ has degree 1 then $\mathcal{G}$
contains no M\"obius fold circles.

\subsection{Cusps and bifurcation diagrams}

In applications it is common to plot
bifurcation diagrams in which the
position of the rest points is plotted
against a single bifurcation parameter
$\param_1$.  For example for bistable
systems one obtains a S or Z shaped
multivalued graph over the parameter. 
If there is another
parameter $\param_2$ involved in the
system it is then of interest to know
whether the 2-parameter bifurcation
system contains cusps. 

Now we extend the cusp index for use in
such a situation to determine the number
of cusps.  We restrict to fold
bifurcations between attractors and
index 1 saddles but it will be clear
that it applies to other fold
bifurcations.

\begin{figure}[H]
	\centering
	\includegraphics[width=\linewidth]{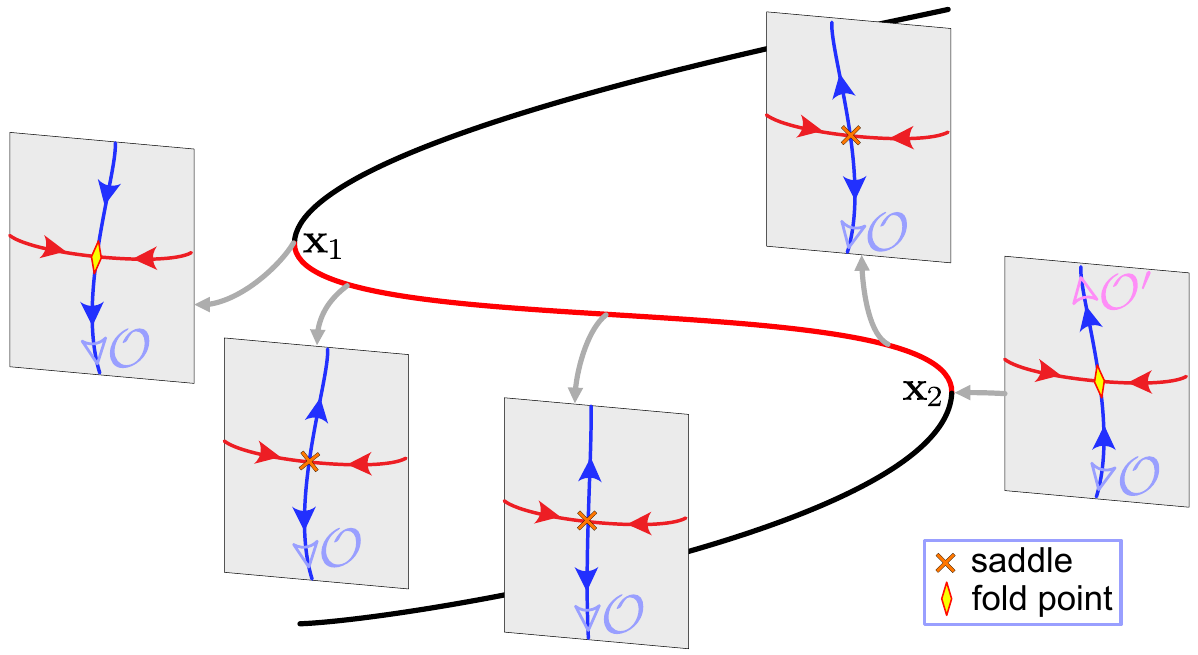}
	\caption{The black and red curve is a
		curve in the catastrophe manifold such
		as that over the line AB in Fig.\ \ref{fig:bluecusp}.
		The red part corresponds to saddles.  The
		left hand fold point determines the
		orientation on the center manifold shown
		using the hollow arrow.  This induces
		the corresponding orientation on the
		saddle points in the systems
		corresponding to the points along the
		red curve.  Then all the systems on the
		red curve close to the right hand fold
		point have the opposite orientation to
		that fold point.  This implies that on
		any fold curve joining the two fold
		points shown there are an odd number of
		cusps.
	}
	\label{fig:along_bcurve}
\end{figure}

Suppose that $\Gamma (t)$, $0\leq t\leq
1$ is a smooth curve in $\catm$ that
connects two fold points $\x_i$, $i=1,2$, and
except at its end-points consists of
index 1 saddles.  Suppose moreover that
the tangent $E^u (\x )$ to the
unstable manifold of these saddles $\x$
varies smoothly as $t$ varies.  

Given this, we can compare the
orientation of $\ell (\x_1)$ with that
of $\ell (\x_2)$.  Use the orientation
of $\ell (\x_1)$ at the fold point
$\x_1$ to put an orientation on 
$E^u (\Gamma (t))$ for $t$
close to $0$ using the fact that $E^u
(\Gamma (t) )$ converges to $\ell (\x_1)$ as
$t$ tends to $0$ and extend this to all
$0\leq t\leq 1$ by continuity.  Since
$E^u (\Gamma (t) )$ converges to $\ell
(\x_2)$ as $t$ tends to $1$ this
produces an orientation $\mathcal{O}$ on
$\ell (\x_2)$ and the question is
whether this is the same as that
$\mathcal{O}'$ induced by the fold at
$\x_2$ (see Fig.\ \ref{fig:along_bcurve}).

\begin{figure}[H]
	\centering
	\includegraphics[width=\linewidth]{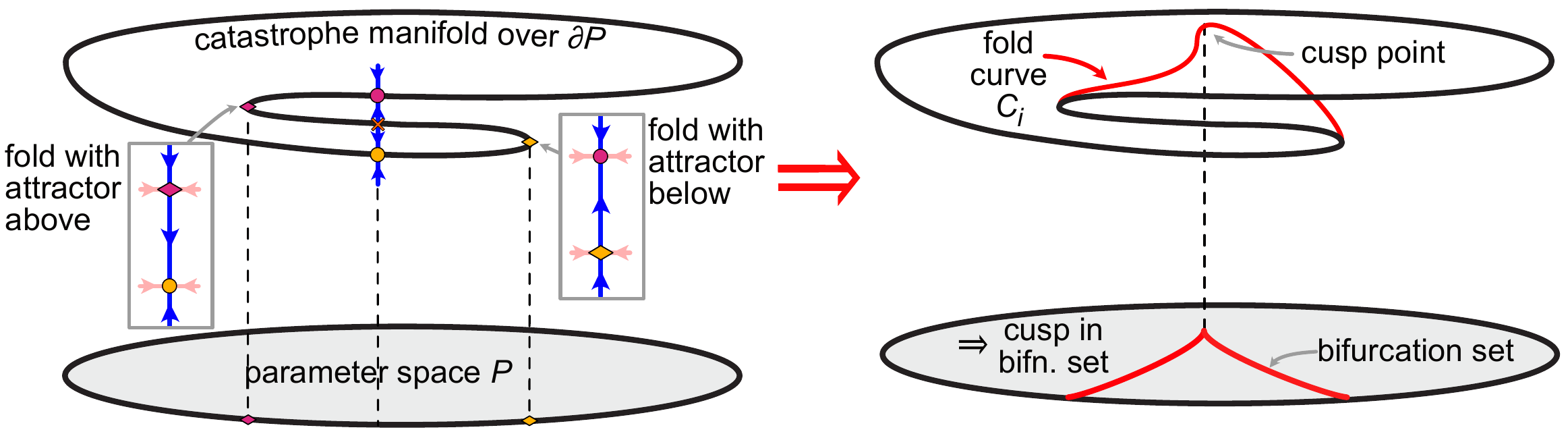}
	\caption{Over part of the
		boundary $\del P$ of a 
		2-dimensional domain of parameter
		space $P$ we assume we
		have an S-shaped bistable section and
		that the system is monostable over the
		rest of $\del P$.  Then if 
		Condition (*) holds it follows that
		inside $P$ there must be at least one
		cusp.  This is because there must be a
		fold curve joining the two folds shown
		and on this fold curve the two folds
		shown have opposite orientations.
	}
	\label{fig:cuspexistssup}
\end{figure}

\begin{proposition}\label{prop:cuspexist1}
	If  $\,\x_1$ and $\x_2$ are on a common
	fold curve $C$ and the orientations 
	$\mathcal{O}$ and $\mathcal{O}'$
	of $\ell (\x_2)$ are incoherent
	then there is an odd number of cusps on $C$
	between $\x_1$ and $\x_2$.
\end{proposition}

In a number of applications of CT (e.g.\
\cite{zeeman1977CT}) one is given the behaviour on the
boundary of the parameter domain and one
wants to deduce the bifurcation
behaviour inside.  By far the most
considered situation in this regard is
where the boundary behaviour is as in
Fig.\ \ref{fig:cuspexistssup}(Left) i.e.\
one has a parameter space $P$ which is a
2-dimensional disk-like region and the
catastrophe manifold over the boundary
$\partial P$ is bistable over part of
$\partial P$.  In this
case one would like to know if there is a
cusp inside $P$.

Suppose the boundary $\partial P$ is
given by $\theta=\theta(s)$, $0\leq
s\leq 1$, and that the interval where
there is bistable behaviour is given by
$\theta(s)$, $s_1\leq s\leq s_2$, so
that there are folds at $s=s_1$ and
$s=s_2$. To obtain the
required result about cusps we need the
following condition on the two
attractors involved in the two folds: (*)
for at least one of these attractors,
the attractor varies smoothly with $s$
in $(s_1,s_2)$ and for all $s$ in
$(s_1,s_2)$ the unstable manifold of the
saddle point intersects the basin of
this attractor.  This condition is
automatic if the state space is
1-dimensional but not so in higher
dimensions.  To understand why see Fig.\
\ref{fig:bistable_but_no_switch} in
Appendix \ref{sect:just}.  Note that we
are not assuming that the saddle and
attractors that are involved in the fold
bifurcations are all the rest points in
the system.  If they are then Condition
(*) is redundant.  An example of a
bistable 1-parameter system that swaps
orientation but where the two folds are
not on a common fold curve is given by
taking a straight line in the parameter
space in Fig.\ \ref{fig:CEU} which is
roughly parallel to one of the boundary
curves and does not meet the deltoid
$\mathcal{B}$.  Then we have the following
theorem:

\begin{theorem}\label{corollary:1}
	If the two fold points in $\catm$
	corresponding to the two folds at
	$\theta=\theta(s_1)$ and
	$\theta=\theta(s_2)$ are on a common fold
	curve in $\catm$ and condition (*) holds
	then generically there are an odd number
	of cusps in $P$.
\end{theorem}

\proof
By Condition (*), as $s$ moves between
the folds at $s_1$ and $s_2$ the
unstable manifold of the saddle must
intersect the basin of the smoothly varying attractor. 
The identity of this attractor remains constant for all $s$ and we call it $A$. 
The unstable manifold can
also be involved in heteroclinic
bifurcations but, in generic systems,
for all but a finite number of $s$
values, it will also intersect the basin of
a different attractor of the system at
$\theta (s)$ which we denote by $B(s)$.
At these finite number of $s$ values 
the identity of $B(s)$ may change.
In the generic case when $s$ in
$(s_1,s_2)$ is close to $s_1$ or $s_2$
the attractors $A$ and $B(s)$ 
vary smoothly and they are connected
by the unstable manifold of the saddle.
If for $\x_1$ in Prop.\
\ref{prop:cuspexist1} we take the fold
involving $A$, then $\mathcal{O}$
points away from $A$ when $s$ is near
the bifurcation value.  Since the
unstable manifold of the saddle and $A$ vary
smoothly, $\mathcal{O}$ remains doing this for all
$s\in (s_1,s_2)$ and therefore is
incompatible with the orientation at the
fold involving $B(s)$ since this orientation points
away from $B(s)$.  Thus the odd number
of cusps follows from Prop.\
\ref{prop:cuspexist1}.
\qed

\noindent{\bf Historical Note.}
Results similar to Theorem.\
\ref{corollary:1} for the case where
$n=1$ were obtained by Stewart in
\cite{stewart1980catastrophesup} using
entirely different methods based on
singularity theory.  Ian Stewart pointed
out to us a paper of Tim Poston
\cite{poston1978deducing} that discusses
the problem in Theorem.\
\ref{corollary:1} and mentions a
preprint by E.\ C.\ Zeeman (\emph{A
boundary value problem involving cusps})
in which this problem is treated.
However, in Poston's discussion the need
to check the incompatibility of the
orientations $\mathcal{O}$ and
$\mathcal{O}'$ is not mentioned.  If it
existed this manuscript seems to have
disappeared and is not listed in the
Zeeman Archive
(www.lms.ac.uk/2015/zeeman\_archive).
One of us (DAR) recalls Zeeman
mentioning that someone else had proved
this result but we cannot trace such a
paper.

\section{Characterising 3-attractor MS components}

\begin{figure}[H]
	\centering
	\includegraphics[width=0.6\linewidth]{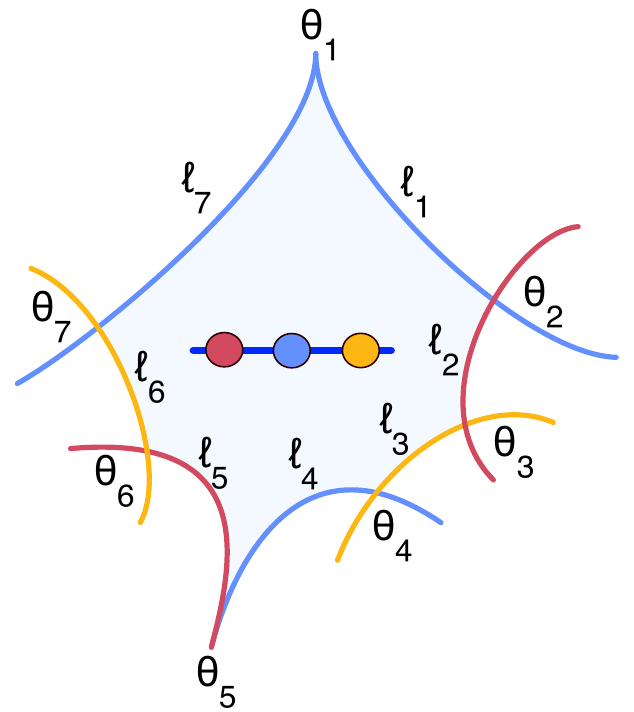}
	\caption{
	}
	\label{fig:aroundD}
\end{figure}

\def\cp{\theta}
Now we consider 3-attractor MS
components $\D$ in 2-parameter systems with compact
dynamics.  Our aim is to understand
the bifurcations in the boundary of $\D$ 
and we will do this by deducing a set of rules for the 
way in which cusps, crossings and flip end points
are disposed around the boundary.

We denote the attractors by
$A,B$ and $C$.  Since they do not
bifurcate in $\D$ they vary smoothly with
the parameters and maintain their
identity.

\def\L{\mathcal{L}}
As one proceeds around the boundary
$\partial \D$ of $\D$ there is a sequence of
codimension 2 points $\cp_1\,\ldots
\cp_N$ with $\cp_N=\cp_1$ as described
in Fig.\  2 of {\bf I}, namely cusps, dual cusps,
and crossings (see Fig.\
\ref{fig:aroundD}).  
(The other relevant codimension 2 points,
namely, the two types of endpoints
of flip curves are dealt with in
Sect.\ \ref{sect:flipends}.)
If we know these
then we know all the bifurcation
structures associated with $\del
\D$.  The points $\cp_i$ and $\cp_{i+1}$
are joined by a smooth fold bifurcation
curve segment $\ell_i$.  The
bifurcations on this segment involve one
of the attractors $X\in\{ A,B,C\}$ and
we denote this by $\ell_i\in\L_X$ or say
that the attractor identity of $\ell_i$
is $X$.  For a sequence $\ell_i,\ldots
,\ell_{i+j}$ with $\ell_k$ having
attractor identity $X_k$ we write
$\ell_i,\ldots ,\ell_{i+j}
\in\L_{X_i,\ldots ,X_{i+j}}$.  There are
strong constraints on what sequences can
occur as we now describe.

\subsection{Crossings of bifurcation curves}

\begin{figure}[H]
	\centering
	\includegraphics[width=0.9\linewidth]{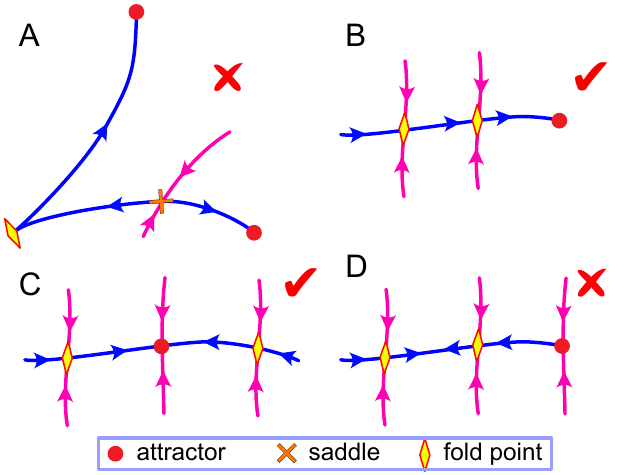}
	\caption{(A) If at
		the central attractor the connector has
		a cuspoidal structure then a fold
		bifurcation of this attractor is
		impossible.  This is proved in Lemma
		\ref{lem:nofold} Appendix
		\ref{App:no_fold} and basically follows
		from the incoherence of the two arrows
		on the branches approaching the fold
		point.   (B-D) This illustrates examples
		of what dynamics on the connector are
		possible and impossible at points where
		bifurcation curves cross.  The
		orientations of two folds at adjacent
		rest points must be the same. 
		There are no restrictions on the way
		the blue connector approaches the red attractor.
		(B) allowed since the arrows are
		consistent, (C) allowed since the folds
		are not adjacent, (D) not allowed as
		center and bifurcating peripheral have
		opposite orientations.
	}
	\label{fig:compat}
\end{figure}

In general when two bifurcation curves
cross, the two bifurcations may be
completely independent.  However, this
is not the case when the corresponding
rest points are both on the same smooth
connector and are adjacent.  This is
because, when it is smooth, the connector
is contained in any center manifold of a
fold point on the connector and hence
the fold orientations (Sect.\
\ref{sect:orient}) of the two folds
provide orientations for the connector
that must be the same for the
simultaneous bifurcations to be allowed.
See Figs. 
\ref{fig:compat} and \ref{fig:crossings}.

Now consider the case where the system
has 3 attractors $P_1$, $P_2$ and $X$,
and two saddles, where $X$ is the central
attractor. Assume all these are on a smooth connector.

\def\O{\mathcal{O}}
Firstly, note that there is only one
possibility for the orientation of a
fold involving a peripheral attractor.
The vector field near the fold point must
point along the connector towards the
other rest points on the connector.  In
particular, the orientations
$\O_1$ and $\O_2$ of folds
involving the two peripheral attractors 
$P_1$ and $P_2$ are opposite.
Given this, when considering a fold curve
or fold bifurcation curve involving the
central attractor $X$ we say \emph{the fold of
	$X$ is compatible with} $P_i$ (with $i=1$
or $2$) to indicate that the orientation
of the fold involving $X$ is
$\mathcal{O}_i$.

Now consider two smooth segments
$\ell_1$ and $\ell_2$ of fold
bifurcation curves that cross each
other.  Suppose that the segment
$\ell_i$, $i=1,2$, involves the
attractor $A_i\in \{ P_1,P_2,X\}$.
Then the following lemma follows 
immediately from
the above discussion.

\begin{lemma}\label{lemma:2}
	The only restrictions on crossings concern the case
	where one of the $A_i$ is the central attractor $X$.
	In this case the folds on
	$\ell_1$ and $\ell_2$ all have the same orientation.
\end{lemma}

\subsubsection{Crossings near cusps}

A key fact in this section is that 
since a bifurcation curve $B_C$ is the 
image of a single fold curve in $\catm$, 
as one moves along $B_C$, either the attractor 
or the saddle that is bifurcating remains constant.

The bifurcation curve through a cusp is
the projection under $\chi$ of a fold
curve in $\catm$.  Near the cusp point
the bifurcation curve in parameter space
$\Omega$ consists of two smooth segments
$\ell$ and $\ell'$ joined at the cusp.
In a neighbourhood of the cusp point in
$\Omega$ there is a smooth center
manifold containing the rest points
involved in the bifurcation because at the
cusp point the central manifold is
normally hyperbolic and hence varies
smoothly with the parameters and
contains the bifurcating rest points.  

\begin{figure}[H]
	\centering
	\includegraphics[width=\linewidth]{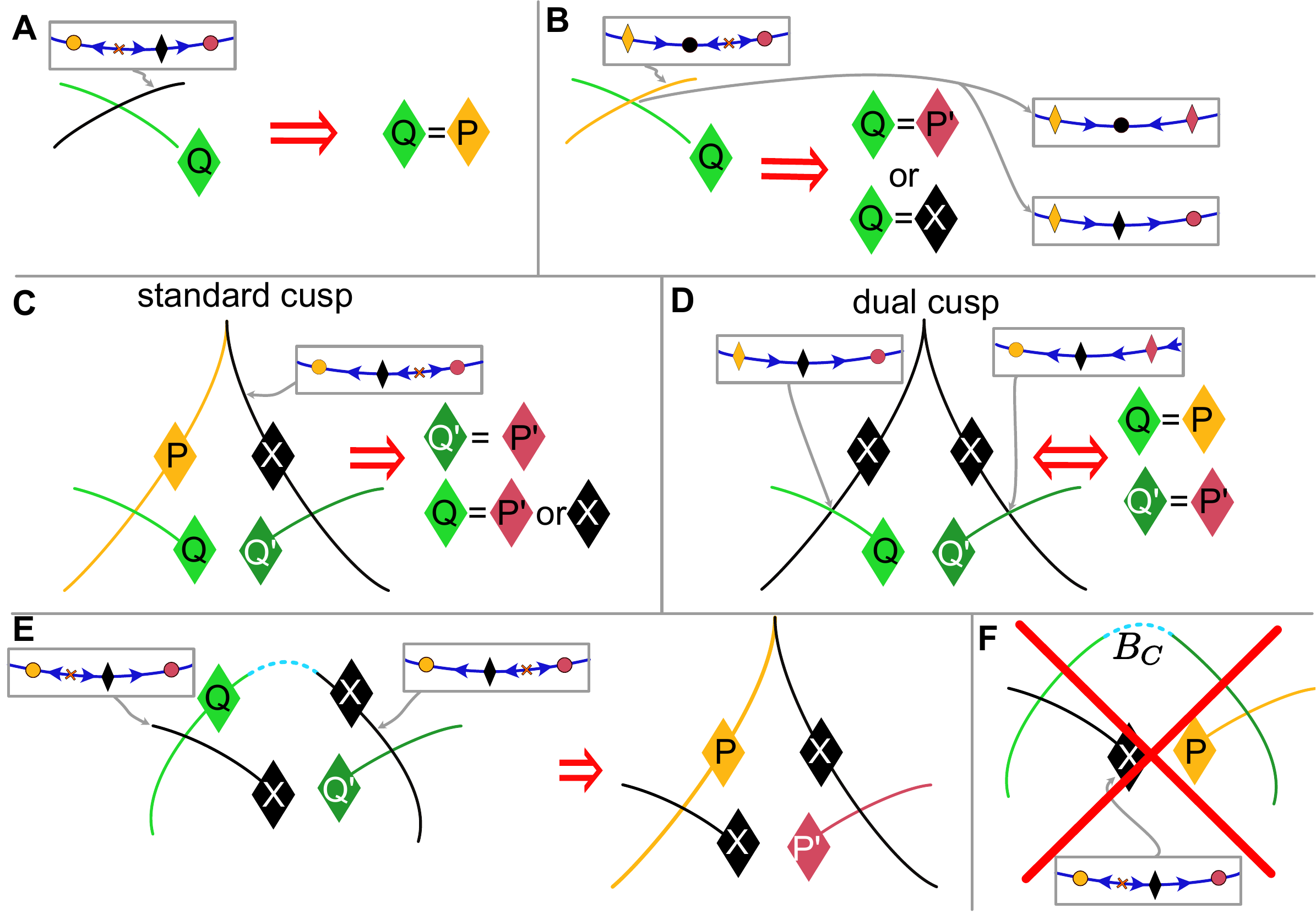}
	\caption{Illustration of Lemma \ref{lemma:2} and Theorem \ref{Prop:7}.
		{\bf A.} The orientation of the fold 
		bifurcation involving the central 
		attractor determines the possible 
		crossings of the fold curve.
		{\bf B.} The peripheral attractor 
		involved in the fold constrains 
		the possible crossings of the fold curve.
		{\bf C.  Standard cusp.} Fold curves 
		crossing the cusp's branches satisfy 
		the conditions in (A) and (B).
		{\bf. D. Dual cusp.} Fold curves 
		crossing the cusp's branches satisfy 
		the conditions in (A) and (B) 
		and the crossings imply the existence 
		of the cusp.
		{\bf E.} The orientations of the folds of 
		the central attractor imply the existence of a cusp.
		{\bf F.} Certain consecutive crossing are not possible.
		{\bf Legend}: A diamond shape denotes a fold 
		bifurcation of the specified attractor.
		Bifurcation curves are coloured accordingly.}
	\label{fig:crossings}
\end{figure}

For a dual cusp the folds along both of
the segments $\ell$ and $\ell'$ involve
the central attractor and they have
opposite orientations on the two
segments (Fig.\ \ref{fig:crossings}D).  
For a standard cusp the two
smooth segments either side of the cusp
correspond to a common saddle and two different attractors. 
One of them is the central attractor
$X\in\{A,B,C\}$ and the other a
peripheral attractor $P\in\{A,B,C\}$
(Fig.\ \ref{fig:crossings}C).

As a corollary of Lemma \ref{lemma:2},
if $C$ is a fold curve with such a
standard cusp involving $P$ and $X$
compatible with the other peripheral
attractor $P'$, then a fold curve
crossing the $P$ side involves the
bifurcation of $P'$ or $X$ compatible
with $P$, and a fold curve crossing the
$X$ side involves the bifurcation of
$P'$ (Fig.  \ref{fig:crossings}C).

Now consider a fold curve $C$ in
$\mathcal{M}$ such that $B_C=\chi (C)$
is contained in the boundary of a
3-attractor MS component and which is
such that moving out of the component
across $B_C$ involves the loss of an
attractor.  Suppose that the attractors
in the MS component are (central) $X$,
(peripheral) $P_1$ and $P_2$.  Suppose
that on $B_C$ there are two crossing
points at $\param_1$ and $\param_2$ and
none in between them and suppose the
fold bifurcation curves which cross at
these points are $\ell_1$ and $\ell_2$
respectively.

\begin{theorem} \label{Prop:7}
	1. $\ell_1$ and $\ell_2$ are
	associated with different peripheral
	attractors if and only if $C$ is a fold
	curve for $X$ and it contains an odd
	number of dual cusps (Fig.
	\ref{fig:crossings}D).\\
	2. If the folds in $\ell_1$ at
	$\param_1$ and in $B_C$ at $\param_2$
	involve the central attractor $X$ and
	the former (resp.\ latter) have
	orientation compatible with $P_1$
	(resp.\ $P_2$) then there is an odd
	number of standard cusps in $B_C$ between
	$\param_1$ and $\param_2$ and the fold
	in $\ell_2$ at $\param_2$ involves $P_2$
	(Fig.  \ref{fig:crossings}E).\\
	3.  It is not possible that the fold in
	$\ell_1$ at $\param_1$ involves $X$ with
	orientation compatible with $P_1$ and
	the one in $\ell_2$ at $\param_2$
	involves $P_1$ (Fig.
	\ref{fig:crossings}F).
\end{theorem}

\proof
Proof of 1.  If $C$ is a fold curve for
$X$ and it contains an odd number of
dual cusps, then the orientation of the
$X$ fold is different at $\x_1=(x, \param_1)$ and
$\x_2=(x, \param_2)$.  Hence, by lemma \ref{lemma:2}
the crossing curves involve the
bifurcation of different peripheral
attractors.

Conversely, assume that the crossing
curves involve different peripheral
attractors and in particular that
$\ell_1$ at $\param_1$ involves $P_1$.
Then, $B_C$ at $\param_1$ involves
either $X$ compatible with $P_1$ or
$P_2$.  Assume that it is $P_2$.  Since
the fold in $\ell_2$ at $\param_2$
involves $P_2$ there must be a cusp in
$C$ that changes the bifurcating
attractor.  This cusp is a standard cusp
that involves $X$ compatible with $P_1$,
which again can not cross $\ell_2$ at
$\param_2$.  Therefore, $C$ at
$\param_1$ involves $X$ compatible with
$P_1$.  Similarly, $C$ at $\param_2$
involves $X$ compatible with $P_2$.
Therefore, $C$ is a curve of $X$ folds
with different orientations at
$\param_1$ and $\param_2$ and therefore
it contains an odd number of cusps
between these two points.

Proof of 2.  By lemma \ref{lemma:2} 
the fold along $B_C$
near $\param_1$ involves $P_1$ and the 
fold along $\ell_2$ near $\param_2$ involves $P_2$.
Therefore $B_C$ contains an
odd number of dual cusps between these two
points.

Proof of 3. 
The condition at $\param_1$ implies that
at that point in $B_C$ the saddle between 
$P_1$ and $X$ bifurcates with $P_1$.
Thus there cannot be a crossing by $P_1$
at $\param_2$ as this would also have
to involve the same saddle to create the additional fold.
\qed

\subsection{Heteroclinic bifurcations
	(flips) and flip curves}\label{sect:flipends}
Before proceeding we need to discuss
flip curves and how they end.  This relates to the
question of whether or not the connector
is smooth.  As one moves along a
bifurcation curve such smoothness can
change. This happens at certain endpoints
of flip curves.

\begin{figure}[H]
	\centering
	\includegraphics[width=\linewidth]{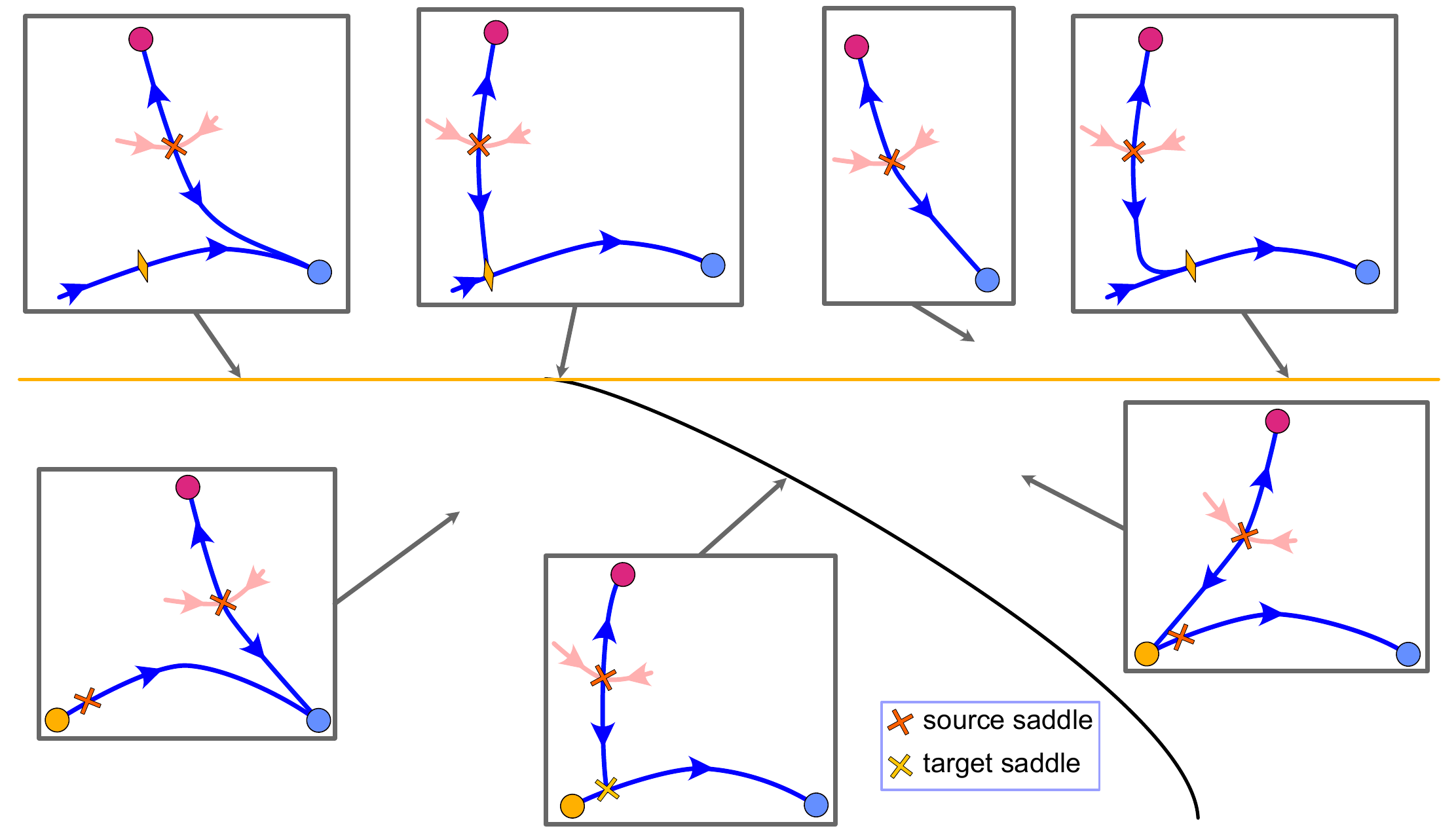}
	\caption{This shows how a generic flip
		curve (black) meets a fold curve (yellow)
		when the fold occurs at the central
		attractor.  See \cite{Vegter82}.}
	\label{fig:fold+flip_point}
\end{figure}

\subsubsection{Target saddles}

The first way that flip curves can end is by a bifurcation of the corresponding target saddle.

\begin{proposition}\label{prop:9}
	Consider a smooth fold bifurcation
	segment $\ell $ involving the attractor
	$X\in\{ A,B,C\}$.  Suppose that at a
	point $\theta_0$ on $\ell $ the
	connector changes from smooth to
	non-smooth.  Then, generically, $X$ is
	central at $\theta_0$ which is the
	end-point of a flip curve where the
	target saddle is destroyed in a fold
	bifurcation with $X$.  There is a
	neighbourhood $U$ in $\ell$ of such a
	point $\theta_0$ such that for $\theta$
	in one component of
	$U\setminus\{\theta_0\}$ the connector
	is smooth and $X$ is central, while for
	$\theta$ in the other component the
	connector is non-smooth and another
	attractor $Y$ is central.  The third
	attractor $Z$ is the source attractor.
	In this situation the only fold bifurcation curves
	that can cross the $U\setminus\{\theta_0\}$
	must involve the attractor $Z$.
\end{proposition}

\proof
For the first part of the proposition
see Theorem I.3 and Corollary I.4 of
\cite{Vegter82}.  Strictly speaking the
results quoted are for gradient systems
but it is clear that the arguments used
apply to our situation.  According to
this, the generic situation is as
follows.  On a neighbourhood of
$\param_0$ there are coordinates $(
\mu_1 ,\mu_2)$ such that $\param_0$ is
given by $\mu_1=0, \mu_2=0$, the fold
bifurcation curve through $\param_0$ is
given by $\mu_2=0$ and the flip curve is
given by $\mu_2=\mu_1^2$, $\mu_1\geq 0$.
Along $\mu_2=0$ the connector is smooth
for $\mu_1>0$ and non-smooth otherwise.

To see the result about crossing fold bifurcation
curves note that in the component of
$U\setminus\{\theta_0\}$ where the
connector is smooth, the orientation at
the fold is compatible with $Z$ and not
$Y$.  In the other component, the
central attractor is $Y$ which is
therefore not able to undergo a fold
bifurcation.  Thus any crossing must be
by a curve involving $Z$.
\qed

It follows that if there are no flip
curves then the connector is always
smooth or always non-smooth.  Also a
corollary is that in terms of its
connections with crossing points and
cusps such a bifurcation segment with a
single such target flip curve endpoint
behaves exactly as one that has a smooth
connector all along its length.

The above result raises the question of
how the transition from a smooth to a
cuspoidal connector takes place in the
region where there are three attractors.
If the attractors involved are as in
Fig.\ \ref{fig:fold+flip_point} 
then the transition involves
the yellow attractor i.e.\ the attractor
that is close to bifurcation.
When the system is
sufficiently close to the fold curve the
yellow attractor $x$ has a center manifold
and a strong stable manifold $W^{ss}(x)$
as described in Sect.\ 
\ref{para:pseudo-hyperbolic}.  The transition
takes place when the unstable manifold
from the source saddle $s_S$ intersects
$W^{ss}(x)$ and flips from joining
smoothly at $x$ to the unstable manifold
of the other saddle, to having a cuspoidal
joint to it at $x$.

\subsubsection{Source saddles}
The other way a flip curve can end is via 
a fold bifurcation that destroys
the source saddle. This is more straightforward
and needs no special consideration. It
meets the
fold curve transversally.

\subsubsection{Crossings sequences}

Recall that if the connector is not smooth at
the point where two fold bifurcation curves cross
then the two crossing curves
must involve both peripheral attractors
as a fold bifurcation of the central
attractor is impossible in this case
(see Appendix \ref{App:no_fold}).

\begin{lemma}\label{lem:PXP}
	Consider an adjacent
	pair of crossing points involving the smooth
	segments $\ell_{i-1}$, $\ell_i$ and
	$\ell_{i+1}$ where these respectively
	involve the attractors $Y$, $X$ and $Z$. 
	Assume that for each segment, if it
	contains more than one target flip curve
	endpoint then they all involve the same
	target saddle.  
	Then if $X$ is central at some point on
	$\ell_i$, $Y=Z\neq X$.	
\end{lemma}

\proof
Firstly, suppose that there are no flip
curves along $\ell_i$.  Then as you move
along the fold curve $\ell_i$ the
central attractor remains the same and
the orientation $\mathcal{O}$ of the
fold at the central attractor stays
constant.  At a crossing of $\ell_i$ by
a fold bifurcation curve corresponding to $P$
the orientation induced by
the bifurcation of $P$ must be
compatible with $\mathcal{O}$ and this
is the case for only one peripheral
attractor.  If there is a flip curve
ending then the result follows from
Prop.\ \ref{prop:9}.
\qed

\begin{lemma}
	With the same assumption as in the
	previous lemma, suppose that
	$\ell_i,\ldots ,\ell_{i+j}$ is a
	sequence of crossing points and $X_i,
	\ldots ,X_{i+j}$ is the corresponding
	sequence of attractors.  Then if we
	replace all triples of the form $PXP$
	(where $X$ is central and $P$ peripheral)
	by $P$, the resulting sequence just
	alternates between $P$ and $P'$ where
	$P'$ is the other peripheral attractor.
\end{lemma}

\proof
This follows from the fact that
by Lemma \ref{lem:PXP} if $\ell\in\L_X$
has $X$ central for some $\theta\in\ell$
then the crossings are of the form
$PXP$.  Thus replacing the $PXP$ triples
by $P$s removes all $X$s.				 Moreover, a
$PXP$ triple is adjacent either to a
$XP$ pair or a $P'$.  So repeating this
process produces a sequence alternating
between $P$ and $P'$.
\qed

\subsection{Simplest/minimal MS components} 

Given the constraints outlined above it
is clear that  for a given
configuration of cusps the first four of the
following give the minimal complexity
examples.  The fourth is more complex
(see Sect.\ \ref{sect:topcharge}).  In
the first three the connector is always
smooth and there are no flip curves.
The first two can be obtained as a
2 dimensional section through the butterfly
catastrophe.
\begin{enumerate}
	\item With a dual cusp and no other cusps: {\bf I} Fig.\ 5A. 
	\item With a dual cusp and a standard cusp: {\bf I} Fig.\ 5A(inset).
	\item With a standard cusp and no other cusps: {\bf I} Fig.\ 5B.
	\item With a standard cusp and a flip curve: {\bf I} Fig.\ 5C.
	\item With no cusps and only smooth bifurcation segments: {\bf I} Fig.\ 5D \& see Sect.\ \ref{sect:topcharge} below.
\end{enumerate}

\section{Topological charge 
	and the elliptic umbilic}\label{sect:topcharge}

Our goal in this section is to explain
why the boundary behaviour in Fig.\ \ref{fig:CEU}
necessitates the complex dynamics
inside.  This is a particularly striking
boundary value problem and we introduce
the notion of a phase or \emph{topological
	charge} into the analysis.  In physics an
optical vortex has a topological charge.
This is observed by interfering an
optical vortex with a plane wave of
light.  This reveals the spiral phase of
the vortex as concentric spirals.  The
number of arms in the spiral equals the
topological charge.

In our case the topological charge of a
fold circle is the number of cusps it
contains.  It is topological because
when measured mod 2 the even and odd
cases correspond to the topology of the
various line bundles we have introduced.
One can recognise the presence or not of
mod 2 charge by topology.  Lastly,
non-zero mod 2 charge is rigid and
cannot be perturbed or homotoped away
while zero charge fold curves can.

We are concerned  with a system
where the boundary behaviour is as shown
in Fig.\ \ref{fig:CEU}, and described as follows:

\noindent{\bf Compact Elliptic Umbilic boundary 
	behaviour assumption} is that there are three smooth
curves $\ell_A$, $\ell_B$, and $\ell_C$
which intersect as shown in Fig.\ \ref{fig:CEU}
and on which
the indicated attractor undergoes a fold
bifurcation that destroys the attractor.
On a neighbourhood of the three curves we
have the bistable and monostable behaviour
shown in Fig.\ \ref{fig:CEU}.

We denote by $\Delta$ the open region 
enclosed by $\ell_A$, $\ell_B$, and $\ell_C$.
Clearly, it is diffeomorphic to the 2-dimensional
disk.
Since we are interested in characterising
3-attractor MS components we also assume that
inside $\Delta$ 
we have three attractors 
$A$, $B$ and $C$ that
have smooth dependence upon the parameters.
\label{sect:CEUboundary}

We show that given this boundary
behaviour the system in $\Delta$
contains all the ingredients of the
compact elliptic umbilic (CEU) that
we now introduce.

\subsection{The compact elliptic umbilic}
\label{sect:CEU}
\begin{figure}[H]
	\centering
	\includegraphics[width=\linewidth]{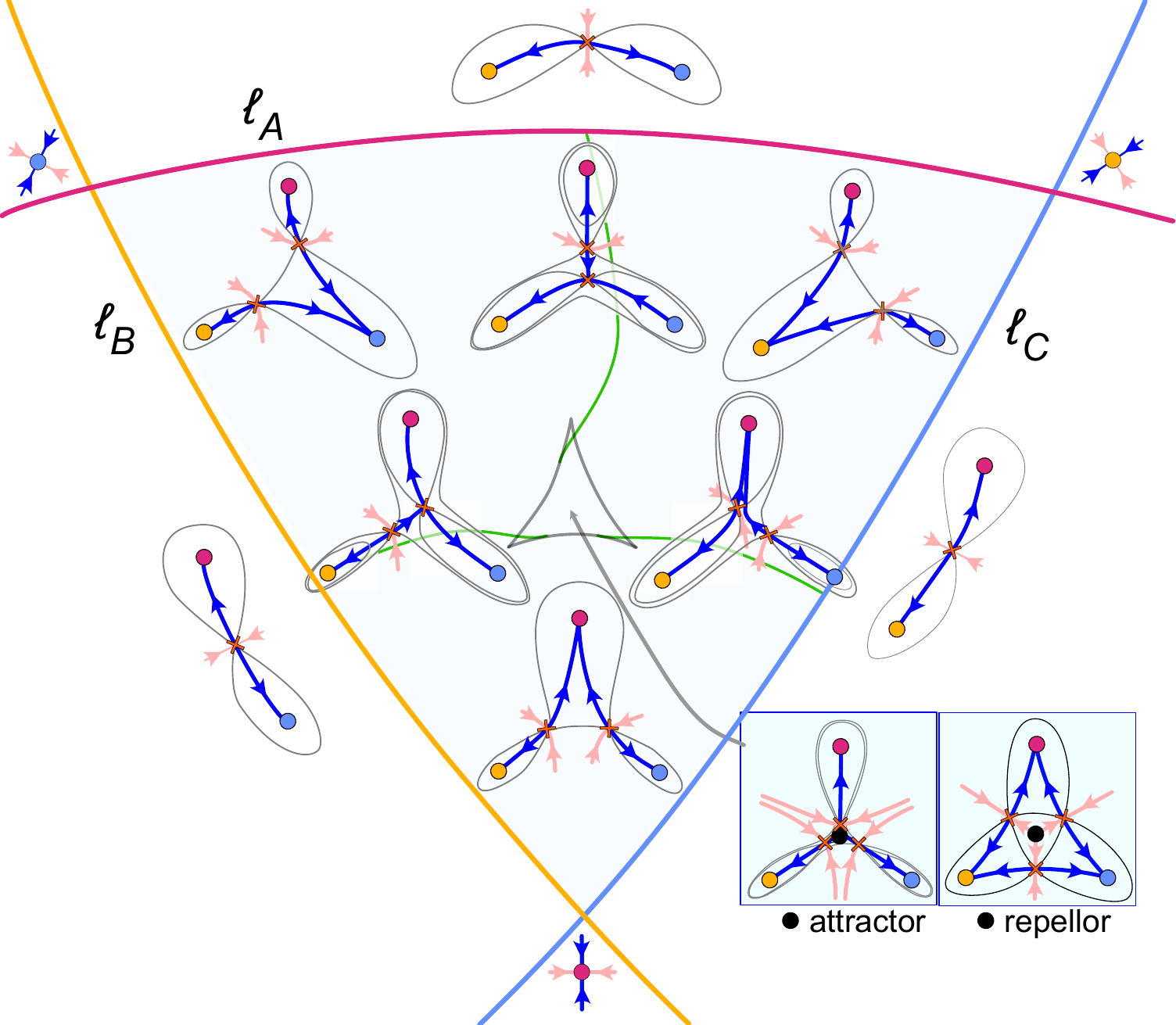}
	\caption{The
		compact elliptic umbilic. See text for description.
	}
	\label{fig:CEU}\label{fig:3-curves}
\end{figure}
This is the set
of gradient systems
associated with the potential
\begin{equation}
	f_\param (x_1,x_2)=x_1^4+x_2^4 +x_1^3-2x_1x_2^2
	+\param_3(x_1^2+x_2^2)+\param_1 x_1+\param_2 x_2.
\end{equation}
These parameterised families
of gradient systems are given by
$\dot{x}_i =-\sum_j g^{ij}{\partial
	f_\param}/{\partial x_j} $ where
$g_{ij}$ is a Riemannian metric and
$(g^{ij})=(g_{ij})^{-1}$.

We illustrate the case where $\param_3$ is
fixed at 0.1 but we will see this
provides a ubiquitous landscape
structure.  There are three curves
$\ell_A$, $\ell_B$, and $\ell_C$ in the parameter
space $\reals^2$ of fold bifurcations as
shown in Fig.\ \ref{fig:CEU}.  
For $X\in\{ A,B,C\}$
the curve $\ell_X$ is where the attractor
$X$ is destroyed in a fold bifurcation.
These curves enclose a
triangular shaped region which we denote
by $\Delta$.  There is a hypocycloid
$\B$ with three cusps (a \emph{deltoid}) inside
$\Delta$ such that for $\param$ inside
$\Delta$ but outside $\B$ the system has
three attractors that we label $A$, $B$,
and $C$ plus two saddles.  Inside $\B$
it has an extra saddle and either an
extra attractor (if $\param_3 >0$) or an
extra repellor. 

For a given Riemannian metric $g_{ij}$
the corresponding gradient system also
has three flip curves in $\Delta$ as shown on
Fig.\ \ref{fig:CEU} on which heteroclinic
bifurcations occur.  These start on
$\ell_A$, $\ell_B$, and $\ell_C$ and end on the
hypocycloid $\B$.  The configurations
shown are for a generic metric; for the
standard one the situation is not generic
and the  curves end at a cusp
point \cite{Khesin90sup}.

Let $\catm$ be the catastrophe manifold
over $\Delta$ i.e.\ $\catm=\{ \x
=(x,\param) : x \text{ a critical point
	of } f_\param , \param\in \Delta \}$.
Since the attractors $A$, $B$ and $C$
vary smoothly, the components $\catm_A$,
$\catm_B$ and $\catm_C$ of $\catm$
corresponding to them are graphs over
$\Delta$ that form disjoint components.
Thus we focus on $\catm_0$ which is the
remaining component of $\catm$ after
$\catm_A$, $\catm_B$ and $\catm_C$ have
been removed.  Let $\chi_0=\chi
|\catm_0$.

If $\B^\text{int}$ denotes the interior
of $\B$ then
$\catm_1=\chi_0^{-1}(\Delta\setminus
\B^\text{int})$ is an annulus that under
$\chi_0$ double covers the annulus
$\Delta\setminus \B^\text{int}$ i.e
$\chi_0|\catm_1$ is equivalent to the
map: $\mathbb{S}^1\times (-1,1)\ni
(\param ,t)\mapsto (2\param \text{ mod }
2\pi ,t)\in \mathbb{S}^1\times (-1,1)$.  

To obtain $\catm_0$ we take the inner
boundary of $\catm_1$ which is a circle
and glue in to it a disk correponding to
the new non-saddle rest point that is
created by the fold bifurcation that
occurs on $\B$.  This $\catm_0$ is
diffeomorphic to a disk but the mapping
$\chi_0$ is non-trivial.

\subsection{Boundary conditions}
Now we show that if we have the same
boundary conditions as for the CEU then
we have dynamics closely similar to
those of the CEU inside.  Indeed the
only change is the possible addition of
one or more simple fold circles inside
the M\"obius fold circle corresponding
to $\mathcal{B}$ that then add extra
rest points to the dynamical system that
bifurcate in and out maintaining the
architypal CEU structure.

For the
rest of this section we assume the CEU
boundary behaviour defined above in Sect.\
\ref{sect:CEU}.

Since there is a neighbourhood $U$ of
$\partial \Delta$ that contains no
bifurcation points in $U\cap \Delta$, 
then by Proposition \ref{prop:disk}, each
connected component of $\catm$
over $\Delta$ is
diffeomorphic to a disk.  
As above, since the attractors $A$, $B$ and $C$
vary smoothly, the components $\catm_A$,
$\catm_B$ and $\catm_C$ of $\catm$
corresponding to them are graphs over
$\Delta$ that form separate components.
We denote by $\catm_0$ the
remaining component of $\catm$ after
$\catm_A$, $\catm_B$ and $\catm_C$ have
been removed.  Let $\chi_0=\chi
|\catm_0$. This is the connected component
of the catastrophe manifold that involves
the saddles.

In the
following sections we will deduce the
following results which mirror similar
results for the CEU.  We now show that
close to $\partial \Delta$ it has the
same double covering annular structure
as this CEU.  This will involve analysis
of the flip curves in $\Delta$.  From
this we will deduce the existence of a
M\"obius fold circle $C$ in $\catm$.
Further analysis of the flip curves will
show that $C$ is unique and that
$B_C=\chi(C)$ plays the role of the
deltoid $\B$ of the CEU and has a 3-fold
cusp structure.

\subsection{Using flip curves}

\subsubsection{Flips ending on
	$\ell_A$, $\ell_B$ and $\ell_C$}

In the following subsections
$X$, $Y$ and $Z$ are always 
distinct elements of $\{A,B,C\}$.

\begin{proposition}
	The CEU boundary behaviour assumption implies that
	\begin{enumerate}
		\item there is a neighbourhood
		$U$ of $\ell_Y\cap \ell_Z$ such that
		if $\param\in U\cap\Delta$ then
		$X$ is the central attractor.
		\item  there is a neighbourhood $V$
		of $\ell_X$ such that for all $\param\in
		V\cap\Delta$ either there is a
		heteroclinic connection or the system is
		MS with $X$ a peripheral
		attractor.
	\end{enumerate}
\end{proposition}

\proof
Firstly, note that the CEU boundary behaviour 
assumption implies that there is a neighbourhood $U$ of the
boundary such that if $\param\in U\cap\Delta$
then $X_\param$ has three attractors and two saddles
all of which vary smoothly with $\param\in U\cap\Delta$.

Since the curves $\ell_X$, $\ell_Y$, and $\ell_Z$
are smooth and since $\ell_Y$ meets both
$\ell_X$, and $\ell_Z$ then $Y$ is peripheral
along $\ell_Y$ by Lemmas \ref{lemma:2}
and \ref{lem:PXP}.  Similarly, $Z$
is peripheral along $\ell_Z$.  Thus at
$\ell_Y\cap \ell_Z$, $X$ is central.  It
follows that $X$ is also central close
to $\ell_Y\cap \ell_Z$ for $\param\in \Delta$
and $X$ is peripheral for all $\param$
in $\Delta$ near to $\ell_X$.
\qed

\begin{corollary}\label{cor:5}
	The CEU boundary behaviour assumption implies 
	that for $X\in \{ A,B,C\}$
	there are generically an odd number of flip curves
	starting on the intersection of
	$\ell_X$ with the closure of $\Delta$. On each of these
	the central attractor changes from $Y$ to
	$Z$ or vice-versa.
\end{corollary}

\proof
The
result follows because in moving along $\ell_X$ from
$\ell_X\cap \ell_Y$ to $\ell_X\cap \ell_Z$ is to
move from $Z$ being the central
attractor to $Y$ being central.
\qed


\subsubsection{Algebra of three-way flips}\label{sect:alg}
\begin{figure}[H]
	\centering
	\includegraphics[width=\linewidth]{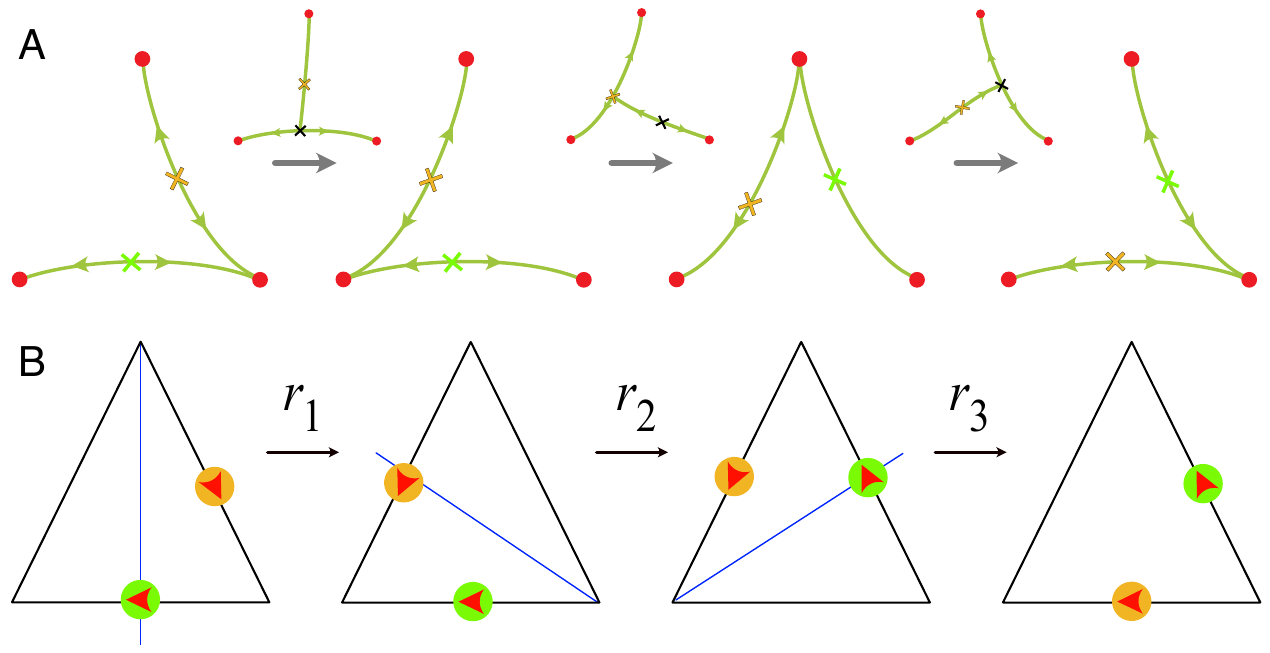}
	\caption{A series of flips that results in the saddles swapping.
		The $r_i$ are all reflections in the thin blue line that is shown.}
	\label{fig:equilaterals}
\end{figure}
To each generic configuration 
of three attractors and two saddles 
as above we can
associate a triangle with vertices
marked $A$, $B$ and $C$.  We label the
saddles $g$ (green) and $y$ (yellow).  
See Fig.\ \ref{fig:equilaterals}A.
Then we associate to each MS 3-attractor
system the corresponding marked
equilateral triangle as in Fig.\
\ref{fig:equilaterals}B.  The vertices
correspond to the attractors $A$, $B$
and $C$ and we mark an edge with green or yellow
if the corresponding unstable manifold
of $g$ or $y$ joins the corresponding two
attractors. The meaning of the red arrows
in Fig.\ \ref{fig:equilaterals} is explained below.

To each flip we can associate a
corresponding reflection as shown in
Fig.\ \ref{fig:equilaterals}.  This is
an element of the Dihedral group $D_3$
which can be represented as the group of
symmetries of the equilateral triangle
consisting of reflections and rotations
through 2$\pi/3$ radians.

Thus, consider an oriented loop
$\gamma(t)$ in $\Delta$ with a given
orientation and starting point (not in a 
flip curve), that does
not meet any fold or cusp points and
which is transversal to the flip curves.  To
this we can associate a sequence
$\sigma_\gamma$ of elements of $D_3$.
This is the product $r_k r_{k-1}\cdots
r_1$ where $r_i$ is the reflection that
occurs when the curve crosses the $i$th
flip.  The reduction of this product is
$r_{i_r}r_{i_{r-1}}\cdots r_{i_1}$ which
is what is left when we iteratively
delete all products in the product of
the form $r_ir_i^{-1}$ or $r_i^{-1}r_i$.
It is then easy to see that the
resulting product is
\[
r_{\sigma(1)}r_{\sigma(2)}r_{\sigma(3)}\cdots 
r_{\sigma(1)}r_{\sigma(2)}r_{\sigma(3)} = 
(r_{\sigma(1)}r_{\sigma(2)}r_{\sigma(3)})^k
\]
for some cyclic permutation $\sigma$ of
$\{ 1,2,3\}$ and some $k\geq 0$.  This is because
the starting and ending configurations are the same
and since
once all the terms $r_ir_i^{-1}$ or $r_i^{-1}r_i$
have been removed the sequence of indices
must be periodic of the form
$\cdots \sigma(1) \sigma(2) \sigma(3)
\sigma(1) \sigma(2) \sigma(3)\cdots$
since only two $r_i$s can be applied
to any given configuration and if that
configuration has been achieved then
the inverse of one of these was applied to
get it.

We
call $(\sigma,k)$ the \emph{flip index} of
$\gamma$.  Note that $k$ does not depend
on the starting point and that, for $k$
even, when evaluated the product is the
identity and otherwise is a non-trivial
reflection.

When the unstable manifolds of the
saddles undergo the transformations
shown in Fig.\ \ref{fig:equilaterals}
the unstable manifolds undergo
interesting changes in orientation.
Similar things occur for the stable
manifolds but we only consider unstable
manifolds here as we are interested in
index 1 saddles and then the unstable
manifold is 1-dimensional even in
$n$-dimensional systems.

Put an orientation on the two
saddles.  This is indicated in Fig.\
\ref{fig:equilaterals} by the red arrow.
As the flips occur this orientation is
propagated as shown. 
(Note that this is
not what you get by applying the
geometric reflections.  Instead follow
what is happening in the top row.)  
After applying them once, the two saddles have switched and 
one of the unstable manifolds has opposite orientation.
From
this we see that applying the flips
corresponding to
$r_{\sigma(1)}r_{\sigma(2)}r_{\sigma(3)}$
twice
we reverse the orientation of  both unstable manifolds 
and the saddles switched twice. 
Thus we have the following result.

\begin{proposition}
	Suppose that $\gamma$ is a closed curve in $\Delta$
	that contains no fold or cusp bifurcation points.
	Put an orientation on the unstable manifolds
	of the two saddles. The flip index of
	$\gamma$ has $k$ odd if and only if
	this orientation is reversed in one of the unstable
	manifolds
	when $\gamma$ is traversed once.
\end{proposition}

Consider a simple closed curve
$\gamma(t)$ in $\Delta$ that has a lift
$\Gamma (t)=(x (t), \param (t))$ $0\leq
t\leq 1$ to a smooth closed curve in
$\catm_0$.  Assume that it meets no fold
points.  Then the unstable manifold
$W^u(x(t))$ of the saddle $x(t)$ varies
smoothly with $t$.  Thus as
$t\rightarrow 1$ from below $W^u(x(t))$
converges as a point set to $W^u(x(0))$.
Thus if we fix an orientation on
$W^u(x(0))$ and use the smooth
dependence of $W^u(x(t))$ on $t$ to
propagate this orientation to all
$W^u(x(t))$ we can ask whether as
$t\rightarrow 1$ from below the
orientation on $W^u(x(t))$ converges to
that on $W^u(x(0))$.  Clearly, this is
the case if and only if the saddle
bundle of $\Gamma$ is a cylinder.

On the other hand, from the above
discussion, this is not the case if the
flip index of $\Gamma$ has $k$ odd and
is the case otherwise.  Thus we deduce
the following result

\begin{proposition}\label{prop:indexodd}
	The flip index of $\Gamma$ is odd
	if and only if the bundle $E_\Gamma^c$ is M\"obius.	
\end{proposition}

\proof
By the above, the flip index is odd if and only if
a complete revolution around $\Gamma$ causes any orientation
put on  the unstable manifolds of the saddles to be reversed
and this is the case if and only if the bundle $E^c_\Gamma$
is M\"obius.
\qed

\subsubsection{Flips at $\partial\Delta$
	imply a M\"obius fold circle}

Cor.\ \ref{cor:5} above tells us that
there are an odd number of flip curves
along each of the edges $\ell_X$,
$X\in\{A,B,C\}$, and these flip curves
induce the same element or its
inverse when calculating the flip index
of a simple closed curve $\gamma$ close
to $\partial \Delta$. 
{We denote the corresponding element $r_X$.} 
If the curve is
sufficiently close to $\partial \Delta$ then the elements
must alternate between $r_X$ and
$r_X^{-1}$ with a total contribution of
$r_X$.  Thus, in the notation  analogous to  Cor.\
\ref{cor:5}, the resulting product for
the whole curve is of the form 
$r_{\sigma(Z)}r_{\sigma(Y)}r_{\sigma(X)}$ and we
deduce that $k=1$.

Since $\gamma$ is a closed curve we can
parameterise it as $\param=\gamma(t)$,
$t\in\reals$, with $\gamma(t+1)\equiv
\gamma(t)$.  Since there are no bifurcation points
in $\Delta$ close to $\partial \Delta$,
we can lift this to a closed
curve $\x=\Gamma(t)$ in $\catm$ so that
if $\Gamma(t)=(x(t),\param(t))$ then
$x(t)$ is one of the two saddles.  If
$0\leq t<1$ it follows from the above
discussion that $\Gamma(t)$ and
$\Gamma(t+1)$ are two distinct points in
$\catm$ that are interchanged when $t$
is increased by one.  This is because
$k=1$ and therefore going once round
$\gamma$ flips the saddles as explained
in Sect.\ \ref{sect:alg}.  Let $\catm_0$
be the connected component of $\catm$
containing $\Gamma$ and $\chi_0=
\chi|\catm_0$.

\begin{proposition}
	There is a neighbourhood
	$U$ of $\partial \Delta$ such that 
	$ U\cap \Delta$ and 
	$\chi_0^{-1}(U\cap \Delta)$ are annuli
	and the latter is a double cover of
	the former on which $\chi_0$
	has degree 2.
\end{proposition}

\proof
Take a small tubular neighbourhood $N_X$
of each curve $\ell_X$, $X\in\{A,B,C\}$
so that (i) there are no bifurcation
points in $N$ other than in $\partial
\Delta$ and (ii) so that if $N=N_A\cup
N_B\cup N_C$ then $U=N\cap\Delta$ is an
annulus and such that for any point
$\param\in U$ there is a curve $\gamma$
as above with $\param\in\gamma$.  Then
$\chi$ is a local diffeomorphism and,
from the above discussion every point in
$U$ has two preimages in $V=\chi_0^{-1}(U)$
under $\chi_0$.  Put an orientation on
$V$ so that the degree of $\chi$
is non-negative. Then  if $\x_1$ and $\x_2$ are the two
preimages of $\param$,  since we can
move from $\x_1$ to $\x_2$ by moving
along a closed curve in $U$, $\chi_0$ is
orientation preserving at both points.
Thus the degree of $\chi_0|V$ is two.
\qed

Now we can prove one of our main results.

\begin{theorem}
	Suppose that the configuration of 
	the smooth bifurcation curves $\ell_A$, $\ell_B$, and
	$\ell_C$ is as in the CEU boundary behaviour assumption
	(cf.\ Fig.\ \ref{fig:3-curves}) and that $\Delta$ is
	the open region enclosed by these curves.
	Then $\Delta$ contains $B_C=\chi(C)$ where
	$C$ is a M\"obius fold circle.
\end{theorem}

\proof
Let $\gamma$ be a smooth simple closed curve
in $\Delta$ close to the boundary of $\Delta$.
Lift
$\gamma$ to a closed curve $\Gamma$ in $\catm_0$.
Then by Cor.\ \ref{cor:5} the flip index of $\gamma$ is odd
and therefore the same is true for $\Gamma$.
Therefore, by Prop.\ \ref{prop:indexodd}, the
bundle $E_\Gamma^c$ is M\"obius. It follows by Theorem \ref{thm:1}
that there is a M\"obius fold circle $C$ in the interior
of $\Gamma$. Since $C$ is in the interior
of $\Gamma$, $B_C$ is contained in $\Delta$. 
\qed

\subsubsection{Flip curves in $\catm$}

In order to understand the way
in which flip curves end on fold
bifurcation curves in the parameter space
we need to lift the flips to
$\catm_0$.  Whereas in $\catm$
a lift of a flip curve can only 
meet fold curves at its end points, in
parameter space
a flip curve can meet a fold
bifurcation curve at other points
(see e.g.\ Fig.\ \ref{fig:CEU}).

If $\param\in\Delta$ is a
flip point then we associate to it two
points $\x =\x(\param ) =(x,\param
)\in\catm_0$ where $x$ is either the
target saddle $x_T$ or the source
saddle $x_S$.  Thus a flip curve
$\theta = \varphi(t)$, $0\leq t\leq 1$,
in $\Delta$ can be lifted to two smooth
curves $\x=\Phi_T(t)$ and $\x=\Phi_S(t)$
in $\catm_0$ that we respectively call
\emph{the target flip curve} and
\emph{source flip curve}.

If such a flip curve in $\catm_0$ meets a
fold curve in $\catm_0$ it involves a fold
in which the target or source saddle is 
destroyed.  It follows that a flip
curve in $\catm_0$ only meets fold curves
at its end points.  As we will see this
is different from flip curves in
$\Delta$ which can cross fold
bifurcation curves. We call a flip curve a target-source one if
one end point is a target one and the other
a source one. Source-source and target-target
curves are possible but here we are only interested
in  target-source ones

If $\Gamma$ is a lift to $\catm_0$ of a
closed curve $\gamma$ in $\Delta$ then
there is a 1-1 correspondence of points
where they meet target flip curves and
the intersections are ordered
identically as in $\gamma$.  
Therefore, if one defines the
flip index of such a closed curve $\Gamma$
in $\catm_0$ analogously to that of $\gamma$
but only using the target flip curves in $\catm_0$,
then the flip indices of $\Gamma$ and $\gamma$ are
equal. Of course, instead of using target flip curves one
could use source flip curves.

Now suppose that the closed smooth curve
$\Gamma^\prime$ is obtained from
$\Gamma$ by a smooth homotopy and that
both are transverse to the flip curves.
Assume also that the homotopy does not
cross any fold curves.

\begin{proposition}\label{prop:12}
	The reduced products 
	$(r_{\sigma(1)}r_{\sigma(2)}r_{\sigma(3)})^k$
	obtained from $\Gamma$ and $\Gamma^\prime$ are the same.
\end{proposition}

\noindent{\bf Proof.}
The homotopy $\Gamma_t$ can be chosen so
that the curve $\Gamma_t$ is always
transversal to the flip curves except at
a finite number of $t$ and at these it
touches a single flip curve
quadratically.  The only changes in the
sequence of flips occur at such $t$ but
then this just adds or removes a term of
the form $r_i^{-1}r_i$ and therefore
does not change the reduced form.  \qed

Consider a closed curve  $\Gamma$
in $\catm$ that contains a fold
circle $C$ in its interior. Suppose,
moreover, that $C$ is
the only fold circle contained
in the interior of $\Gamma'$
except possibly for fold circles
contained in the interior of $C$.
We call such a curve $\Gamma$ a
\emph{primary flip container} for $C$.

\begin{proposition}
	The fold circle $C$
	is M\"obius if and only if 
	there is a primary flip container $\Gamma$ for $C$
	for which the
	flip index has $k$ odd.
\end{proposition}

\proof
Let $N$ be a tubular neighbourhood of a
fold curve and $r$ be the associated
retraction.  One can use a homotopy to
move $\Gamma$ into $N\setminus C$ so
that the new curve $\Gamma'$ has degree 1
with respect to $r$.  Moreover, by
Prop.\ \ref{prop:12}, $\Gamma'$ has $k$
odd if and only if $\Gamma$ does.

If $N$ is sufficiently small, $C$ is
M\"obius $\iff$ its cusp bundle is a
M\"obius band $\iff$ $E_{\Gamma'}^u$ is
a M\"obius band $\iff$ $k$ is odd.
\qed

\begin{theorem}\label{prop:remove_flip}
	Suppose that $\Gamma$ is a smooth simple
	closed curve in
	$\catm$ consisting of index 1 saddles.
	If its flip index is odd then
	in the generic situation, $\Gamma$
	contains an odd number of M\"obius fold
	circles in its interior.  Moreover, if
	$C$ is one of these M\"obius fold
	circles one can find a primary flip
	container $\Gamma'$ for $C$ for which the
	flip index has $k$ odd.
\end{theorem}

\proof
One uses a construction as in Appendix \ref{sect:E}
to show that if $\Gamma$ is a simple closed curve in
$\catm$ then it contains at least one fold circle
and that $\Gamma$ can be modified so that it only contains any
one of these that is M\"obius.
One firstly removes
the fold circles with $k$ even and then pairs of
fold circles with $k$ odd. Having modified $\Gamma$ in this
way one can use a homotopy to move it arbitrarily close to $C$
and use Prop.\ \ref{prop:12}.
\qed

\begin{corollary}\label{cor:4}
	Under the CEU boundary behaviour assumption,
	$\Delta$ contains a unique M\"obius fold circle.
\end{corollary}

\proof
By Cor.\ \ref{cor:5} the lift $\Gamma$ to $\catm_0$ of
any closed curve in $\Delta$ close to
$\partial \Delta$ has $k$ odd.  Thus by
Theorem \ref{prop:remove_flip} there is a M\"obius fold
circle $C$ in the interior of $\Gamma$
and a primary flip container $\Gamma'$
for $C$ for which the flip index has $k$
odd.  This means that for each $X\in\{
A,B,C\}$ there is a flip curve
$\ell_X^\prime$ connecting the boundary
component $\ell_X$ to $\Gamma'$.  If
$C'$ were another M\"obius fold circle
in the interior of $\Gamma$ it would
also have a primary flip container
$\Gamma''$ with similar connections
$\ell_X^{\prime\prime}$ to $\ell_X$.
Since $\ell_X^\prime$ and
$\ell_X^{\prime\prime}$ cannot meet as
they are flip curves, this is
impossible.  Thus $C$ is unique.
\qed

%% file: DecisionStructures.tex


\section{Decision structures}

We identify a cell state with an attractor in the dynamical system. We say that a cell takes a decision when it changes state and transitions to another attractor. This transition happens when the cell crosses an index 1 saddle and follows the corresponding unstable manifold towards another attractor. Therefore, we identify a \emph{decision} with an index 1 saddle and its unstable manifold, which connects two attractors: the two options for the decision. 

The complete dynamical system can be represented by a DAG as explained in ${\bf I}$. Since we want to focus only on the decisions, we will simplify the DAG in order to retain only the information on the possible decisions as per our definition above (that is, attractors and index 1 saddles). 

\begin{figure}[ht!]
	\begin{center}			
		\includegraphics[width=\linewidth]{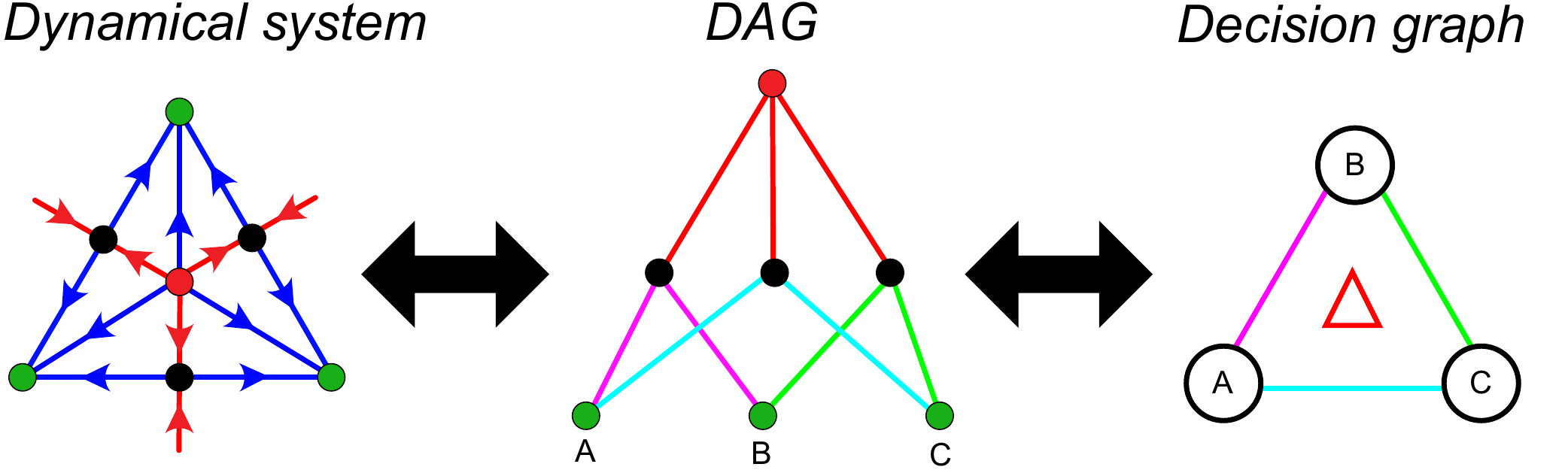}
		\caption{Example of the correspondence between Dynamical systems, DAGs and decision graphs.}
		\label{fig:Dynamics_DAG_DecisionG}
	\end{center}
\end{figure}

Given a DAG we define the \emph{decision graph} (see Fig. \ref{fig:Dynamics_DAG_DecisionG}) as the undirected graph with as many nodes as attractors in the dynamical system and an edge connecting two nodes (or a node to itself) if there is a saddle whose unstable manifold connects the corresponding two attractors (or the attractor to itself). The loops in this graph correspond to index 2 saddles in the DAG (in the dynamical system).

If the edges in the decision graph define a loop, then there is an index two saddle in the dynamical system that connects to the index 1 saddles that define it. This a consequence of the Morse-Smale inequalities for the $n$-sphere. A compact dynamical system (with a repeller at infinity) needs to satisfy $M-s+m\geq 1$ where $M$ is the number of index 2 saddles, $s$ is the number of index 1 saddles and $m$ is the number of attractors. If a dynamical system defines a loop in the decision graph, then $s=m$ and therefore $M\geq 1$. Hence,there is an index two saddle and in particular, if the system is planar, there is a repeller that connects to those saddles. We mark with a triangle the loops in the graph to emphasize the existence of such a higher index rest point.

\begin{figure}[ht!]
	\begin{center}		

	\includegraphics[width=0.4\textwidth]{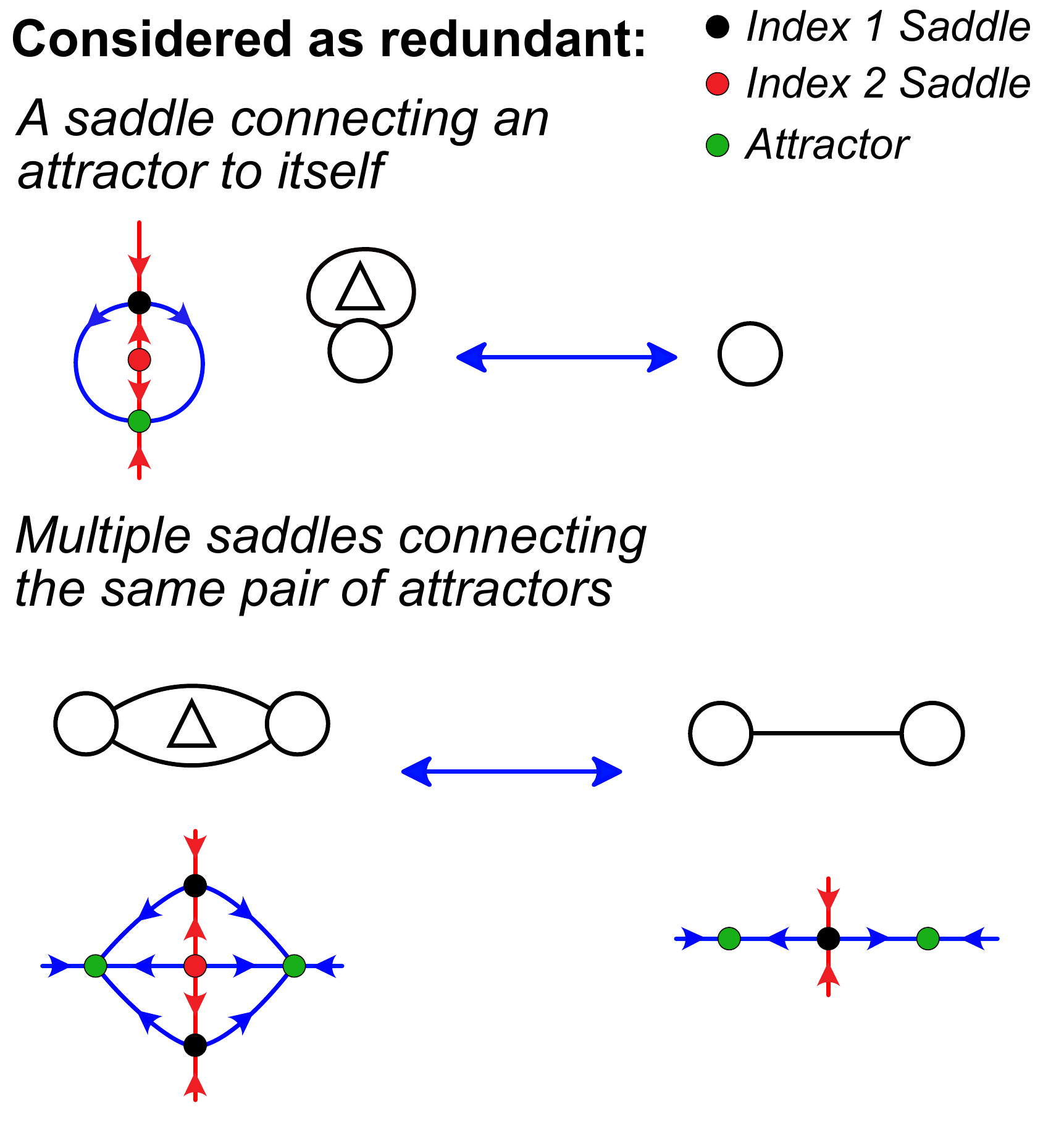}
	\caption{Types of edges modified when computing a decision structure from a decision graph or DAG.}
	\label{fig:RedundantEdges}
	\end{center}
\end{figure}

The decision graph may not be a simple graph, that is, it may have self-edges and parallel edges. We can remove the self-edges and keep only one edge from a set of parallel edges (see Fig. \ref{fig:RedundantEdges}) and obtain the simple graph associated to a decision graph. An edge connecting a node to itself does not define a proper decision as we defined it since there is no change of states in such transition. Several parallel edges are defining the same decision, so they are redundant in this context. 

\begin{figure}[ht!]
	\begin{center}		
		
		\includegraphics[width=\linewidth]{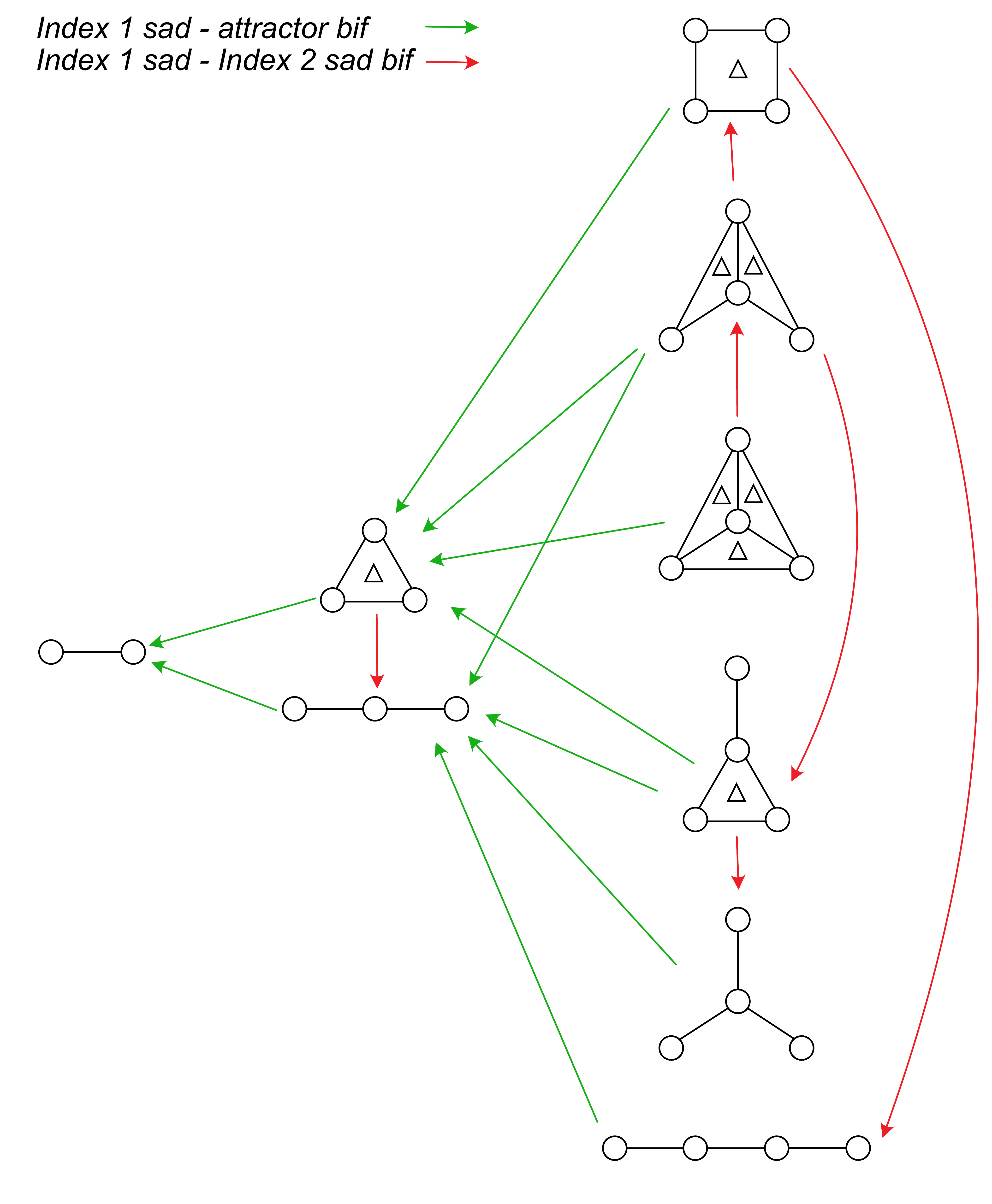}
		\caption{Decision structures with 2, 3 or 4 nodes. The possible decision structures with a given number of attractors are in one-to-one correspondence with the connected simple graphs with that number of nodes.}
		\label{fig:DecisionStructures234}
	\end{center}
\end{figure}

We call a \emph{decision structure} a simple decision graph. Its nodes correspond to attractors, the edges to unstable manifolds connecting them, and the loops correspond to index 2 saddles in the dynamical system. It encodes the essential information regarding the decision making process.  Note that we will not specify the correspondence between attractors and nodes, hence the same decision structure corresponds to any dynamical system with three attractors and two saddles irrespective of the particular central attractor.  Any decision graph can be reduced to a unique decision structure graph as explained above. The possible decision structures given a number of attractors are in one-to-one correspondence with the connected simple graphs with that number of nodes (see Fig. \ref{fig:DecisionStructures234}). Decision structures define equivalence classes for decision graphs and they are the simplest representatives of the different classes.

\begin{figure}[ht!]
	\begin{center}		
		
		\includegraphics[width=0.9\linewidth]{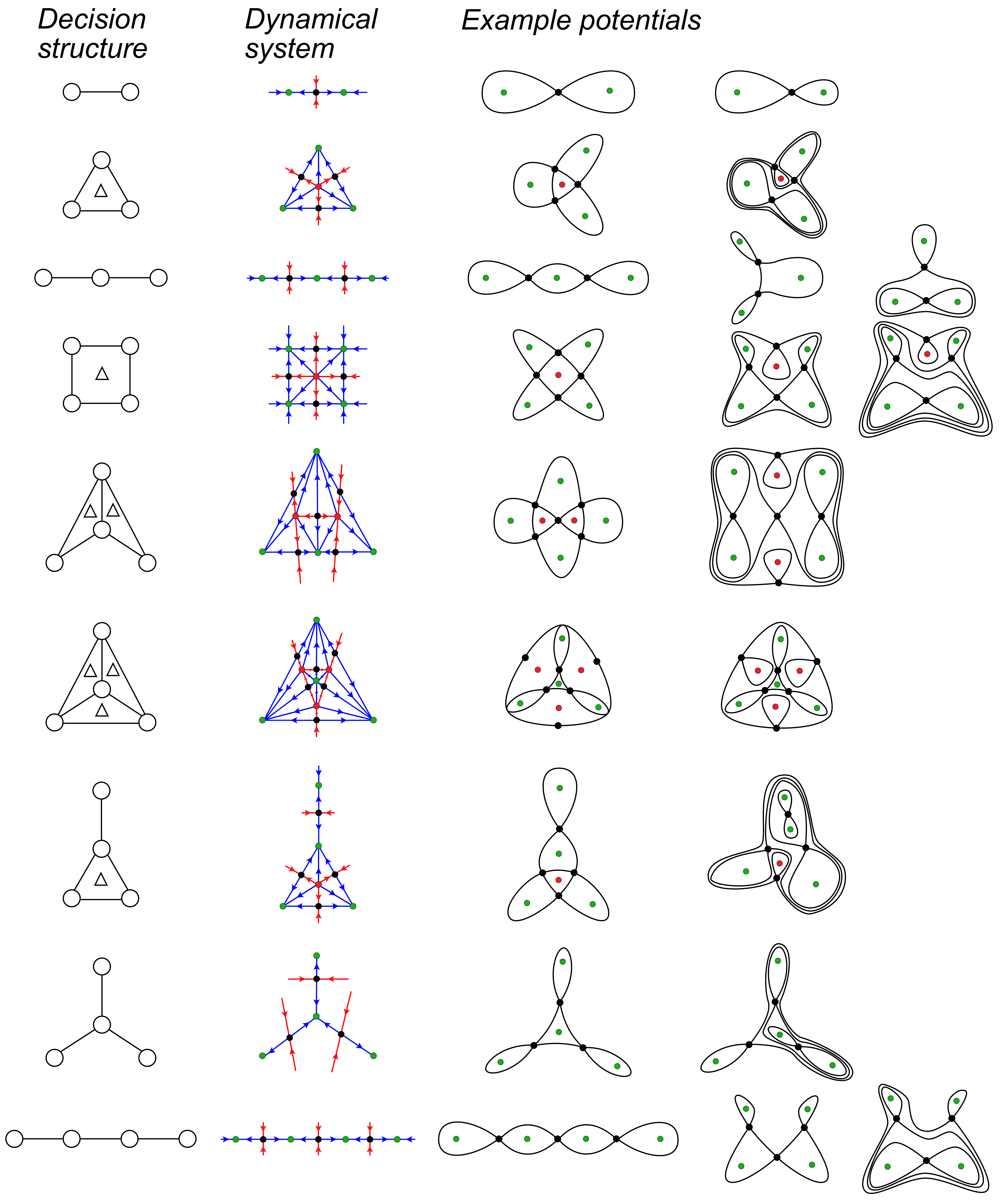}
		\caption{Dynamical structure and potential functions examples for the decision structures with 2, 3 or 4 nodes. Legend as in Figure \ref{fig:RedundantEdges}.}
		\label{fig:DynamicsandStructures}
	\end{center}
\end{figure}

The different decision structures correspond to a dynamical system structure that can correspond to several potential functions. The decision graph is independent of the specific saddle hierarchy in any possible potential structure associated with it. Note that as shown in \cite{Smale61gradient} we can find a potential function corresponding to any decision structure such that all saddles have the same height (first column in example potentials). Figure \ref{fig:DynamicsandStructures} shows examples of dynamical structures and potentials corresponding to the decision structures with 2, 3 or 4 attractors.

%% file: Appendices_current.tex
\noindent{\bf\large Appendices}

\section{Existence of decision regions $R$}\label{sect:decision}
This result relies on some results in Smale's 
proof of the higher dimensional Poincar\'e conjecture
\cite{smale61Poincare}
and his $h$-cobordism theory \cite{smale1962structure}
as expounded in Milnor \cite{milnor2015lectures}.

It will be important that for the studied vectorfields $X$
there is a region $B$ with spherical boundary $\partial B$
such that the $X$ is transversal 
to $\partial B$ and all trajectories 
cross $\partial B$ into $B$.

The following lemma that is taken from \cite{NewhousePeixoto76sup}
will be particularly useful.
In it we are particularly interested in
controlling the changes in the
connection between the index 1 saddles
$\sigma_i$, $i=1,\ldots ,M$ and the
attractors $\alpha_j$, $j=1,\ldots ,m$.
Consider an index 1 saddle $p=\sigma_i$
with unstable manifold $W^u(p)$.  Then
$W^u(p)\setminus {p}$ has two
components.  Let $\gamma$ be one of these,
and assume $\gamma$ is asymptotic to the attractor
$\alpha_j$.  
We call such curves \emph{connecting
trajectories} of $\sigma_i$ to $\alpha_j$.
The set of all such pairs $(i,j)$ is denoted 
by $\mathcal{C}_X$ and the corresponding connecting
trajectory is denoted by $\gamma_{i,j}$.

\begin{lemma}\label{lem:simplearc}
Suppose that $X$ is a
gradient-like MS vectorfield with
attracting region $B$ as above and that
we are given a small neighbourhood $U_k$
of each rest point $\beta_k$ and a small
neighbourhood $U_{i,j}$ of $\gamma_{i,j}$ for all 
$(i,j)\in \mathcal{C}_X$.
Then for
any $\eps >0$ one can find smoothly
varying MS vectorfields $X_t$, $0<t<1$
with $X_0=X$ such that (i) the $X_t$
all have the same rest points and they all agree
outside the $U_k$ for all $k$;
(ii) for some $0<t_0<1$, $X_0$ only
differs from $X_{t_0}$ inside the $U_i$
and this is by less than $\eps$ in the
$C^1$ topology; (iii) the connections
$(i,j)\in \mathcal{C}_{X_t}$ do not vary with $t$
and the connecting trajectories $\gamma_{i,j}$
are inside $U_{i,j}$ for all $t$; and (iv)
$X_1$ is a
gradient vector field with a smooth
potential.
\end{lemma}

This lemma allows us to modify a
gradient-like MS $X$ in a controlled way
in small neighbourhoods of each rest point so
as to produce a gradient vectorfield, at the same
time controlling how the unstable manifolds of the
index 1 saddles change. If $f$ is a smooth 
Liapunov function for $X$, then one can do the 
modification so that $X_1$ is the gradient flow for $f$
with respect to some Riemannian metric $g$.

\proof
This is Lemma 2 of
\cite{NewhousePeixoto76}.  The fact that
the $X_t$ all agree outside the
neighbourhood $U_i$ is not stated in the
lemma but is clear from the proof.

To see why the results about connectors are true,
consider the process used to modify the
dynamics at the rest point $p$ in the last paragraph of
the proof of Lemma 2 in \cite{NewhousePeixoto76}.
The vectorfields $\xi_t$ constructed
there all have the same stable manifold
inside the neighbourhood $\phi^{-1}U_2$
used there and they are independent of
$t$ outside $\phi^{-1}U_2$.  This is
because (in the notation of the
proof in \cite{NewhousePeixoto76})
in $U_2$, $\phi_*^{-1}\xi_t$ is a
linear combination of $\phi_*^{-1}X_1$
and $X_t$ both of which are tangent to
the curve $x=0$ in the coordinates
$(x,y)$ used there.  As a consequence
the connecting trajectory is independent
of $t$.  Therefore, the only
perturbations to it occur when (in the
notation of \cite{NewhousePeixoto76})
the vectorfield $X_1$ is modified so
that it is equal to its linear part near
$p$ (an $\eps$-small amount in the
$C^1$-topology) and inside the region
when the corresponding attractor is
modified.  Thus we deduce that the
constructed family of vectorfields has
the required property.
\qed

\begin{proposition}
	Suppose that $X$ is a gradient-like MS
	flow on $\reals^n$ with an attractor region
	$B$ as above with a smooth spherical boundary
	on which $X$ is inward transverse.
	There is an open subset $R$ in phase space
	with smooth boundary $\partial R$
	with the following
	properties (i) the flow is inward
	transverse on $\partial R$ which is a
	sphere; (ii) $R$ contains all the
	attractors of the flow and no rest points
	of index $>1$; (iii) the stable
	manifolds of the index 1 saddles  in $R$
	intersect $\partial R$ in disjoint spheres of
	dimension $n-2$. These index 1 saddles escape
	to two distinct attractors contained in 
	distinct connected components of $R\setminus W^s$
	where $W^s$ is the union of the stable
	manifolds of the index 1 saddles in $R$.
\end{proposition}

\proof
Denote the $M$ saddles and $m$
attractors of $X$ by $\sigma_i$ and
$\alpha_j$ as above.

This step is only necessary if the
number $M$ of index 1 saddles is greater
than $m$.  Fix such an index 1 saddle
$s$ of $X$ and let $\Sigma$ be a set of
$M-m+1$ saddles that does not contain
$s$.  By \cite{Meyer68sup} we can find a
Liapunov smooth function $L_\Sigma$ for
$X$ that takes the value $0$ at
attractors, $1$ at the saddles in
$\Sigma$ and $1/2$ at all the saddles
not in $\Sigma$.  Let $R_{\Sigma}$
denote the region in phase space given
by $L_\Sigma<\eta$ where $1/2<\eta <1$
is a regular value of $L_\Sigma$.  Then
$R_{\Sigma}$ contains all the attractors
and the $m-1$ index 1 saddles not
in $\Sigma$ together with their unstable
manifolds.  Moreover, $X$ is transverse
to the smooth codimension 1 submanifold
$\partial R_{\Sigma}$.  Let
$R_{s,\Sigma}$ be the connected
component of the intersection of the
$R_{\Sigma}$ for which $\Sigma$ does not
contains $s$.  This is open and contains
$s$ and its unstable manifold.

We use Lemma \ref{lem:simplearc} to
modify our vectorfield $X$ to obtain the
vectorfields $X_t$ given by Lemma
\ref{lem:simplearc}.  For the index 1
saddles $\sigma_i$ we choose the
$U_{i,j}$, $(i,j) \in \mathcal{C}_X$, so
that they are contained in
$R_{\sigma_i}$ and so that $U_i\subset
U_{i,j}$ and $U_j\subset U_{i,j}$ where
$U_k$ is the modification region of the
saddle $\sigma_i$ and the  
attractor $\alpha_j$ respectively.  This ensures that
the modifications of the saddle $\sigma_i$ and the
attractor $\alpha_j$ do not affect
$\partial R_\Sigma$ for all $\Sigma$ not
containing $\sigma_i$.  We also want to
ensure that the modification of the
rest points do not change the
vectorfields near any of the
submanifolds $R_{\Sigma}$.  So we choose
the $U_{i,j}$ so that they meet none of
these.

Then $X_{1}$ is a MS gradient flow with
the same rest points as $X$ and with the
same connections between index 1 saddles
and attractors as $X$ all of which are
held in a $U_{i,j}$ for some $(i,j) \in
\mathcal{C}$.

Choose a small $n$-disk $D_j$ around
each attractor $\alpha_j$ with boundary
$V_j=\partial D_j$ a smooth $n-1$ sphere
so that the flow is transverse on $V_j$
towards $\alpha_j$ and so that there are
no other rest points in $D_j$.  If the
$U_j$ of Lemma \ref{lem:simplearc} were
chosen sufficiently small we can assume
that the closure of $U_j$ is contained
in $V_j$.  Then since $X$ and $X_1$
agree outside the $U_j$ they are both
inward transverse on the boundary of the
$V_j$.

Let $V'=\partial B$, the flows of $X$
and $X_1$ are also inward transverse on
this since they agree there.  Let
$W=B\setminus D_0$.  Then $W$ has
boundary $V_0\cup V'$ and in the
terminology of \cite{milnor2015lectures}
$(W,V_0,V')$ is a cobordism with trivial
relative (singular) homology
$H_*(W,V_0)=H_*(W,V')=0$.

Now we consider the flow of the gradient
system $X_1$.  Using Theorem 8.1 of
\cite{milnor2015lectures} and its proof
we deduce that for each of the
attractors $\alpha_j$, $j=1,\ldots , m$,
there is an index 1 saddle
$s_j=\sigma_{i(j)}$ such that the
unstable manifold of the saddle
intersects $V_j$.

Since the flow is inward towards the
attractor on each $V_j$ this
intersection point is unique.  Let
$\gamma$ denote the part of the unstable
manifold between the saddle $s_j$ and
$\alpha_j$.  We can then use Morse's
cancellation lemma (Theorem 5.4 of
\cite{milnor2015lectures}) to modify
$X_1$ in an arbitrary small
neighbourhood $W_j$ of $\gamma$
contained in $U_{i(j),j}$ to remove this
attractor-saddle pair so that (i) the
modified flow $X_2$ has no rest points
in $W_j$ and every trajectory that
enters $W_j$ also leaves it, and (ii) so
that the modified flow $X_2$ is a
gradient-like MS flow that equals $X_1$
outside $W_j$.  As noted in
\cite{milnor2015lectures} there is no
restriction on dimensionality for 
connections between index 1 saddles and attractors.
This process essentially
involves a fold bifurcation.

We firstly do this for $j=m$.  As a
result we end up with a MS system with
one less saddle-attractor pair that
equals the original system $X_1$ outside
$U_{i(j),j}$.  Applying Theorem 8.1 of
\cite{milnor2015lectures} again we have
that for each of the attractors
$\alpha_j$, $j=1,\ldots , m-1$ there is
an index 1 saddle such that the unstable
manifold of the saddle intersects $V_j$.
Now we can use Morse's cancellation
lemma again and continue inductively
until all the attractors
$\alpha_1,\ldots \alpha_m$ are removed.
We end up with the single attractor
$\alpha_0$ remaining and this is inside
the ball $D_0$.

If this process does not delete all the
index 1 saddles (i.e.\ $M>m-1$) then
proceed as follows.  Let $\Sigma$ denote
the set of index 1 saddles that have not
been deleted and let $R$ be $R_\Sigma$.
Since all modifications to bifurcate
away the attractors took place inside
sets $U_{i,j}$ where the saddle was not
in $\Sigma$, $X_1$ and $X_2$ agree on a
neighbourhood of $\partial R_\Sigma$.
It follows from Theorem 3.4 of
\cite{milnor2015lectures} that the
cobordism $(R\setminus D_0,\partial
R,\partial D_0)$ is a product cobordism
and therefore that $\partial R$ and
$\partial D_0$ are diffeomorphic.
Consequently, $\partial R$ is a smooth
$(n-1)$-dimensional sphere.

Since all modifications to $X$ to
produce $X_1$ took place in the $U_i$,
$X$ and $X_1$ agree on a neighbourhood
of $\partial R$  and we have found an
appropriate smooth sphere containing the appropriate 
rest points.

If $M=m-1$ we take a Liapunov function
$L$ for $X$ that takes the value $i$ at
each rest point of index $i$ and take
for $R$ the connected component of
$L<\eta$ for $1<\eta <2$ a regular value
of $L$ and demand that the $U_i$ and
$U_{i,j}$ for $(i,j) \in \mathcal{C}$
have closures in $R$.  Then we deduce
that $\partial R$ is a smooth sphere in
a similar fashion.

In either case we obtain a smooth sphere
$\partial R$ on which the flow is inward
transverse and which contains all the
$m$ attractors of the flow, $m-1$ index
1 saddles and no rest points of index
$>1$.  We also obtain a Liapunov
function $L$ for $X$ for which $\partial
R$ is a level manifold.

The stable manifolds of the index 1
saddles in $R$ intersect $\partial R$
transversally in spheres of dimension
$n-2$ since in some neighbourhood of
each index 1 saddle point $p$ the level
curve $L(x)=L(p)+\eps$ intersects the
stable manifold of $p$ in an $n-2$
dimensional sphere $S_\eps$ when
$\eps>0$ is small.  As $t\rightarrow
-\infty$ every trajectory on $S_\eps$
leaves $R$ and hence intersects
$\partial R$ transversally (since
$\partial R$ a level surface of $L$) at
a single point.  The points on the
intersection of the unstable manifold
with $\partial R$ are connected to those
of $S_\eps$ by the trajectories of $X$
and this provides a diffeomorphism
between them.
\qed

\subsection{Four attractor configurations}
When there are four attractors there are
two possible configurations.  There will
be three cancelling saddles and the
stable manifolds of these will intersect
the $(n-1)$-sphere $\partial R$ in three
$(n-2)$-spheres, $S_1$, $S_2$ and $S_3$.  
Each of the $S_i$ will separate $\partial R$ and
clearly for two of them, say $S_1$ and
$S_2$, $\partial R$ is separated into
two components one of which $D_i$,
$i=1,2$, does not intersect $S_3$.  In this case, $D_i$ is
a $(n-1)$-disk and is contained in the
basin of one of the attractors $A_i$
with $A_1\neq A_2$.  The third
$(n-2)$-sphere $S_3$ may or may not have
this property.

If it does, then  there is a $(n-1)$-disk $D_3$
with similar properties.  In this case
the saddle $s_i$ in the stable manifold $W^s_i$
corresponding to $S_i$, $i=1,2,3$
does not escape to $A_j$ for
$j=1,2,3,j\neq i$.  Since every saddle
escapes to two different attractors,
every saddle $s_i$ must escape to the
fourth attractor $A_4$.  Thus the
configuration in phase space is that
there is an attractor $A$ such that each
of the three saddles escapes to $A$ and
also to one other of the three other
attractors.  We call this configuration
\emph{EU-like} since it is what occurs
in the compact elliptic umbilic.

In the other configuration one of the
components of $\partial R\setminus S_3$
contains $S_1$ and the other $S_2$.  Let
$Q_i$ denote the component of $R\setminus W^s_3$ containing
$S_i$, $i=1,2$.  Then $Q_i$ contains the
unique saddle $s_i$ in $W^s_i$, the
attractor $A_i$ and one other attractor
$B_i$ that $s_i$ escapes to.  Since
$B_i\in  Q_i$, $B_1\neq B_2$.  It
follows that the saddle $s_3$ in $W^s_3$ must
escape to the attractors $B_1$ and
$B_2$.  Thus in this configuration the
saddles are ordered so that $A_1\prec
s_1\succ B_1 \prec s_3 \succ B_2\prec
s_2\succ A_2$.  We call this the
\emph{linear configuration}.

\section{No folds at cuspoidal joining attractors}\label{App:no_fold}

\begin{lemma}\label{lem:nofold}
	Consider a 3-attractor system as
	in Sect.\ \ref{sec:sad-att}.
	If the connector is not smooth then
	a fold bifurcation of the central
	attractor is impossible.
\end{lemma}

\proof
We need to check that there cannot be a fold
at a cuspoidal join.
If there is a fold there
it is at a point $\param$
on a fold bifurcation curve $B_C$.
In the generic case we can move $\param$ an
arbitrary small amount in $B_C$
to ensure that the eigenvalues at the
fold point are non-resonant.
if the move is small enough the
connection will still be non-smooth and cuspoidal.

Note that this new fold point in $n$
dimensions for $r>2$ one can find $C^r$
coordinates $(x,y)\in\reals\times
\reals^{n-1}$ in which the differential
equation takes the form of equation (\ref{eqn:ndim}) 
(\cite{arnold2013dynamical}
Sect.\ 5.7).  Without loss of generality
we can assume that this bifurcation takes
place at $\param=0$. 
Consider the case
$+x^2$.  Then for $\param>0$ small there
are two saddles $s_1$ and $s_2$ only one
of which (say $s_1$) is involved in the
bifurcation.  The unstable manifold of
these are asymptotic to the attractor
taking part in the fold bifurcation
where they meet cuspoidally.  Because
they meet in this way we must have that
for points starting on the unstable
manifold of $s_2$, $\dot{x}(t) <0$ as
they converge onto the attractor.
However, from equation (\ref{eqn:ndim})
$\dot{x}(t) \geq 0$ in a neighbourhood
of $x=0,y=0$ which gives a
contradiction.
\qed

\section{Proof that 
	the connected components of $\catm$ are disks}\label{app:2}

\subsection{The interior of a fold circle is a disk}\label{app:3}
If $\x_i=(x_i,\theta_i)$, $i=1,2$ are in $\catm$ write
$\x_1\sim \x_2$ if there exists a connected
neighbourhood $U$ in $\catm$ containing both
points and not intersecting any fold curves.  
This is clearly an equivalence relation
and the equivalence class of a point
$\x\in \catm$, which we denote by $[\x]$,
is connected.

Take a small tubular neighbourhood $N$ of $C$
and consider the two components of $N\setminus C$.
Any two points in one of these are equivalent under 
$\sim$ and not equivalent if they are in different
components. 

If one of the components $N^\text{ext}$ contains in its
closure a boundary point of $\catm$ 
we say $C$ that is at \emph{depth} 0.
Then, if
$N^\text{int}$ is the other we
denote by $M_C$ and $V_C$
the equivalence classes of
$\x\in N^\text{ext}$ and $\x\in N^\text{int}$.
We are thinking of $V_C$ as being part of the inside of $C$.

Now for an arbitrary fold circle $C$ we define depth, $M_C$
and $V_C$ by
induction: $C$ has depth $n > 0$ if 
for some tubular neighbourhood $N$
the equivalence class of
one of the components of $N\setminus C$
is $V_{C'}$ for a fold curve
$C'$ of depth $n-1$.  Then that
equivalence class is denoted $M_C$ and $V_C$ is
the other equivalence class. 

\begin{proposition}\label{cor:disk}
	Each fold curve $C$ separates $\catm$ into
	two components.  One of them $I_C$ is a
	disk with boundary $C$ and is such that
	$\chi|I_C$ is injective on a neighbourhood of
	$C$ in $I_C$.  The
	other has the property that it contains
	in its closure either boundary points of
	$\catm$ or singularities of $\chi$ which are
	not in $C$.
\end{proposition}

\noindent{\bf Proof.} 
We prove this by induction on height.

Suppose $C$ has height $0$ by which we mean
that $V_C$ contains no other fold points.
Then $\chi:V_C\-> U_C=\chi(V_C)$ is injective.
Consequently, by invariance of domain
$\chi | V_C$ is a homeomorphism between $V_C$
and $U_C$ and $\chi(C)$ separates $\reals^2$ into
two components only one of which is bounded.
It follows that this is $U_C$ and that it and $V_C$ are disks.

Now suppose the height is $n$. Then for
each fold circle $C'$ in the closure of $V_C$ of height $n-1$, $I_{C'}$
is a disk with boundary $C'$ and 
$\chi|I_{C'}$ is injective on a neighbourhood of
$C'$ in $I_{C'}$. Moreover, $V_C$ is a
surface with boundary $C$ and the union of
the $C'$ in $V_C$ of height $n-1$.
Therefore, to obtain $I_C$
one just glues the disks $I_{C'}$
into $V_C$ to obtain a disk with boundary $C$.  
\qed

\subsection{Proof of Prop.\ \ref{prop:disk}}

Consider a connected component $\catm_0$
and let $J(\x )$ be the Jacobian of
$\chi$ at $\x$ and let $\catm_1$ denote
the set of points $\x\in\catm_0$ where
$J>0$.  Since $J>0$ on a neighbourhood
of the boundary, it follows by Prop.\
\ref{cor:disk} that, in the generic
case, the sets where $J<0$ are disks
bounded by fold circles.  Therefore, if
$\catm_0$ is not orientable $\catm_1$
must contain a M\"obius band.  But that
is impossible as then $\chi$ must have a
singularity in $\catm_1$ contradicting
$J>0$.

To show that $\catm_0$ is a disk we must
prove that $\catm_1$ does not contain
any handles.  Then the result
follows from the classification of
surfaces (e.g.\ \cite{Hirsch2012}).

Suppose that there is a handle.  Then
there is an embedding $f: S^0\times
D^2\rightarrow \catm_0$ such that the
handle is given by using $f$ to glue
the cylinder $K=D^1\times S^1$ into
$\catm_0\setminus f(K)$.  But then the
induced orientation on $K$ induces an
orientation on $1\times D^2$ that is
opposite to that on $-1\times D^2$.
Hence the two circles bounding these
have opposite orientations.  It follows
that $\chi|\catm_0$ has singularities as
otherwise it would map the handle to an
annulus and induce opposite orientations
on its inner and outer boundary circles.
\qed

\section{Bifurcation curves:
	Proof of Prop.\ \ref{prop:BsubC}}\label{app:1}

We use the fact that every connected component
of $\catm$ is diffeomorphic to a disk.
One can prove the results of this section without this assumption
using ideas such as those employed in Sect.\ \ref{app:3}.
From this assumption it follows that we can talk about the
interior and exterior of a fold circle and we have 
that the restriction of $\chi$ to the 
interior of a fold circle is injective.

That
$B_C$ is smooth except at cusp points
follows from the generic local structure of fold
points as described above. 

Suppose that $B_C$ has a
self-intersection.  By genericity the
intersection will be transverse.  Then
there are two points $\x_1,\x_2\in C$
such that $\chi(\x_1)=\chi(\x_2)$ and
that there is an open segment $\gamma$ in
$C$ that has $\x_1$ and $\x_2$ as its end
points  such that $\chi|\gamma$ is an
embedding.  $\gamma$ might contain cusp
points.  Let $N$ be a small tubular
neighbourhood of $C$ and let
$N^\text{int}$ be the connected
component of $N\setminus C$ contained
in the interior of $C$. If $N$ is small enough
$\chi|N^\text{int}$ is injective.
Because of the
transversality of the intersection the
thin ribbon $\chi(N^\text{int})$ in $\reals^2$
has self-intersection and so there are
distinct points in $N^\text{int}$ with the same
image under $\chi$.  This
contradicts the injectivity of $\chi$ on
$N^\text{int}$.

Now the result follows from the 
Jordan-Brouwer Separation Theorem
which implies that $\reals^2\setminus B_C$
has exactly two components, one of 
which is bounded and the other not.
\qed

\section{Proof of Prop.\ \ref{prop:remove}}\label{sect:E}

\begin{figure}[H]
	\centering
	\includegraphics[width=0.9\linewidth]{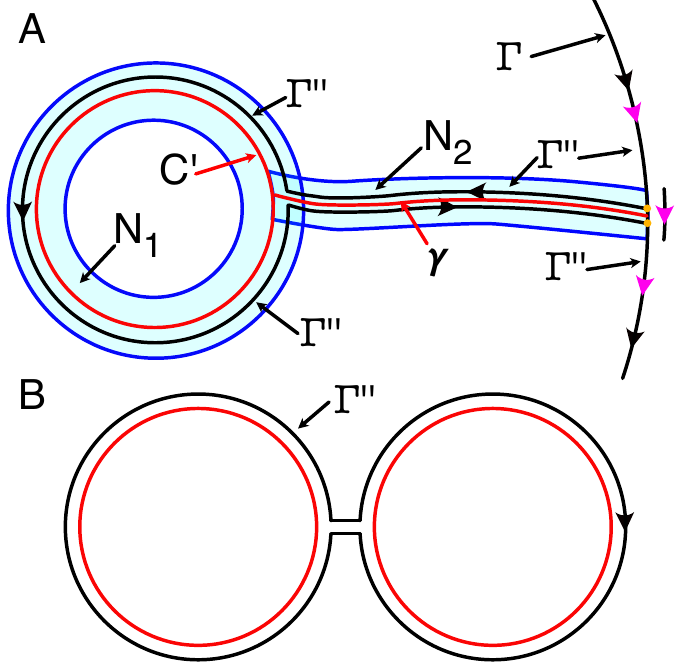}
	\caption{(A) We connect $C'$ (red) to
		$\Gamma$ (black) and then take a small
		neighbourhood $N_1\cup N_2$ as shown (light blue).
		Then we take the black curve in $N_1\cup
		N_2$ as shown that joins the two orange
		points on $\Gamma$ either side of
		$\gamma$.  By following the black arrows
		instead of the pink ones we get a loop
		$\Gamma'''$ that has an odd index as
		$\Gamma$ (because the index round the
		new part is even) but now has $C'$ in the
		exterior part.  (B) How to move two odd
		index fold circles to the exterior.
	}
	\label{fig:Proof1}
\end{figure}

\def\G{\mathcal{G}}
\proof
Let $\G$ be the interior of $\Gamma$. 
Let $C'$ be a fold circle in 
$\G$ that
is not contained in any other fold
circle and is such that $\mathcal{C}_{C^\prime}$ is trivial.
We firstly prove that we can find a
simple closed curve whose interior contains exactly the
fold circles that are in $\G$ except for ${C^\prime}$.

Let $\catm_0$ be the connected component of
$\catm$ that contains $\Gamma$ and let 
$\gamma$ in $\catm_0$ be
a curve that connects ${C^\prime}$ to $\Gamma$
(see Fig.\ \ref{fig:Proof1}A).  Since
all fold points inside $\G$ must be on
fold curves we can choose $\gamma$ so
that it intersects no other fold points.
Also choose $\gamma$ so that the
endpoint on ${C^\prime}$ is not a cusp point.
Let $N_1$ be a small annular tubular
neighbourhood of ${C^\prime}$ and $N_2$ a small
annular tubular neighbourhood of
$\gamma$ and let $N=(N_1\cup
N_2)\setminus({C^\prime}\cup\gamma)$.  Choose a
point in each of the components of
$N\cap\Gamma$ and let $\Gamma'$ denote
the short part of $\Gamma$ between them.
Then there is a curve $\Gamma''$ in $N$
that connects these points.  Let
$\Gamma'''$ be the closed curve obtained
by combining $\Gamma\setminus \Gamma'$
with $\Gamma''$.  Moreover, since $N$
encloses a disk, $\Gamma'''$ separates
$\catm$ into two components as required.
But now ${C^\prime}$ is in the exterior
component.  Now $\Gamma'\cup
\Gamma''$ is a closed curve for which
the saddle bundle is trivial.  Since
$\Gamma'''=
(\Gamma\setminus\Gamma')\cup \Gamma''$ it
follows that if $\Gamma'$ is short
enough the saddle bundle of $\Gamma'''$
is a M\"obius band.  Moreover,
$\Gamma'''$ separates $\catm_0$ into two
components and one of the components
contains less fold circles with
trivial saddle bundle than $\G$ did.

Repeating this process for all such fold
circles we obtain a closed curve which
must contain in its interior component a
fold circle since it has a
M\"obius saddle bundle but which
contains no non-M\"obius fold circles 
at top level.

To see that one can also remove any even number
of M\"obius fold circles note
that one can find a closed curve 
$\Gamma^{\prime\prime}$ around
any two M\"obius fold circles with the
property that its bundle $E^c_{\Gamma^{\prime\prime}}$ is a
cylinder (see Fig.\ \ref{fig:Proof1}B).
Then one can proceed as above
to use $\Gamma^{\prime\prime}$
to adjust $\Gamma$ to obtain a curve
with a M\"obius saddle bundle whose
interior component does not contain
these two fold circles.
\qed

Note that this proof also shows that any
curve $\Gamma_0$ satisfying the conditions 
of the proposition contains an odd number of 
M\"obius fold circles.

\section{Justification of Condition (*)
	for Theorem \ref{corollary:1}}
\label{sect:just}

\begin{figure}[H]
	\centering
	\includegraphics[width=\linewidth]{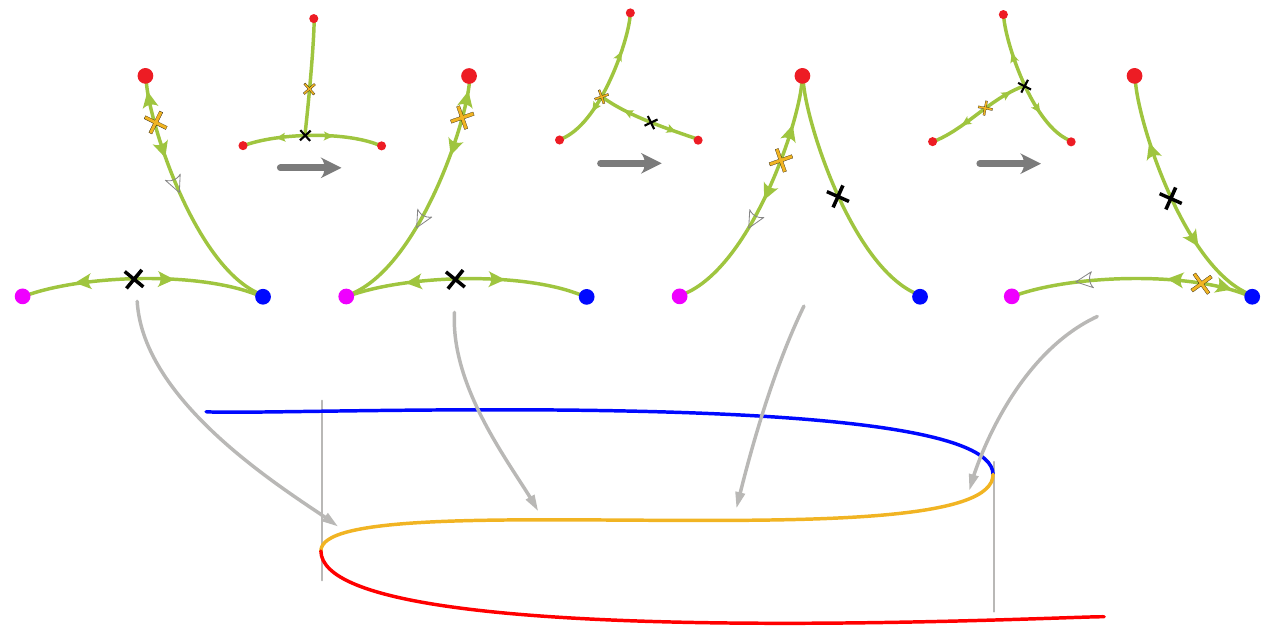}
	\caption{This shows how over  a 
		1-dimensional parameter space one
		can have a seemingly bistable Z shaped system 
		which does not flip the
		orientation of the folds. This involves
		generic flip bifurcations and
		violates
		Condition (*) thus
		illustrating why bistability is not sufficient in
		Theorem \ref{corollary:1}. 
		Note that the lines could cross 
		depending on the projection considered.
	}
	\label{fig:bistable_but_no_switch}
\end{figure}


%% file: examples_SI.tex
\section{Three state system has elliptic-umbillic parameter space}

For three cells or genes $i,j = 1..3$ define mutual repression among them by the rational functions

\begin{equation}
\begin{aligned}
	\dot x_i &= \frac{2}{\prod_{j \neq i} (1+\frac{x_j^n}{k_{i,j}^n})}- x_i, \\
	k_{i,j} &= 1+ c_i - c_j
\label{equ:3cellinhibition}
\end{aligned}
\end{equation}

\noindent
with the Hill exponent $n=4$ ($n=2$ does not generate 7 restpoints at the symmetry point).

Each of the $c_i$ favors $x_i$ by dividing $x_i$ by a number smaller than $1$ and $x_{j \neq 1}$ by a number larger than $1$. 
We confine ourselves to a two dimensional parameter space by imposing $\sum_i c_i = 0$ and then projecting onto two orthogonal directions such that permutation among the $i$ corresponds to a rotation by $2 \pi /3$. 
Rational functions are technically useful in this example since Mathematica routines are guaranteed to find all the rest points, and one can also include the zero Jacobian condition to find the fold lines. We believe the same phase diagram would emerge from Eq.\ref{equ:corson_model} for $N=3$ but the root finding is more difficult with transcendental functions. We again emphasize that imposing the 3-fold symmetry simplifies the parameter search but the phase diagram is invariant in a neighborhood in the complete parameter space. One presumably would find the same phase diagram if the interactions were completely symmetric and the parameters controlled the degradation rates. With an exponent $n=4$ the states are reasonably close to binary, i.e., at the corners A in Fig.\ref{parameterspace2D}b the stable point is at $x_i = (2.00, 0.03, 0.03) + \mathrm{permu}$, while along the inner black curve the stable points are $x_i \sim (2.00, 0.2, 0.2) + \mathrm{permu}$ and the saddle point is at $(1.00, 1.00, 0.44)$

\section{A flow defines a potential}

We use the flows which are later used for the activator-inhibitor pair showing Turing patterns. 
\begin{align}
 \dot{a} &= \frac{a^2}{(1 + a^2)(1 + h)} - \nu a + D_a \partial_x^2 a , \\
 \dot{h} &= \rho \frac{a^2}{1+a^2} - h + D_h \partial_x^2 h .
 \label{GMeqnsfp}
\end{align}
Parameter values, which are also used later, are $\nu = 0.2$ and $\rho = 5$. There are two stable fixed points on either side of the inset. The potential is defined as $\int_0^\infty (\dot{a}^2 + \dot{h}^2) dt$ in accordance with the one dimensional definition i.e. it is the integral of the of the vector field along the trajectory. To define the metric, we need to project the flows given by the potential onto the actual flow. This can be done with a projection operator, which, however is not a proper metric because it has determinant zero. Let $\vec{a} = (\dot{a},\dot{h})$ represent the vector flows.  Let $\vec{u}$ be the flows which derive from the derivative of the constructed potential. Let $\vec{u}_{P}$ be the vector perpendicular to $\vec{u}$ with the same magnitude. Now define the inverse metric as the sum of two projection operators 
\begin{equation}
g = \frac{1}{(\vec{u}.\vec{a})} \left( \vec{a} \vec{a}^T + \vec{u}_P \vec{u}_P^T \right)
\end{equation}
Evidently, when operating on $\vec{u}$, this gives $\vec{a}$. Though these two independently have determinant zero, their sum can be shown to have positive determinant as long as the angle between them is less than $\pi/2$ i.e. $\vec{u}.\vec{a} > 0$. Other linear combinations of these two operators are possible reflecting the redundancy in the definition of the metric. 

\begin{figure}[h]
\centering
	\includegraphics[scale=0.40]{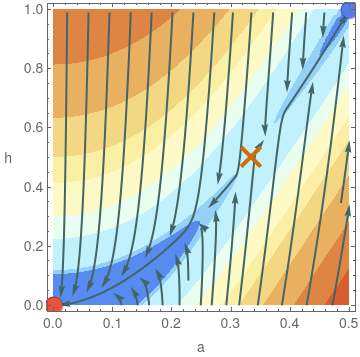}
\caption{ Contours of the potential defined in the text are shown going from low (blue) to high (red). The flow defined by the potential multiplied by the metric is shown in green as streamlines.
}
\label{flowtopotentialmet}
\end{figure}

Near the fixed points, this construction chooses the potential to be the symmetric part of the Jacobian. In two dimensions, it can be shown that a matrix cannot be written as the product of two symmetric matrices where the second matrix is the symmetric part of the matrix being represented. Therefore, the potential and the metric need to be defined independently near the fixed points. Near the fixed point, one can approximate the dynamics by its linear part. The Jacobian matrix is not symmetric but can be written as the product of two symmetric matrices, one of which is positive definite. Hence, we can approximate the metric as constant and the potential as parabolic near the fixed points and glue this together with the global construction above. The gluing is done with a sigmoidal function $1/(1 + \exp(((a-a_0)^2 + (h-h_0)^2 - m)/c)$. $a_0$ and $h_0$ are just the value at the fixed points whereas $m$ and $c$ are chosen to be $0.001$ and $0.00001$ in this example but can be chosen appropriately. 

\section{Patterning by lateral inhibition}


The model in question describes a field of cells, subject to mutual inhibition via the Notch-Delta pathway \cite{corson2017selfsup}. Each cell is described by a single variable $a$ between $0-1$ and when a cell is ON it forces its neighbors to be OFF. The biological data consists of the time course of expression beginning with all cells off and ending when a sparse random subset are on.  An elegant minimal model describing this situation reads:

\begin{equation}
\begin{aligned}
	\tau \dot a_i &= \sigma(a_i - h_i) - a_i , \\
	\sigma(a) &= (\tanh(4 a) + 1)/2 ,  \\
	h_i &= \sum_{j \ne i} K_{i,j} d(a_j) , \\
	d(a) &= 0.05 a+\frac{2.85 a^4}{1+2 a^2}
\label{equ:corson_model}
\end{aligned}
\end{equation}

\noindent where $K_{i,j}$ is a Gaussian function of the spacing between cells $i,j$ and $d(a)$ maps the interval $[0,1]$ onto itself. The system is initialized with all $a_i \sim 0$ and positive. In the terminal configuration only a sparse set of the $a_i \sim 1$ with a spacing set by the $K_{i,j}$ and the rest are zero.  The mean field inhibition $h_i$ from the neighbors of $i$ slides the sigmoid function $\sigma$ along the axis giving rise to two saddle-node bifurcations that destabilize the upper and lower fixed points.

\begin{figure}[h]
\centering
\includegraphics[width=.99\linewidth]{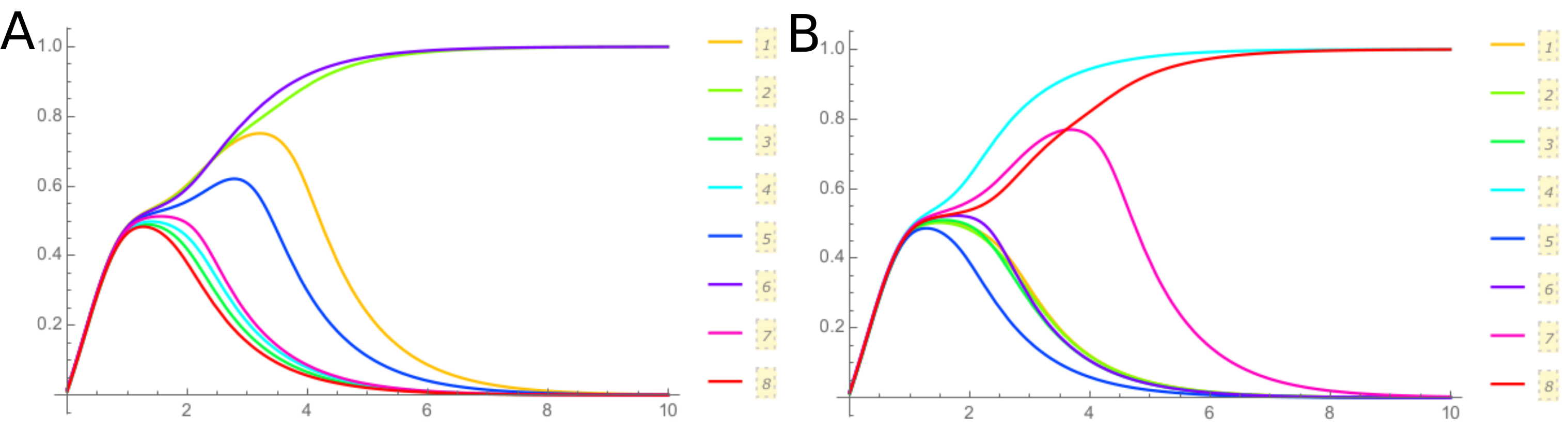}
\caption{Solution $a_i(t)$ for the model of Eq.\ref{equ:corson_model} for 8 cells evenly spaced around a circle with slightly different initial conditions shown in (A) and (B). The range of the inhibitory interactions is adjusted to allow only two cells to be on in the final state, which are always on opposite sides of the circle.
}
\label{fig:corson_ring_t}
\end{figure}

The behavior of the system is most easily understood by solving it in one dimension for cells uniformly positioned around a circle Fig.~\ref{fig:corson_ring_t}. We have omitted additive Gaussian noise to better understand the underlying dynamics.

\begin{figure}[h]
\centering
\includegraphics[width=.99\linewidth]{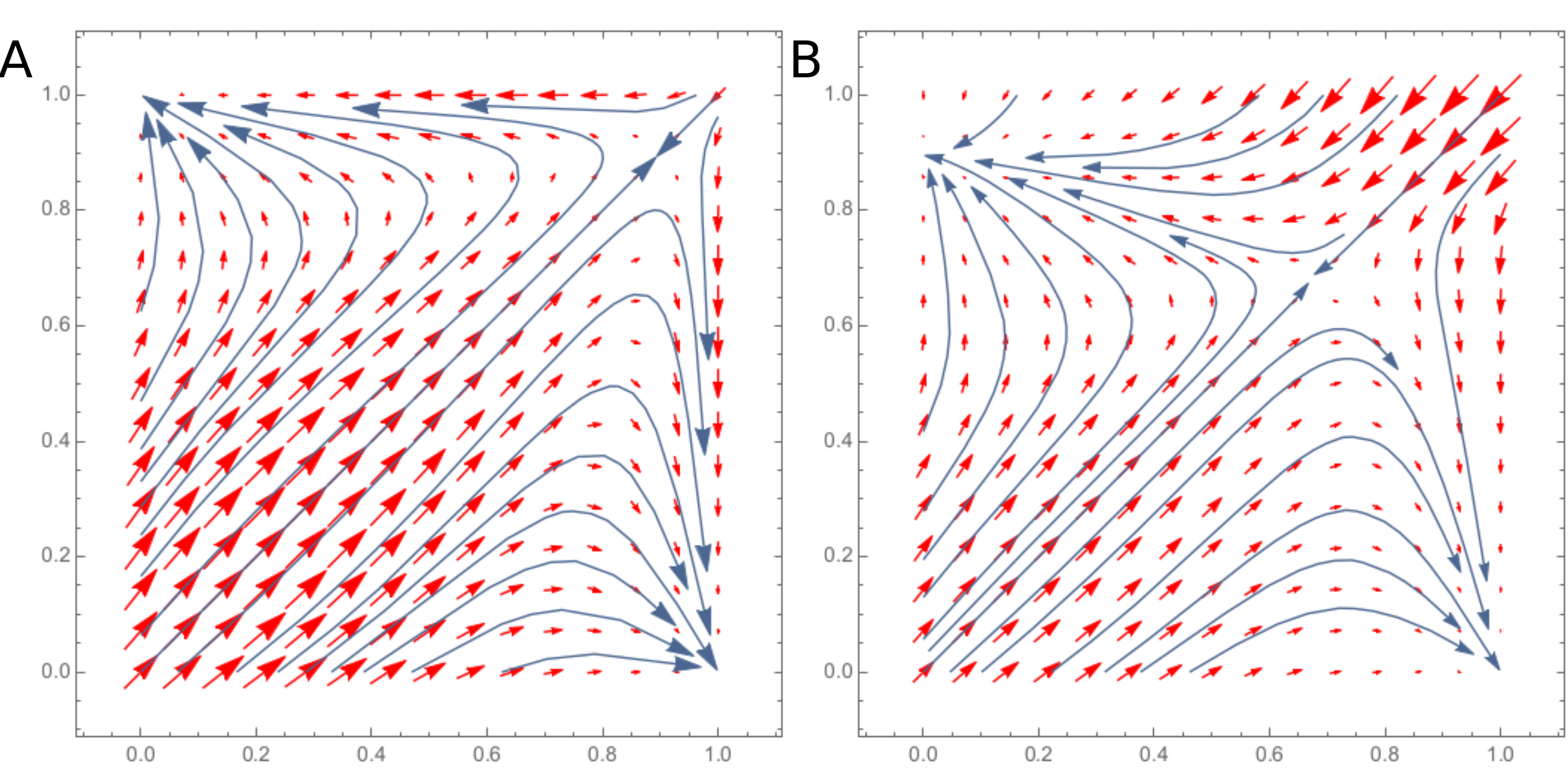}
\caption{Vector field and stream lines from Eq.\ref{equ:corson_model} for (A) two cells and (B) three cells with $a_1$ on the x-axis and $a_2 \equiv a_3$
}
\label{fig:corson_vplot}
\end{figure}

Note the cells all rise together and then when $a \sim 0.5$ cells peal off. Most go to zero but a few persist to larger $a$ when they too peal off leaving the two survivors. The intuitive reason for this is already apparent when observing the flow field with just two cells Fig.~\ref{fig:corson_vplot}. 

In fact for any number of particles there is a saddle point along the diagonal in $a$ space at $a \approx 0.89$ for two cells which decreases down to $a \approx 0.51$ (for $K_{0,1} \approx 0.841$) when the number of cells $N$ exceeds the range of interactions. There is one strongly stable eigenvalue along the diagonal and $\mathcal{O}(N)$ unstable directions. The pattern repeats as suggested in Fig.~\ref{fig:corson_vplot}B, namely for $N=3$ there are secondary saddle points around $\vec a \approx (0, 0.89, 0.89) + \mathrm{permu}$ with two stable and one unstable directions. Thus the dynamics in Fig.~\ref{fig:corson_ring_t} is qualitatively explained by a tree or star-burst of saddle points with an increasing number of stable directions.

Can we capture the essentials of this behavior in a potential model of the form of Eq.\ref{equ:pot_model} where both $K_{i,j}$ and $d(a)$ agree with Eq.\ref{equ:corson_model}?

\begin{equation}
\begin{aligned}
	\dot a_i &= -g(a_i) \nabla_{a_i} F \\
	F &= \sum_i s(a_i) + \frac{1}{2} \sum_{i \ne j} d(a_i) K_{i,j} d(a_j)
\label{equ:pot_model}
\end{aligned}
\end{equation}

\noindent We first impose that the location of the critical point along the diagonal in $a$ space agrees with Eq.\ref{equ:corson_model}:

\begin{equation}
\begin{aligned}
	s'(a) + d'(a) h(a) = 0 & \Leftrightarrow  \sigma(a - h(a)) = a \; \text{where}\\
	h(a) \equiv \sum_{j}K_{0,j} d(a) & = a - \frac{1}{8} \ln(\frac{a}{1-a})
\label{equ:critical_pt}
\end{aligned}
\end{equation}

\noindent where the expression involving $\ln$ in Eq.\ref{equ:critical_pt} is the inverse of $\sigma$ in Eq.\ref{equ:corson_model}. Thus we have an explicit expression for $s'(a)/d'(a)$ in terms of $a$ at the critical point which falls in the range $0.51 - 0.89$ depending on $N$.

To make the Jacobians $J_{i,j}=\partial \dot a_i/\partial a_j $ agree compare the expressions in Eq.\ref{equ:corson_model} and Eq.\ref{equ:pot_model}:

\begin{equation}
\begin{aligned}
	J_{i,j} =& \delta_{i,j} (\sigma'(a - h(a)) - 1)  - (1-\delta_{i,j}) \sigma'(a - h(a)) \partial h_i/\partial a_j \\
	J_{i,j} =& -\delta_{i,j} g(a) (s''(a) + d''(a) h(a)) \\ &- (1-\delta_{i,j}) g(a) d'(a)  \partial h_i/\partial a_j 
\label{equ:jacobian}
\end{aligned}
\end{equation}

\begin{figure}[h]
\centering
\includegraphics[width=.99\linewidth]{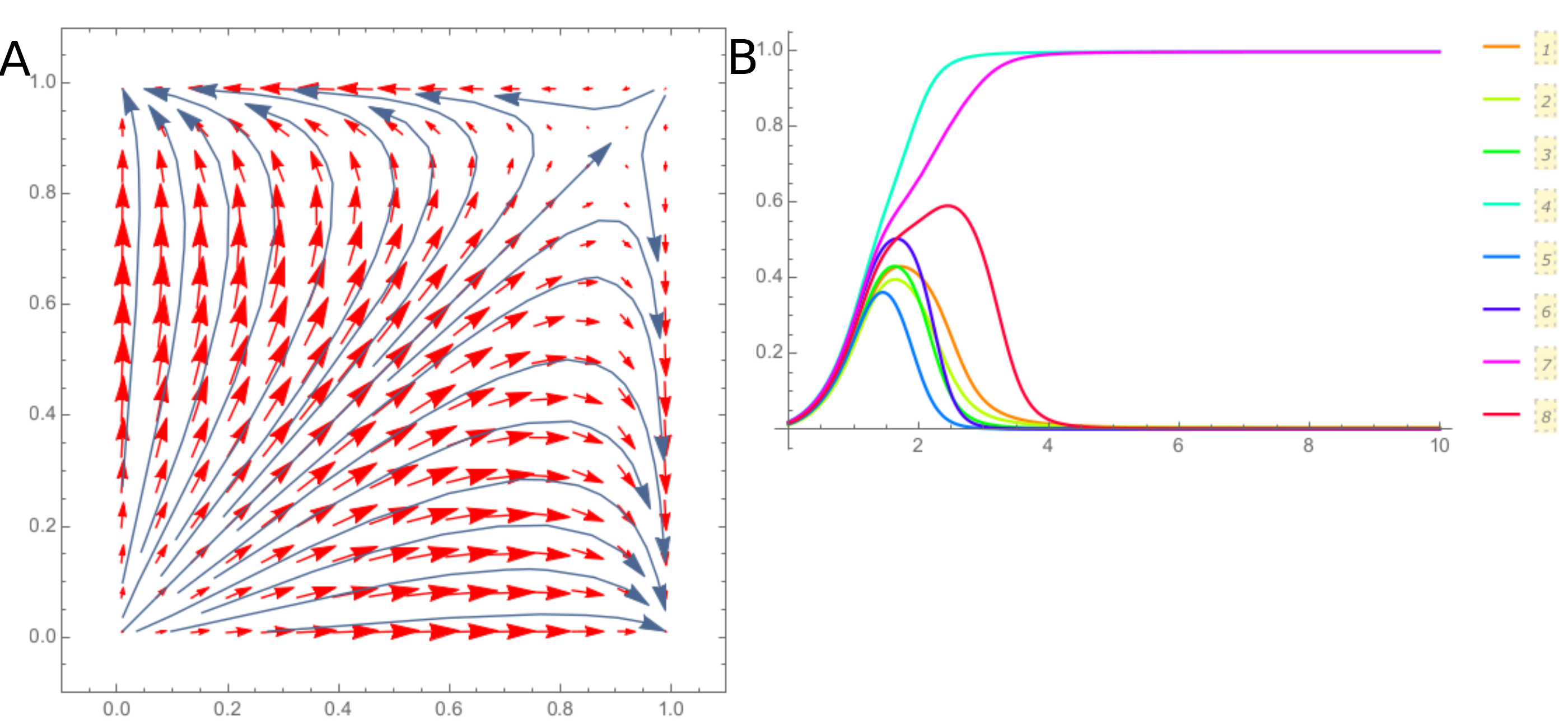}
\caption{The potential model from Eq.\ref{equ:pot_model} with a metric satisfying $g(a) d'(a) = 8 a (1-a)$. (A) The vector field and streamlines (B) The dynamics. Note the dynamics around the origin are not correct as is clear in both (A) and (B).
}
\label{fig:pot_metric0}
\end{figure}

\noindent Note at the critical point the term involving the derivative of the inverse metric vanishes, and we can calculate $\sigma'(a - h(u)) = 8 a (1-a)$. Thus the off diagonal terms are equivalent if $g(a) d'(a) = 8 a (1-a)$, and it can be shown the diagonal ones follow, as we also checked numerically. Thus we have determined both $s'(a)$ and $g(a)$ in Eq.\ref{equ:pot_model} in terms of $d(a)$ and explicit functions of $a$.The vector field for N=2 and the dynamics for N=8 are shown in Fig.~\ref{fig:pot_metric0}. The most obvious problem is that the potential model has vanishing velocity in the neighborhood of the origin since $g(a) \sim a$.

The origin is a completely unstable critical point well removed from the saddle on the diagonal which is restricted to $a > 0.51$. So in particular we can define the metric for $a \geq 0.5$ as before, but modify the definition for $a < 0.5$ so that $\tau \dot a = \frac{1}{2}$ at $a=0$ to agree with Eq.\ref{equ:corson_model}. For instance the following choice illustrates the idea:

\begin{equation}
\begin{aligned}
	g(a) = 8 a (1-a) - (1-2a)^4 (4/\ln(a) + 8 a (1-a)) \eta(\frac{1}{2}-a)
\label{equ:metric}
\end{aligned}
\end{equation}

\noindent where expression involving the step function $\eta$ is $C^4$ around $a=\frac{1}{2}$ ($\eta(x<0)=0$) and the $\ln(a)$ term subtracts the singularity in $s'(a)$ near the origin that's implicit in Eq.\ref{equ:critical_pt}. Its overall coefficient together with the singular expression for $h(a)$ in Eq.\ref{equ:critical_pt}, guarantees the desired value at $a=0$

\begin{figure}[h]
\centering
	\includegraphics[width=.99\linewidth]{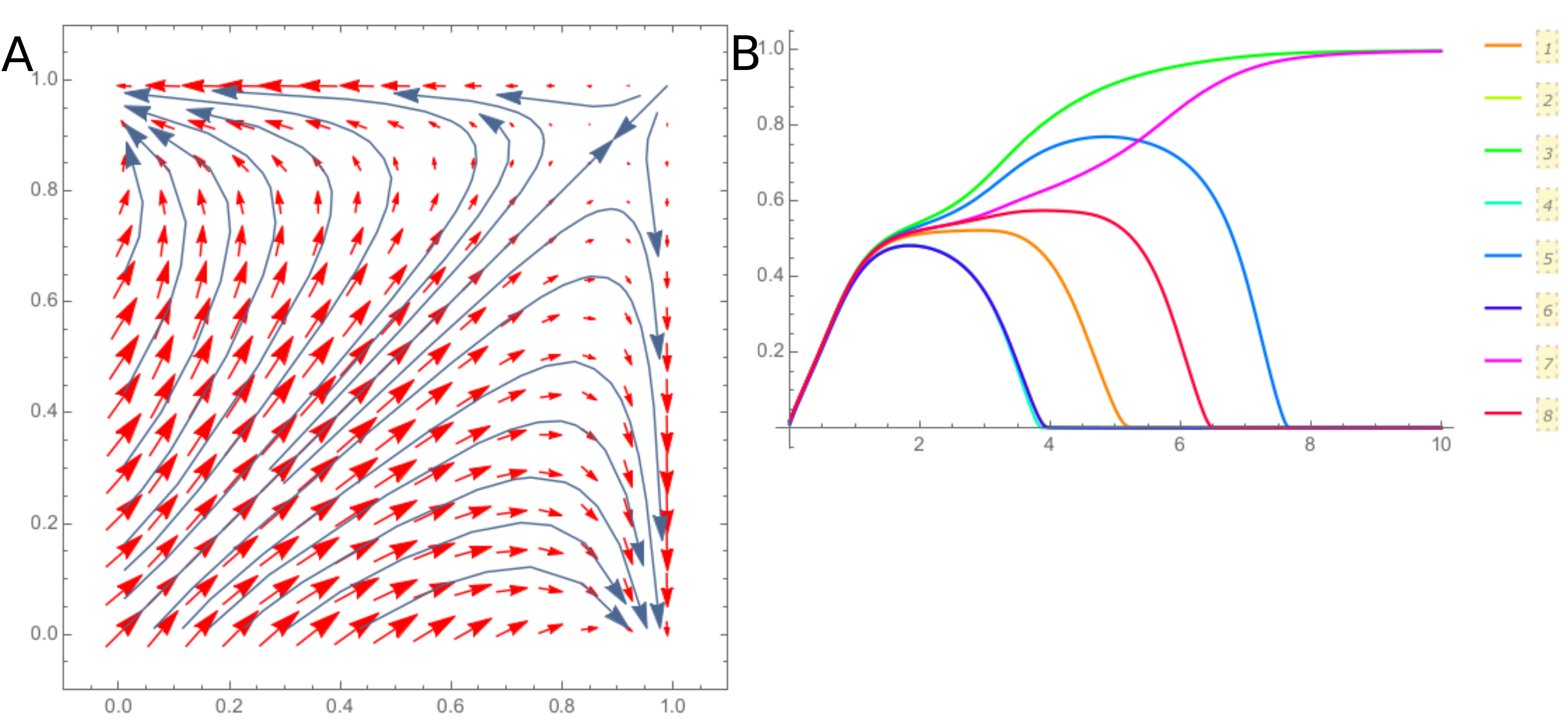}
\caption{The potential model from Eq.\ref{equ:pot_model} with the metric from Eq.\ref{equ:metric}. (A) The vector field and stream lines. (B) The dynamics.
}
\label{fig:pot_metric}
\end{figure}


Inspection of Fig.~\ref{fig:pot_metric} reveals that the velocity field is nearly 2x too large around $\vec a \approx (\frac{1}{2},  \frac{1}{2})$
and that $\dot a_1$ is too large for $\vec a \approx (a_1 < 0.6 ,  1)$ which affects how the curves tend to $0$ for times larger than two. This problem can also be fixed within the context of Eq.\ref{equ:pot_model} if we smoothly interpolate between two values of $s'(a)$ for $a \ll 1$ and $a \sim 1$ as in Eq.\ref{equ:metric}, and make a consistent choice of the two free constants to get appropriate values of $\vec a$ around the origin and $(0,1)$. The choices are now dependent on $N$ and we will not go further in matching a potential model to Eq.\ref{equ:corson_model}, since at this stage the agreement is better than the fit of Eq.\ref{equ:corson_model} to the corresponding data, which unfortunately for the system in question does not yet include quantitative time lapse data. With such data, one can hope to initialize the model specifically to each embryo and then follow the dynamics for some period of time and compare trajectories. We suggest that our potential and metric formulation will be a more principled way to fit dynamic data. Note that we got rather far with a diagonal metric.
 
 \FloatBarrier

\section{The Turing Model in potential form}
\begin{figure}[h]
\centering
	\includegraphics[scale=0.2]{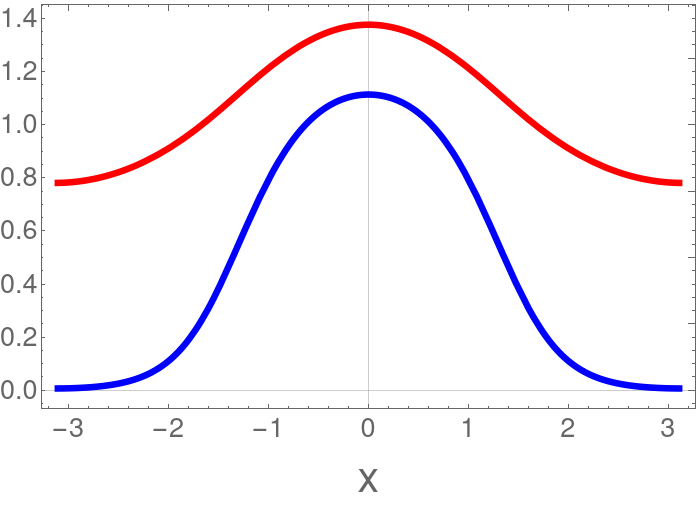}
\caption{The typical behavior of $a$ (blue) and $h$ (red) in the model. Parameter values are $\nu = 0.2$, $\rho = 5$, $D_a = 0.02$, $D_h = 2$. 
}
\label{ahstable}
\end{figure}

The model we use relies on local activation by an activator $a$ and long range inhibition by an inhibitor $h$. The dynamical equations are
\begin{align}
 \dot{a} &= \frac{a^2}{(1 + a^2)(1 + h)} - \nu a + D_a \partial_x^2 a , \\
 \dot{h} &= \rho \frac{a^2}{1+a^2} - h + D_h \partial_x^2 h .
 \label{GMeqns}
\end{align}
We have modified the classic Gierer-Meinhardt form of the model~\cite{meinhardt1982models} to remove the singularity at $h = 0$ and saturate the production of $a$. Parameters have been rescaled or set to $1$ so that the behavior of the model depends only on two parameters $\nu$ and $\rho$ along with the diffusion coefficients $D_a$ and $D_h$. The typical behavior of the model is to produce a bump in $a$ and $h$ as shown in in Figure~\ref{ahstable}.

This matrix can not be written as the product of a metric and a potential without introducing $k^2$ into the metric. The expression for the Jacobian matrix at the homogenous fixed point is quite involved. The essence of the argument can be seen by taking the limit of small $\nu$ (i.e. dynamics of $h$ being quicker than dynamics of $a$). 

\begin{equation}
\begin{pmatrix}
 2 \nu^2 - D_a k^2 && - 2 \nu^2 \\
1/(2 \nu) && -1 - D_h k^2
\end{pmatrix}
\end{equation}

It is easy to show that the opposite signs on the off-diagonal elements implies an off-diagonal metric. One can therefore take the generic form of a metric inverse times a potential 
\begin{equation}
g.V = 
\begin{pmatrix}
 g_{11} && g_{12} \\
g_{12} && g_{2 2}
\end{pmatrix}
\begin{pmatrix}
 V_{11} + U_{11} k^2 && V_{12} + U_{12} k^2 \\
V_{12} + U_{12} k^2 && V_{2 2} + U_{22} k^2
\end{pmatrix}
\end{equation}

Setting this expression to be equal to the Jacobian and solving the equations and matching the $k^2$ terms leads to a contradiction: there is no solution of these equations with the components of $g$ independent of $k^2$ with $D_a \neq D_h$ and $g_{1 2} \neq 0$ which we require to fix the sign problem. Making the metric a function of $k^2$ is not disallowed but it is an overly cumbersome route which we do not pursue. 

Another alternative is to saturate the production of $h$ which gives $h = \rho \oint a^2/(1 + a^2)$. Treating this equation as a constraint, one can write down a long range potential for $a$
\begin{align}
V(a(x), h) =& \int dx ( \nu \frac{a(x)^2}{2} + H(a(x)) + \frac{D_a}{2} (\partial_x a(x))^2 \nonumber \\ &- \frac{1}{2} a(x)^2 h + 2 \rho a(x)^2 \oint dx' a(x')^2 ) + \nu \frac{h^2}{4} 
\label{eqlongrangepotential}
\end{align}
where $H(a) = \int da \, a^2/(1 +  a^2)$. What this does in essence is to put a long range anti-ferromagnetic potential to compensate for the $h$ which is introduced into the potential with the wrong sign thus resolving the sign difficulty. This works in practice but is biologically implausible and does not get us much more than the original formulation of the problem.

Finally, the solution as discussed in the main text is to focus only on the unstable manifold where it is much easier to write down a potential formulation. We can do the analysis in a general form~\cite{haken1978analytical} writing the equations as
\begin{equation}
\dot{\vec{a}}(k) = M(k) \vec{a}(k) + \vec{f}(\vec{a(k)}),
\end{equation}
where we have combined the variables into the vector $\vec{a} \equiv (a-a^*, h-h^*)$. We assume that we are working in coordinates around the homogeneous fixed point and the linearization is given by the matrix $M(k)$ which is dependent on the Fourier mode $k$. $\vec{f}$ contains all nonlinearities. We can find eigenvectors of the matrix $M(k)$ which we denote by $\vec{u}^j_k$ where the index on top denotes the eigenvector and $k$ denotes the Fourier coefficient. Let us make the decomposition
\begin{equation}
\vec{a}(x) = \sum_j \sum_k \vec{z}^j_k e^{i k x} .
\end{equation}
$\vec{a}$ is real, so $\vec{z}^j_k = (\vec{z}^j_{-k})^*$. Then, putting this form in the expression above, we get
\begin{equation}
\sum_{j, k} \dot{\vec{z}}^j_k e^{i k x} = M(k) \sum_{j, k} \vec{z}^j_k e^{i k x} + \vec{f}(\vec{a(k)}) .
\end{equation}
We have two eigenvectors for each value of $k$, one of which has a large and negative eigenvalue (corresponding to relaxation of $h$) and the other has an eigenvalue which is relatively small but either positive (if the mode is unstable) or negative (if the mode is stable). We can neglect the large and negative eigenvalue and convert the above vector equation into a scalar equation with the understanding that scalar $z$ and scalar $f$ refer to the dominant eigenvectors and the component of the nonlinearity in its direction respectively. Then we get the equation
\begin{equation}
\sum_k \dot{z}_k e^{i k x} = \sum_k \lambda_k z_k e^{i k x} + f(\sum_{k'} z_{k'} e^{i k' x}) .
\end{equation}
We can project out the coefficient $k$ with the operator $1/(2 \pi) \oint dx e^{-i k x}$ on both sides
\begin{equation}
\dot{z}_k = \lambda_k z_k + \frac{1}{2 \pi} \oint dx e^{-i k x} f (\sum_{k'} z_{k'} e^{i k' x}) .
\end{equation}
Let the nonlinearity be of the form $f(x) = \tilde{f}_2 x^2/2 + \tilde{f}_3 x^3/3$ to cubic order. Further, using the fact that $\frac{1}{2 \pi} \oint e^{- i k x} dx = \delta_{k}$, we get 
\begin{align}
\dot{z}_k =& \lambda_k z_k + \sum_{k', k''} \frac{\tilde{f}_2}{2} z_{k'} z_{k''} \delta_{k' + k'' - k} \nonumber \\ &+ \sum_{k', k'', k'''} \frac{\tilde{f}_3}{3} z_{k'} z_{k''} z_{k'''} \delta_{k' + k'' + k''' - k} .
\end{align}
Consider now a single mode $k = 1$ to be unstable with all other modes stable. Evidently there are no quadratic terms possible. A single cubic term is possible with $k' = 1, k'' = -1, k''' = 1$ and corresponding permutations. This gives the equation
\begin{equation}
\dot{z}_1 = \lambda_1 z_1 + \tilde{f}_3 z_1 |z_1|^2 .
\end{equation}
Now, suppose there are two unstable modes $k = 1, 2$. At cubic order it is now possible to have $k' = 1, k'' = -2, k''' = 2$ with permutations in the equation for $z_1$. There are twice as many permutations of this than with $k' = 1, k'' = -1, k''' = 1$. At the quadratic order, it is possible to have $k' = 2, k'' = -1$ in the equation for $z_1$. One can thus count all possibilities to get the equation
\begin{align}
\dot{z}_1 &= \lambda_1 z_1 + \tilde{f}_3 z_1 (|z_1|^2 + 2 |z_2|^2) + 2 \tilde{f}_2  z_1^* z_2 , \\
\dot{z}_2 &= \lambda_2 z_2 + \tilde{f}_3 z_2 (|z_2|^2 + 2 |z_1|^2) + \tilde{f}_2 z_1^2 .
\end{align}
Notice that this happens to be potential, not requiring a metric. We can rename parameters to put them in the same form as in the main text. 

\begin{equation}
\begin{aligned}
	\dot{z}_1 &= \lambda_1 z_1 - \beta z_1 (|z_1|^2 + 2 |z_2|^2) - 2 \gamma  z_1^* z_2 , \\
	\dot{z}_2 &= \lambda_2 z_2 - \beta z_2 (|z_2|^2 + 2 |z_1|^2) - \gamma z_1^2 .
\label{equ:turing2modessup}
\end{aligned}
\end{equation}

For our simulations, we start with initial conditions close to the homogenous state with a few Fourier modes added with small random coefficients. The $a(x,t)$ obtained for several different initial conditions is linearly projected on to $\exp(i x)$ and $\exp(2 i x)$ to obtain our numerical approximation to $z_1$ and $z_2$. We then fit to the parameters $\lambda_1$, $\lambda_2$, $\beta$ and $\gamma$ which are independent of the initial conditions by fitting our projection to the dynamics in Equation~\ref{equ:turing2modessup}.  These parameters are not difficult to fit for they are directly related to the linear growth rate and the height of the final stable equilibrium. With these values of the parameters, we can then approximate the continuous dynamics as the real part of $z_1(t) \exp(i x) + z_2(t) \exp(2 i x)$. Note that we are using a very simple linear projection on to Fourier modes as an approximation to the unstable manifold which is quite flat. For early times, our approximation suffices.

For mid to late times, the inaccuracies in our Fourier approximation have to do with the other modes, and higher order terms in Equation~\ref{equ:turing2modessup}.  There are several different possibilites to improve the fit. The final stable state is given by a Jacobi Elliptic function which also have a natural interpolation from  trigonometric functions. It is possible to use some such interpolation but in practice, we find that an adequate solution is simply to saturate our sum of two modes with a simple sigmoidal function. The sigmoidal function is made a function of the real radial part of $z_1$ and $z_2$ which we denote by $r_1$ and $r_2$. It is defined as $\sigma(y, r_1, r_2) = \sigma_0(r_1, r_2)/(1 + \exp(-\Lambda (r_1, r_2) (y - y_0 (r_1, r_2)))+ 1)$ whose parameters $\Lambda$, $\sigma_0$ and $y_0$ are functions of $r_1$ and $r_2$ and are fit as explained below. Note that $y_0$ and $\sigma_0$ simply reflect the center and the height of the function. Hence our final approximation to the function $a(x)$ is $\sigma(\mathrm{Re}[z_1(t) \exp(i x) + z_2(t) \exp(2 i x)])$ with parameters in $\sigma$ a function of $r_1$ and $r_2$. The sigmoidal function ensures that the function remains positive and was empirically found to fit well. While the final form of the fit may look complicated to do exactly, it can be approximated in practice as follows. 

First we take the linear form of the fit mentioned above for one initial condition which gives us an approximation $a_{\textrm{l}} (x, t) = \mathrm{Re}[z_1(t) \exp(i x) + z_2(t) \exp(2 i x)]$. We then plot the actual value of the function $a(x, t)$ for the same initial condition as a function of $a_{\textrm{l}}(x, t)$ for a particular time point $t$. This data can be fit to a sigmoidal function $\sigma(y)$ as defined above. We can then find the parameters of the sigmoidal function as a function of $t$ for discrete time points chosen appropriately. The parameters can then be plotted as a function of $r_1$ and $r_2$ by using our known form for $r_1(t)$ and $r_2(t)$ from our linear fit. We can then approximate the parameters as functions of $r_1$ and $r_2$. In practice we found that $\sigma_0$ could be taken to be constant. The parameter $y_0$ was either kept constant (done when fitting either 1 or 2 modes in one spatial dimension) or taken to linearly vary with radius (done when fitting 1 mode in 2 spatial dimensions). Finally the parameter $\Lambda$ was also fit as a polynomial in $r_1$ and $r_2$. Details for the particular cases considered in the paper are given in the captions. We emphasize that our fits are only meant for the precision of a few percent which suffices for the biology under consideration. It is possible that other functional forms may work equally well or even better but in essence we require a mapping from the abstract coordinates $z_1$ and $z_2$ to the physical coordinates where the only time-dependence enters through $z_1$ and $z_2$ themselves. For the case of just one unstable mode, the discussion is entirely analogous to the one above. In summary, we have reduced a very high dimensional problem to a relatively simple fitting exercise in two dimensions involving only Fourier coefficients and a simple sigmoidal saturation function.

If a different set of Fourier modes are unstable, then one could work out the equations as above. Some couplings between the two modes depend on the specific wave number but others are allowed for all pair of modes.

\begin{figure}[h]
\centering
\includegraphics[width=.99\linewidth]{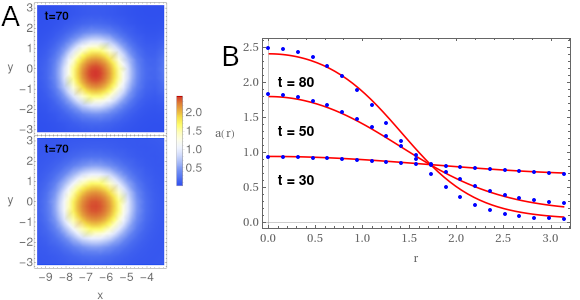}
\caption{A comparison is shown between the actual dynamics and the potential fit to the two dimensional Turing system. A) The first blob is shown on top at time $t=70$ along with the comparison to the potential fit at the same time below it. B) The radial cross-section is shown at three different times. The actual data is shown at discrete time point in blue and the potential fit is shown in red. The function fit is of the form $0.8+r_1(t) \sigma(1.5 J_0(r))$ where $r$ is the radial coordinate in physical space, $r_1(t)$ is the abstract radius and $\sigma$ is the saturation function as defined in the text. For the saturation function we use $y_0 = 1.26+0.5 r_1(t)$ and $\Lambda = 1 + 0.2 r_1(t)$. 
}
\label{fig:spotsfitcom}
\end{figure}

For the two spatial dimensional version of the problem, we use a slightly different set of equations 

\begin{align}
 \dot{a} &= \frac{a^2}{(1 + K a^2)(1 + h)} - \nu a + D_a (\partial_x^2 a + \partial_y^2 a) - \alpha a s(y) , \\
 \dot{h} &= \rho(x, y) a^2 - h + D_h (\partial_x^2 h + \partial_y^2 h) - \alpha h s(y)
 \label{GMeqns2d}
\end{align}

equal diffusion coefficients in $x$ and $y$. The system has length $L_x = 6 \pi$ in the $x$ direction and $L_y = 2 \pi$ in the $y$ direction both with the center at 0. There are periodic boundary conditions in the $x$ and $y$ directions albeit with a sink at the $y$ boundaries to localize the blobs. Parameter values are $D_a = 0.032, D_h = 5, \nu = 0.2, K = 0.2, \alpha = 0.005, s(y) = \exp(-(L_y^2 - y^2))$ and $\rho(x,y) = 5 + \epsilon \sum_{n=5}^{7} a_n/n \cos(2 \pi n x) \exp(-y^2)$ with $\epsilon = 0.1$ and $a_n$ randomly chosen from a uniform distribution between $-1$ and $1$. For the projection and fitting, we first average over the angular variable from the center of each blob. We then fit the function to a Fourier-Bessel series keeping only the first mode. The comparison between the fit and the actual dynamics is shown in Figure~\ref{fig:spotsfitcom}.

\begin{figure}[h]
\centering
\includegraphics[width=.99\linewidth]{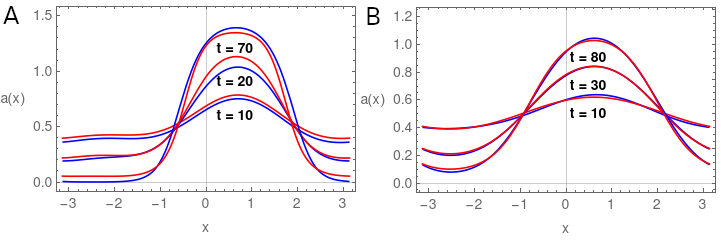}
\caption{A comparison between having one unstable mode or multiple unstable modes shown for random initial conditions. The simulated Turing system is shown in blue and the potential fit is shown in red. (A) There are three unstable modes in the actual system with $D_a = 0.01$ but a model with 2 unstable modes does a good job of fitting it. Other parameter values are $\nu = 0.2$, $\rho = 5$, $D_h = 20$. The parameters for the potential fit are $\alpha = 0.128, \alpha_2 = 0.0796, \gamma = 0.042, \beta = 0.125$. The saturation function is chosen to have $\sigma_0 = $ maximum of $a(x)$, $y_0 = \sigma_0/2$, $\Lambda(r_1, r_2) = 1, \sqrt{r_1^2 + r_2^2} <= 0.52$ and $\Lambda(r_1, r_2) = 1+ 4 (\sqrt{r_1^2 + r_2^2}-0.52)^2$ otherwise. (B) There is only one unstable mode with $D_a = 0.05$. The parameters for the potential fit are $\alpha_1 = 0.061$, $\beta = 0.155$. Here the saturation function has parameters values $\sigma_0 = 1.27$, $y_0 = 0.66$ and $\Lambda = 1.26 + 0.46 r_1$. 
}
\label{fig:imgcomp2}
\end{figure}

 \FloatBarrier

\section{Spreading the pattern by a wave}
The model solved in Fig.~\ref{flyeye} is defined by

\begin{equation}
\begin{aligned}
	u(a,b) =& \int_a(a(a-s)(a - p_1))+ p_2(b - a^2)^2/2 - p_3a b, \\
	U =& \sum_{i=1}^N u(a_i,b_i) + D_b \sum_1^{n+1}(b_i - b_{i-1})^2/2 \\&+  \frac{p_4}{2} \sum_{i \ne j} d(a_i) K_{i,j} d(a_j),\\
	K_{i,j} =& \exp (-(i-j)^2/(2 \sigma^2)).
\end{aligned}
\label{equ:flyeyeSI}
\end{equation}

\noindent
where $s=0.3, p_{1,2,3,4} = (2, 0.3, 0.2, 0.5), D_b=4, \sigma^2=2$.

There are a total of 7 possible parameters modulo rescalings in the functional form of Eq.\ref{equ:flyeyeSI} and it is fairly simple to adjust them. The origin is a stable point of $u(a,b)$ so we impose $s p_1 p_2 - p_3^2 > 0$; $\sigma $ controls the spacing between the cells with $a \sim 1$; $p_1>1$ increases the value of $a$ in active cells and thus makes the potential in the symmetry broken state more negative (making $p_3$ larger has the same effect but is limited by the determinant condition. The $p_2$ parameter weights the induction of $h$ by $a$. Too large a value damps $b$ and prevents the wave from propagating, a low value of $0.05$ activates extra sites.  A larger value of $s$ destroys the wave since cells do not get over the saddle to activate, too small a value and the sites well ahead of the furrow are susceptible to noise. The diffusion constant should be increased until the wave begins to propagate. The parameters shown in Fig.~\ref{flyeye} were adjusted by inspection following these rules. 

\begin{figure}[h]
\centering
	\includegraphics[scale=0.40]{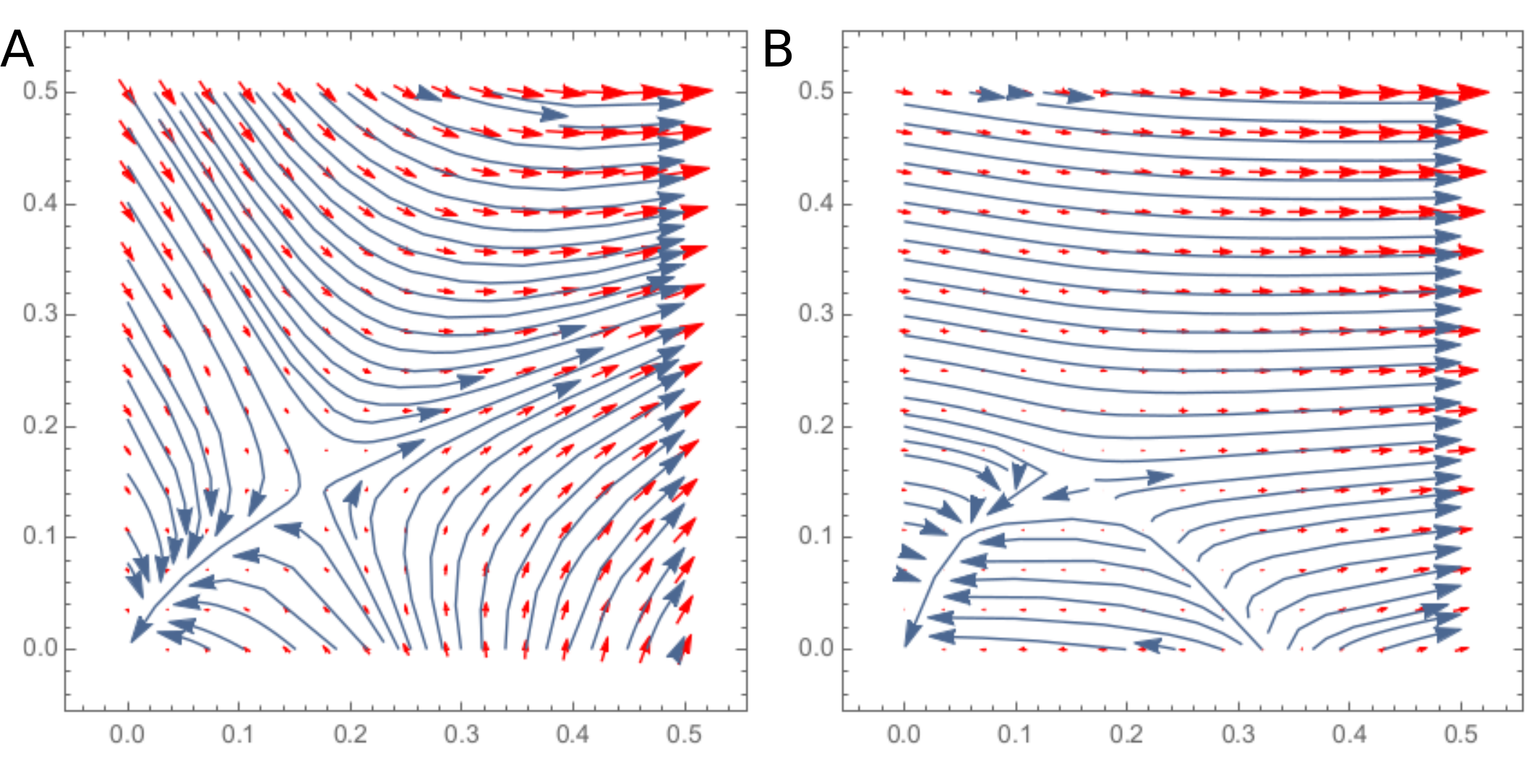}
\caption{The flow around the saddle point defined by $u(a,b)$ in Eq.\ref{equ:flyeyeSI} for two values of the metric, $g=(1,1)$ in (A), and (10,1) in (B) for the same potential. The saddle point occurs at $a,b \sim (0.17,0.14)$ and the upper fixed point for $u$ at $(2,7, 9.2)$. In the coupled system $a,b \lessapprox (1.5,1)$ so the diffusion has a substantial impact on the realized levels.
}
\label{flyeyeSIstreamAH}
\end{figure}

The flow around the saddle controls how the diffusive spreading of $b$ interacts with the repulsion from the previous active cell to select the next active cell. The inset to the saddle can be substantially manipulated by a constant metric, Fig.~\ref{flyeyeSIstreamAH}. If we lower its intersection with the $b$ axis for $a \sim 0$ less diffusive transport of $b$ is required to trigger the ON state. 

\FloatBarrier
 
\section{Morphogen models}
An embryo has to solve the problem of expressing the genes that define cellular fates in the correct position. A classic model for establishing a one dimensional pattern posits a static protein gradient that activates target genes as a function of concentration (aka "French flag") \cite{wolpert2015principles} . Within this simple paradigm are several different dynamical models as we now enumerate in the minimal case of three target genes, $X,Y,Z$ that are positioned from high to low morphogen respectively.  

\begin{figure}[h]
\centering
\includegraphics[width=.99\linewidth]{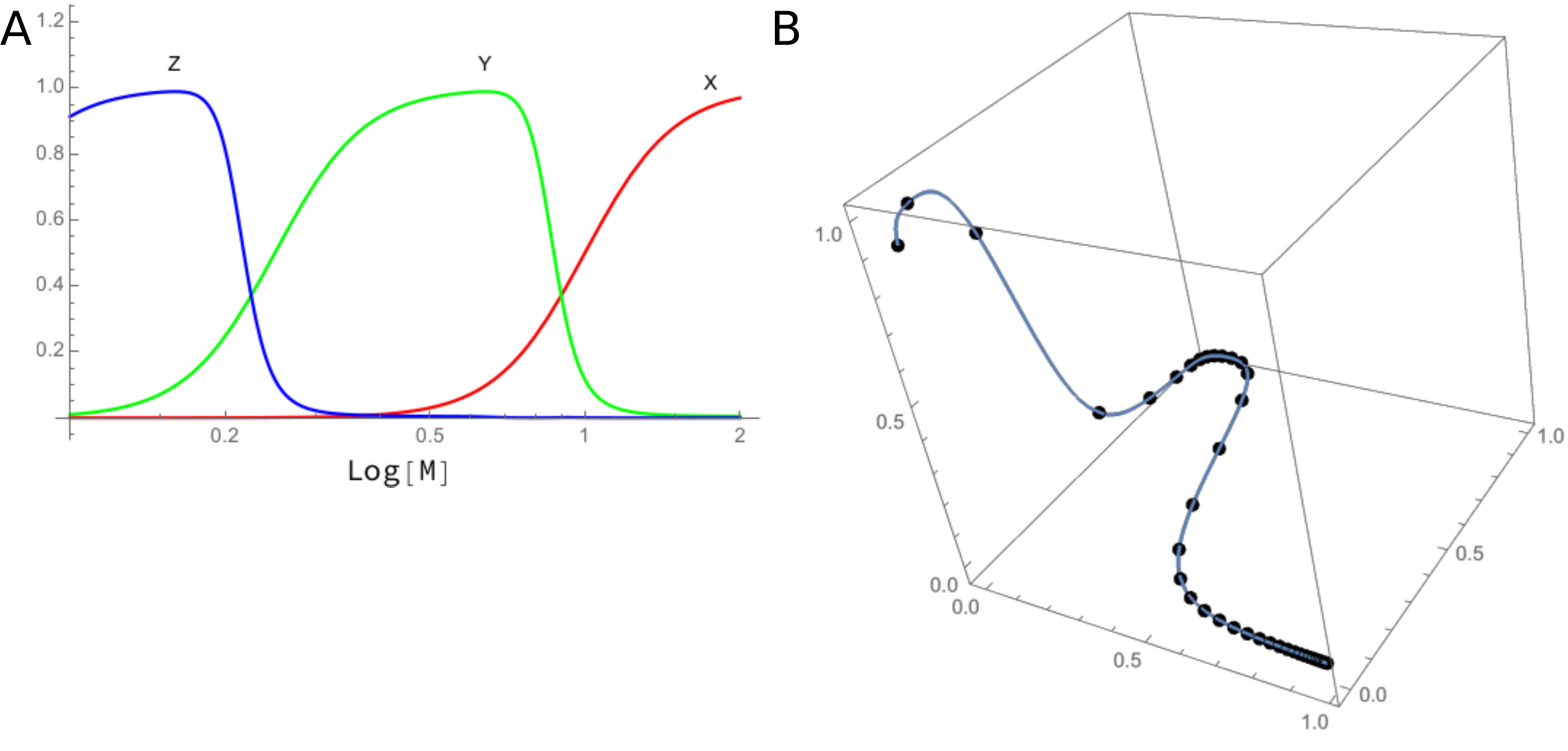}
\caption{Expression of three genes in an exponential morphogen gradient. (A) Expression as a function of position which is equivalent to log of the morphogen. (B) The same data as in (A) shown in three dimensions with the points uniformly spaced in morphogen level from 0.1 to 2 giving a sense of the `speed' of the orbit in the configuration space.
}
\label{morphogenMonoStable}
\end{figure}

The simplest model assumes the morphogen activates all three genes with different thresholds and in addition $X$ inhibits $Y,Z$ and $Y$ inhibits $Z$ i.e., 

\begin{equation}
\begin{aligned}
	\dot x &= \frac{M^n}{M^n+1} - x, \\
	\dot y &= \frac{M^n}{M^n+\lambda^n} \frac{1}{1+(\frac{x}{\lambda})^n} - y, \\
	\dot z &= \frac{M^n}{M^n+\lambda^{2n}} \frac{1}{(1+(\frac{x}{\lambda^2})^n)(1+(\frac{y}{\lambda})^n)} - z.
\end{aligned}
\label{equ:morphogenMonoStableSI}
\end{equation}

\noindent
that we solve with $n=5$ and $\lambda = \frac{1}{4}$. The inhibition is necessary in Eq.\ref{equ:morphogenMonoStableSI} so that the gene most sensitive to the morphogen is only expressed furthest away from the morphogen source. The scale factors are organized to make the domains of $x,y,z$ occupy equal regions in $\log(M)$ or position assuming an exponential profile of morphogen as has been accurately shown for  Bicoid in the fly blastoderm \cite{gregor2007stability}. Thus the scale at which $x$ inhibits $z$ is the square of the point at which $x$ inhibits $y$.
The result is shown in Fig.\ref{morphogenMonoStable},

Dynamically the model in Fig.\ref{morphogenMonoStable} is very boring, there is a single fixed point that is pushed around in three dimensions by the morphogen. New phenomena arise when the interactions among targets of the morphogen interact to generate bi or multi-stability. One such model for the mouse neural tube, \cite{balaskas2012gene}, considers three genes patterned as a function of distance from a ventral signaling source. To fit their data which includes mutating the gene Y positioned at intermediate morphogen levels, they needed to include more cross repressions that resulted in bistability at intermediate morphogen levels. As a result they observed hysteresis, so that the boundaries in the final pattern depend on the temporal profile of the morphogen. Their model is,

\begin{equation}
\begin{aligned}
	\dot x &= \frac{5 M}{M+1}  \frac{1}{1+y+z}- x, \\
	\dot y &= \frac{5 M}{M+1} \frac{1}{1 + x^5} - y, \\
	\dot z &= \frac{3}{1+x^6+y^2} - z.
\end{aligned}
\label{equ:morphogenBiStableSI}
\end{equation}

\noindent
where $x,y,z$ represent the genes Nkx2.2, Olig2, and Pax6.

\begin{figure}[h]
\centering
\includegraphics[width=.99\linewidth]{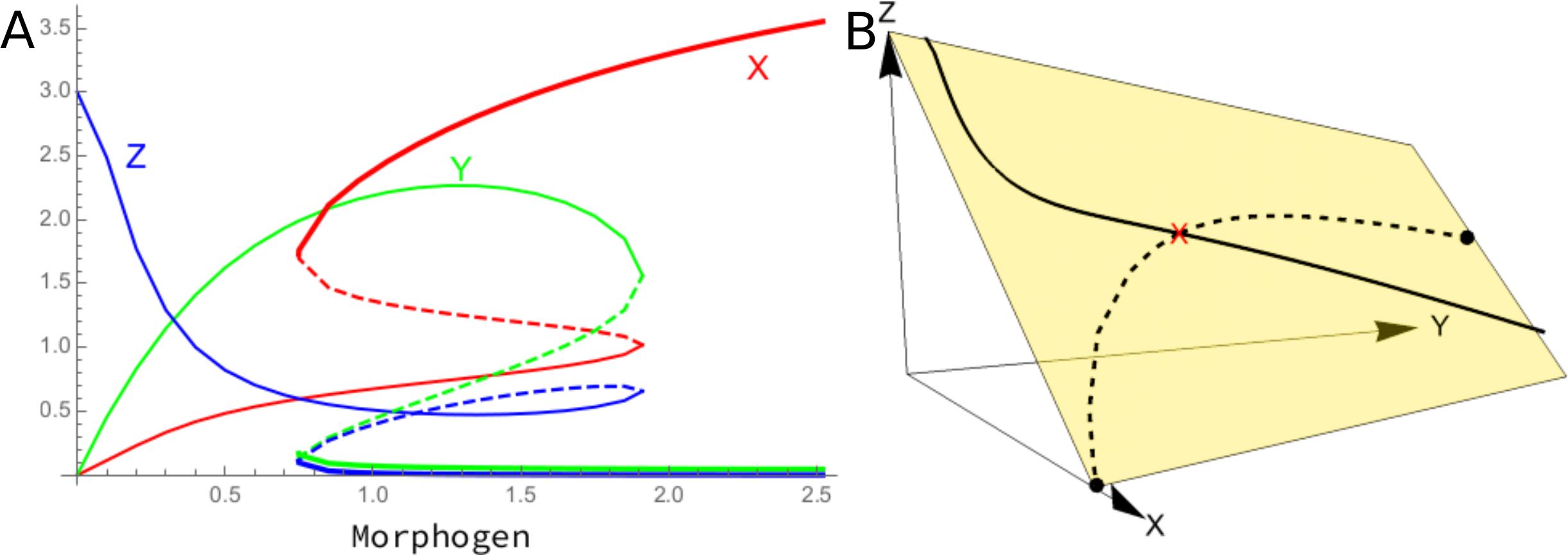}
\caption{Three genes in a morphogen gradient with bistability, Eq.\ref{equ:morphogenBiStableSI}. (A) The branch extending from low morphogen disappears at 1.91, while the upper branch disappears at 0.75. The unstable solution is shown dashed. (B) For the morphogen equal to 1.5, the two fixed points (dots) and the saddle (X) define a plane. The 1D unstable manifold of the saddle (dashed), and the most stable mode of the saddle (black) are projected onto the plane.
}
\label{morphogenBiStable}
\end{figure}

Figure \ref{morphogenBiStable} represents the solution. The bistability is a function of the coupled system, hence in a plot of gene level vs morphogen the saddle node bifurcations appear at the same abscissa for all genes. From the plot we infer that large $X$ strongly inhibits $Y,Z$ and $Y$ inhibits the other two. We can directly replot plot Fig.\ref{morphogenBiStable}A in 3D as three curves parameterized by the morphogen. However it is more informative to represent the the two dimensional stable manifold of the saddle point, which divides the basins of the high and low $X$. Then as the morphogen changes, this surface and the two attracting fixed points move as does the state of the cell, which may transit between basins. An impression for the basins of attraction at fixed morphogen is conveyed by intersecting the two dimensional inset to the saddle with the plane defined by the saddle and two fixed points. It does not suffice to compute the stable and unstable manifold of the saddle restricted to this plane since in the case shown in Fig.\ref{morphogenBiStable}B the planar equations have 9 restpoints in contrast to the 3 restpoints in the complete system. We thus adapt the expedient of just showing a projection.

The segmental patterning of the blastoderm in {\it Drosophila} is the paradigm gene regulatory network, and the control of the gap genes by the anterior morphogen Bicoid has been the subject of several modeling papers \cite{jaeger2004dynamical, papatsenko2011drosophila, petkova2019optimal}. Only the second of these makes any mention of bistability. We consider multi-stability or its absence a salient feature of any gene network model.


\FloatBarrier

\section{Adaptive systems}
One would be hard pressed to find a static morphogen gradient in the early developmental stages in any vertebrate system. 
However, in the embryo positional information can perfectly well be transmitted to the cell by morphogen spreading from a source if the cellular response is adaptive, i.e., a smoothed time derivative \cite{sorre2014encodingsup, yoney2018wnt, heemskerk2019rapid}.

In mathematical terms an adaptive system is one with a single attracting fixed point at which one of the variables is independent of a parameter representing the external input and the others act as a buffer. A simple example from \cite{francois2008case} will fix the ideas,

\begin{equation}
\begin{aligned}
	\dot a &= 1 - s a, \\
	\dot h &= s a - h.
	\end{aligned}
\label{eqn:adaptation}
\end{equation}

\noindent
where $s$ is a time dependent signal and $h$ is the adaptive response and $a$ the buffer. When $s$ is constant, $h=1$ irrespective of $s$. The system is linear so by matrix manipulations it can be written as the product of a positive definite metric times the gradient of a quadratic polynomial. 
But in the case of Eq.\ref{eqn:adaptation} rewriting it in this way will put $s$ in both the metric and potential, thereby obscuring the simplicity of Eq.\ref{eqn:adaptation}. Better is to write families of adaptive systems based on Eq.~\ref{eqn:adaptation} by using the potential $U = a - s a^2/2 + (1-h)^2$ and a metric with an off diagonal term. The potential naturally makes the fixed point of $h$ independent of $s$, and the metric couples $a$ and $h$ when they vary.
